\documentclass[preprint,eqsecnum,showpacs,floatfix]{revtex4}
\usepackage{graphicx}
\usepackage{epsf}
\usepackage{amssymb}
\usepackage{bm}
\usepackage[hypertex]{hyperref}
\def\he#1{$^{#1}$He}

\def\bra#1{\bigl\langle {#1}\bigr|}
\def\ket#1{\bigl| {#1} \bigr\rangle}
\def\rvec {{\bf r}}
\def\pvec {{\bf p}}
\def\hvec {{\bf h}}
\def\qvec {{\bf q}}

\def\qG {\tilde\Gamma_{\!\rm dd}}
\def\qX {\tilde X_{\rm dd}}
\def\vec#1{{\bf #1}}				

\def\perz#1{\alpha_{#1}^{\dagger}}
\def\pver#1{\alpha_{#1}^{\phantom{\dagger}}}       
\def\qerz#1{a_{#1}^{\dagger}}
\def\qver#1{a_{#1}^{\phantom{\dagger}}}            
\newcommand{\I}    [0] {{\rm i}}            
\newcommand{\qkf}  [0] {k_{\rm F}}          
\newcommand{\omP}  [0] {\omega_\mathrm{p}}  
\newcommand{\Vph} [1] 
           {{\widetilde V_{\rm\scriptscriptstyle p_{\bar{\ }\!}h\!}(#1)}}
\newcommand{\SF}      [1] {S_{\rm\scriptscriptstyle F}(#1)}
\newcommand{\SFthree} [0] {S_{\rm\scriptscriptstyle F}^{(3)}}
\begin{document}

\title{Dynamic Many-Body Theory. II. Dynamics of
          Strongly Correlated Fermi Fluids}

\author{H.\ M.\ B\"ohm$^\dagger$, R.\ Holler$^\dagger$,
E.\ Krotscheck$^{\dagger +}$ and M. Panholzer$^\dagger$}

\affiliation{$^\dagger$Institut f\"ur Theoretische Physik, Johannes
Kepler Universit\"at, A 4040 Linz, Austria}
\affiliation{$^+$Department of Physics, University at Buffalo, SUNY
Buffalo NY 14260}

\begin{abstract}
  We develop a systematic theory of multi-particle
  excitations in strongly interacting Fermi systems. Our work is the
  generalization of the time-honored work by Jackson, Feenberg, and
  Campbell for bosons, that provides, in its most advanced
  implementation, quantitative predictions for the dynamic structure
  function in the whole experimentally accessible energy/momentum
  regime. Our view is that the same physical effects -- namely
  fluctuations of the wave function at an atomic length scale -- are
  responsible for the correct energetics of the excitations in both
  Bose and Fermi fluids.  Besides a comprehensive derivation of the
  fermion version of the theory and discussion of the approximations
  made, we present results for homogeneous \he3 and electrons in three
  dimensions. We find indeed a significant lowering of the zero sound
  mode in \he3 and a broadening of the collective mode due to the
  coupling to particle-hole excitations in good agreement with
  experiments. The most visible effect in electronic systems is the
  appearance of a ``double-plasmon'' excitation.
\end{abstract}

\maketitle
\clearpage

\section{Introduction}
\label{sec:intro}

This paper is concerned with a systematic theory of multi-particle
excitations in Fermi systems. We utilize an equations of motion method
that has been used in the past as a vehicle for many purposes: the
derivation of the time--dependent Hartree--Fock (TDHF) theory
\cite{KermanKoonin,KramerSaraceno,ThoulessBook}, its analog for
strongly interacting systems \cite{LDavid,rings}, and for studying
single-- and multi--particle correlations in strongly interacting Bose
liquids \cite{JaFe,eomI}.

The simplest way to deal with excitations is to assume that the
low--lying excited states of a quantum fluid can be characterized by
the quantum numbers of a single particle. This is the core idea of
Landau's quasiparticle picture of ``normal'' quantum fluids
\cite{LandauFLP1,LandauFLP2} as well as of Feynman's theory of
collective modes in the helium liquids \cite{Feynman}. It is
appropriate for many long wavelengths excitations such as sound waves
in Bose fluids or plasmons in an electron liquid.

Already Feynman realized that this concept is insufficient to describe
higher--lying excitations, most prominently the ``roton'' in
\he4. Intuitively appealing, he introduced ``backflow'' correlations
\cite{FeynmanBackflow}.  These are recognizable as a new type of
excitations, depending on two particles: {\it pair
  fluctuations}\/. The notion is plausible: For excitations at
wavelengths comparable to the interparticle distance, the
time--dependence of a system's {\it short--ranged\/} structure is
expected to be relevant.

The presently state-of-the-art theory for Bose liquids originates from
pioneering studies by Jackson, Feenberg
\cite{JaFe,JaFe2,FeenbergBook,Jackson71,JacksonSkw,JacksonSumrules},
and Campbell and collaborators \cite{Chuckphonon}. Recently, a
complete solution of the pair equation of motion has been accomplished
in \he4 \cite{eomI}, showing that the ``uniform limit approximation''
of
Refs.~\onlinecite{JaFe,JaFe2,FeenbergBook,Jackson71,JacksonSkw,JacksonSumrules,Chuckphonon}
is surprisingly good. Consequently, theoretical improvement must be
sought in three-body and higher-order fluctuations \cite{eomII}.

Although quite successful for bosons, there exists to-date no fermion
version of the theory. We therefore develop here the generalization of
the equation of motion method for pair fluctuations to fermions. We
calculate the fermionic density--density response function
$\chi(\rvec\!-\!\rvec';\,t\!-\!t')$, relating the induced density
fluctuation $\delta\rho(\rvec;t)$ to a weak external perturbation
$h_{\rm ext\!}(\rvec;t)$.  In a homogeneous system this is written in
momentum space as
\begin{equation}
\delta\rho({\bf q};\omega) 
	= \rho\> \chi(q;\omega)\, \tilde h_{\rm ext}({\bf q;\omega}) \,,
\label{eq:chidef}
\end{equation}
where $\rho$ is the particle number $N$ per volume $\Omega$. We choose 
Fourier transforms
\begin{equation}
   f(\rvec; \omega)
   \>\equiv\> \displaystyle\frac{1}{N}\displaystyle\!\sum_\qvec 
             e^{-\I \qvec\cdot\rvec}\, \tilde f(\qvec;\omega) 
\label{eq:FTdef}
\end{equation}
to have the same dimension in $\qvec$- and in $\rvec$-space. 

The imaginary part of $\chi(q;\omega)$ is the experimentally 
accessible dynamic structure factor, 
\begin{equation}
 S(q;\omega)=-\frac{\hbar}{\pi}\Im m[\chi(q;\omega)]\,\theta(\omega) \;.
 \label{eq:FD}
\end{equation} 
The dynamic structure factor satisfies, amongst others, the
sum rules
\begin{eqnarray}
 m_0=S(q)&=&\int_0^{\infty}d\hbar\omega \;S(q; \omega), 
 \label{eq:m0}\\
 m_1=\frac{\hbar^2q^2}{2m}&=&\int_0^{\infty}d\hbar\omega \;\hbar\omega \,
 S(q; \omega)
 \label{eq:m1} \,,
\end{eqnarray}
where $S(q)$ is the static structure factor.

We develop our theory with the following objectives:
\begin{itemize}
\item {} Technically, the extension of the Jackson--Feenberg--Campbell
  theory to Fermi systems amounts to including time--dependent {\it
    two--particle--two--hole\/} excitations. We require that the
  fermionic $\chi(q;\omega)$ reduces to that of the boson theory in
  the appropriate limit.
\item {} For bosons, neglecting pair- and higher order fluctuations
  yields the famous Bijl-Feynman spectrum \cite{Feynman} 
  \begin{equation}
    \varepsilon(q) = \frac{\hbar^2 q^2}{2m S(q)} \equiv \frac{t(q)}{S(q)}\,.
  \label{eq:efeyn}
  \end{equation}
  Its fermionic counterpart is the random--phase approximation (RPA),
  formulated in terms of effective interactions \cite{ALP78}.  We
  require that our theory reduces to the RPA if pair fluctuations are
  ignored. This implies, in particular, that we obtain in this case a
  response function of the form
  \begin{equation}
   \chi(q; \omega) = {\chi_0(q; \omega)\over 1 - \Vph{q}\,
   \chi_0(q; \omega)} \, .
  \label{eq:RPAresponse}
  \end{equation}
  Here, $\chi_0(q; \omega)$ is the
  Lindhard function and $\Vph{q}$ an appropriately defined static
  ``particle--hole interaction'' or ``pseudo-potential''.
\end{itemize}

One of the tasks of microscopic many--body theory is to justify and
calculate effective interactions such as $\Vph{q}$, as far as this is
possible. Using Jastrow--Feenberg correlation functions
\cite{FeenbergBook} to tame the microscopic hard--core repulsion, it
has been shown \cite{rings} under what assumptions a density response
function of the RPA form (\ref{eq:RPAresponse}) can be obtained, and a
microscopic expression for the static effective interaction $\Vph{q}$
was derived. Under what conditions a form (\ref{eq:RPAresponse}) is
meaningful at all will be discussed in depth below.

A phenomenological approach to define a particle-hole interaction or
``pseudo-potential'' for \he3 and electrons was introduced by Aldrich,
Iwamoto, and Pines \cite{ALP78,IP84}.  They determined the physically
intuitive and necessary requirements for $\Vph{q}$, postulating that
the dynamic response is given by the RPA form (\ref{eq:RPAresponse}).
Reflecting the same physics, the $\Vph{q}$ derived from microscopic
many-body theory \cite{rings} is very similar to the
Aldrich-Iwamoto-Pines pseudopotentials. The microscopic derivation
leads to a $\Vph{q}$ that is uniquely determined from the static
structure function by the two sum rules
(\ref{eq:m0})-(\ref{eq:m1}). Defining the RPA this way leads for
bosons to the Feynman approximation (\ref{eq:efeyn}) for the spectrum
of collective excitations. From here on, we will use the term ``RPA''
and ``Feynman spectrum'' in this sense.

Our work is organized as follows: Section \ref{sec:basics} introduces
the basic quantities and the most important tools of variational and
correlated basis function (CBF) theory.  For details, the reader is
referred to review articles \cite{Johnreview} and pedagogical material
\cite{KroTriesteBook}; a brief outline of our notations and
definitions is given in appendix \ref{asec:ground_state}.  Section
\ref{sec:EOMs} is the core of our work; it provides the derivation of
the equations of motion, including pair fluctuations.  We show
that the theory can be mapped onto a set of TDHF equations
\cite{ThoulessBook} with {\it energy-dependent, effective\/}
interactions. Thus, our work provides the logical generalization of
Ref.~\onlinecite{rings}, where single-particle fluctuations led to a
TDHF theory with {\it static\/} effective interactions.

Section \ref{sec:localapprox} focuses on the practical implementation
of our theory. We formulate, among others, the ``convolution
approximation'' for fermions. In Section \ref{sec:chi} we derive the
density-density response function $\chi(q;\omega)$ and discuss its
features.

Modern techniques of many-body theory are robust against the details
of the interparticle interaction. We can therefore use the methods
developed here to examine the dynamics of two very different systems:
The very strongly interacting \he3 whose interaction is characterized
by a repulsive hard core and a short-ranged attraction, and electrons
with their rather tame but long-ranged Coulomb interaction.  Section
\ref{sec:applications} implements our method for bulk \he3 and the
electron liquid. In \he3, we compare with neutron scattering
experiments carried out at the Institut Laue Langevin (ILL) in the
group led by R. Scherm \cite{SGFS87,Fak94,GFvDG00}. The
energetics of the collective mode as well as the width of the spectrum
at high momentum transfers are significantly improved compared to RPA
predictions.  In the homogeneous electron liquid the pair-excitation
theory predicts plasmon damping as well as double-plasmon excitations.
Experimental verification of the double-plasmon excitation in recent
inelastic X-ray scattering measurements \cite{SHV05,HSS08} has added
new interest in studying the dynamics of electrons.

Our results are summarized in Sec.~\ref{sec:Summary} where we also
discuss the directions of future work.

Appendices \ref{asec:ground_state}--\ref{asec:sum_rules} give further
details on the derivations, and Appendix \ref{asec:recipe} a very
brief summary of the minimal implementation of our theory.

\section{Theory for strongly interacting fermions}
\label{sec:basics}

\subsection{Variational theory}
\label{ssec:FHNC}

Microscopic many-body theory starts with a phenomenological
Hamiltonian for $N$ interacting fermions,
\begin{equation}
H = -\sum_i\frac{\hbar^2}{2m}\nabla_i^2 + \sum_{i<j}
v\left(\left|\rvec_i\!-\!\rvec_j\right|\right) \;.
\label{eq:Hamiltonian}
\end{equation}
For strong interactions, CBF theory \cite{FeenbergBook} has proved to
be an efficient and accurate method for obtaining ground state
properties. It starts with a variational wave function of the form
\begin{equation}
  \vert \Psi_{\bf o} \rangle = {F \; \vert \Phi_{\bf o} \rangle \over
  \langle \Phi_{\bf o} \vert\,  F^{\dagger} F^{\phantom{\dagger}}
    \vert \Phi_{\bf o} \rangle^{1/2} } \;,
\label{eq:JastrowWaveFunction}
\end{equation}
where $\Phi_{\bf o}(1,\ldots,i,\ldots,N)$ is a model state, normally a
Slater--determinant, and ``$i$'' is short for both spatial and $\nu$
discrete (spin and/or isospin) degrees of freedom. The {\it
  correlation operator}\/ $F(1,\ldots,N)$ is suitably chosen to
describe the important features of the interacting system. Most
practical and highly successful is the Jastrow--Feenberg
\cite{FeenbergBook} form
\begin{eqnarray}
	F(1,\ldots,N) &=&
	\exp\left\{{1\over2}\left[\sum_{1\le i<j\le N} u_2({\bf r}_i,{\bf r}_j)
	+ \sum_{1\le i<j<k\le N}u_3({\bf r}_i,{\bf r}_j,{\bf r}_k)
	+ \ldots\right]\right\} \,.
\label{eq:JastrowCorrelations}
\end{eqnarray}
The $u_n({\bf r}_1,\ldots,{\bf r}_n)$ are made unique by requiring
them to vanish for $|{\bf r}_i\!-\!{\bf r}_j| \to\!\infty$
(``cluster property''). 

From the wave function (\ref{eq:JastrowWaveFunction}),
(\ref{eq:JastrowCorrelations}), the energy expectation value
\begin{equation}
\label{eq:energy}
H_{\bf o,o}
\equiv \left\langle \Psi_{\bf o} \right| H \left| \Psi_{\bf o}\right \rangle
\end{equation}
can be calculated either by simulation or by integral equation
methods. The hierarchy of Fermi-Hypernetted-Chain (FHNC)
approximations is compatible with the optimization problem, {\it
  i.e.\/} with determining the optimal {\it correlation functions\/}
$u_n({\bf r}_1,\ldots,{\bf r}_n)$ through functionally minimizing the
energy
\begin{equation}
{\delta H_{\bf o,o}\over \delta u_n({\bf r}_1,\ldots,{\bf r}_n)} = 0 \,.
\label{eq:Euler}
\end{equation}
Due to the multitude of exchange diagrams, the Fermi-HNC (FHNC) and
corresponding Euler equations can be quite complicated \cite{polish};
the simplest approximation of the Euler equations (\ref{eq:Euler})
that contains the important physics is spelled out in
App.~\ref{assec:fhnc0}.

The optimization of the correlations also facilitates making
connections with other types of many-body theories, such as
Feynman-diagram based expansions and summations \cite{parquet1}.

\subsection{Correlated Basis Functions}
\label{ssec:CBF}

Although quite successful in predicting ground state properties of
strongly interacting systems, the Jastrow-Feenberg form
(\ref{eq:JastrowCorrelations}) of the correlation operator $F$ has
some deficiencies. The most obvious problem is that the nodes of the
wave function (\ref{eq:JastrowWaveFunction}) are identical to those of
the model state $\ket{\Phi_{\bf o}}$. To improve upon the description
of physics, CBF theory \cite{Johnreview,KroTriesteBook,polish} uses
the correlation operator $F$ to generate a complete set of correlated
and normalized $N$-particle basis states through
\begin{equation}
  \vert \Psi_{\bf m} \rangle = {F \; \vert \Phi_{\bf m} \rangle \over
  \langle \Phi_{\bf m} \vert  F^{\dagger} F \vert \Phi_{\bf m}
  \rangle^{1/2} } \;,
\label{eq:States}
\end{equation}
where the $\{\vert \Phi_{\bf m} \rangle\}$ form a complete basis
of model states.
Although the $\vert \Psi_{\bf m} \rangle$ are not orthogonal,
perturbation theory can be formulated in terms of these states
\cite{MF1,FeenbergBook}. We review here this method only very briefly,
details may be found in Refs. \onlinecite{Johnreview} and
\onlinecite{KroTriesteBook}; the diagrammatic construction of the
relevant ingredients is given in Ref. \onlinecite{CBF2}.

For economy of notation, we introduce a ``second--quantized''
formulation of the correlated states. The Jastrow--Feenberg
correlation operator in (\ref{eq:JastrowCorrelations}) explicitly
depends on the particle number, {\it i.e.\/}
$F=F_{\!\scriptscriptstyle N}(1,\ldots,N)$ (whenever unambiguous, we
omit the corresponding subscript). Starting from the conventional
$\qerz{k}, \qver{k}$, creation and annihilation operators
$\perz{k},\pver{k}$ of {\em correlated states\/} are defined by their
action on the basis states:
\begin{eqnarray}
        \bigl|\perz{k}\,\Psi_{\bf m}\bigr\rangle
        &\equiv\>&  F_{\!\!_{N+1}} \qerz{k} \,\ket {\Phi_{\bf m}} \,\Big/
        \bra {\Phi_{\bf m}} \qver{k} F_{\!\!_{N+1}}^\dagger 
         F_{\!\!_{N+1}}^{\phantom{\dagger}}
         \qerz{k}\ket {\Phi_{\bf m}}^{1/2} \, ,
\label{eq:creation}\\
        \bigl|\pver{k}\,\Psi_{\bf m}\bigr\rangle
        &\equiv\>&  F_{\!\!_{N-1}} \qver{k}\,\ket {\Phi_{\bf m}} \,\Big/
        \bra {\Phi_{\bf m}} \qerz{k} F_{\!\!_{N-1}}^\dagger 
        F_{\!\!_{N-1}}^{\phantom{\dagger}}
        a_{k}\ket {\Phi_{\bf m}}^{1/2} \,.
\label{eq:annihilation}\end{eqnarray}
According to these definitions,
$\alpha_{k}^\dagger$ and $\alpha^{\phantom{\dagger}}_{k}$ obey the same
(anti--) commutation rules as their uncorrelated cousins, {\it but
  they are not Hermitian conjugates.\/} If $\ket{\Psi_{\bf m}}$ is an
$N$--particle state, then the state in Eq.~(\ref{eq:creation}) must carry an
$(N\!+\!1)$-particle correlation operator, while that in
Eq.~(\ref{eq:annihilation})
must be formed with an $(N\!-\!1)$--particle correlation operator.

In general, we label ``hole'' states, which are occupied in $\vert
\Phi_{\bf o} \rangle$, by $h$, $h'$, $h_i\;, \ldots\,$, and unoccupied
``particle'' states by $p$, $p'$, $p_i\;,$ \textit{etc}.  To display the
particle-hole pairs explicitly, we will use alternatively to
$\bigl|\Psi_{\bf m}\bigr\rangle$ the notation $\bigl|\Psi_{p_1 \ldots
  p_d\, h_1\ldots h_d}\bigr\rangle $.  A basis state with $d$
particle-hole pairs is then
\begin{equation}
\bigl|\Psi_{p_1 \ldots p_d\, h_1\ldots h_d}\bigr\rangle 
=\perz{p_1}\cdots\perz{p_d}\pver{h_d}\cdots\pver{h_1}\ket{\Psi_{\bf o}}
 \,.
\label{eq:psimph}
\end{equation}

The execution of the theory needs the matrix elements of the
Hamiltonian, the unit operator, and the density operator.  Key
quantities are diagonal and off-diagonal matrix elements of unity and
$H'\!\equiv H\!-\!H_{{\bf o},{\bf o}}$
\begin{eqnarray}
M_{\bf m,n} &=& \langle \Psi_{\bf m} \vert \Psi_{\bf n} \rangle
\equiv \delta_{\bf m,n} +  N_{\bf m,n}\;,
\label{eq:defineNM}
\\
H'_{\bf m,n} &\equiv &
W_{\bf m,n} + {1\over2}\left(H_{\bf m,m}+H_{\bf n,n}-2H_{\bf o,o}
\right)N_{\bf m,n} \,.
\label{eq:defineW}
\end{eqnarray}
Eq. (\ref{eq:defineW}) defines a  natural decomposition \cite{CBF2,KroTrieste} 
of the matrix elements of $H'_{\bf m,n}$.

The ratios of normalization integrals,
$I_{\bf m,m} \equiv
\langle \Phi_{\bf m} \vert F^{\dagger} F \vert \Phi_{\bf m} \rangle $, 
define the factors
\begin{equation}
  z_{p_1\ldots p_d\, h_1\ldots h_d} \equiv\; z_{\bf m} \;\equiv
  \sqrt{I_{\bf m,m}/I_{\bf o,o}} \;.
\end{equation}
For large particle numbers and $d\!\ll\! N$ these factorize as
\begin{equation}
  z_{\bf m} 
 \;=\; 
 \frac{z_{p_1}\ldots z_{p_d}}{z_{h_1}\ldots z_{h_d}} \>+ {\cal O}(N^{-1})\,.
 \label{eq:zfac}
\end{equation}

Likewise, to leading
order in the particle number, the {\it diagonal\/} matrix elements of
$H'\!\equiv H\!-\!H_{{\bf o},{\bf o}}$ become additive, so that for
the above $d$-pair state we can define the CBF single particle
energies
\begin{equation}
 \bra{\Psi_{\bf m}} H' \ket{\Psi_{\bf m}} \>\equiv\>
 \sum_{i=1}^d e_{p_ih_i}  + {\cal O}(N^{-1}) \;,
\label{eq:CBFph}
\end{equation}
with $e_{ph} = e_p - e_h$.

For the off--diagonal elements $O_{\bf m,n}$ of an operator $O$
(specifically the Hamiltonian, the unit-, density- and current-operator) 
we sort the quantum numbers $m_i$ and $n_i$ such that $|\Psi_{\bf m} \rangle$ 
is mapped onto $\left|\Psi_{\bf n}\right\rangle$ by
\begin{equation}
  \label{eq:defwave}
  \left|\Psi_{\bf m}\right\rangle = \perz{m_1}\perz{m_2}
  \cdots 
     \perz{m_d} \; \pver{n_d} \cdots \pver{n_2}\pver{n_1}  
     \left|\Psi_{\bf n} \right\rangle \;.
\end{equation}
From this we recognize that, to leading order in $N$, any $O_{\bf
  m,n}$ depends only on the {\it difference\/} between the states
$\vert \Psi_{\bf m} \rangle$ and $\vert \Psi_{\bf n} \rangle$ and {\it
  not\/} on the states as a whole.  Consequently, $O_{\bf m,n}$ can be
written as matrix element of a $d$-body operator
\begin{equation}
  \label{eq:defmatrix}
  O_{\bf m,n} \equiv \langle m_1\, m_2 \, \ldots m_d \,| 
     {\cal O}(1,2,\ldots d) \,|n_1\,
  n_2 \, \ldots n_d\rangle_a \;.
\end{equation}
(The index $a$ indicates antisymmetrization.) 
According to (\ref{eq:defmatrix}),
$W_{{\bf m},{\bf n}}$  and $N_{{\bf m},{\bf n}}$ define 
$d-$particle operators ${\cal N}$ and ${\cal W}$, {\em e.g.\/}
\begin{eqnarray}
N_{{\bf m},{\bf o}} &\equiv& N_{p_1p_2\ldots p_d \,h_1h_2\ldots h_d,0}\equiv 
  \langle p_1p_2\ldots p_d \,|\, {\cal N}(1,2,\ldots,d)\,
  |\,h_1h_2\ldots h_d \rangle_a  \;,\nonumber\\
W_{{\bf m},{\bf o}} &\equiv& W_{p_1p_2\ldots p_d \,h_1h_2\ldots h_d,0}\equiv 
  \langle p_1p_2\ldots p_d \,|\, {\cal W}(1,2,\ldots,d)\,
  |\,h_1h_2\ldots h_d \rangle_a  \;.
\label{eq:NWop}
\end{eqnarray}
Diagrammatic representations of ${\cal N}(1,2,\ldots,d)$ and ${\cal
  W}(1,2,\ldots,d)$ have the same topology \cite{CBF2}.  In
homogeneous systems, the continuous parts of the $p_i,h_i$ are wave
numbers ${\bf p}_i,{\bf h}_i$; we abbreviate their difference as ${\bf
  q}_i$.  The highest occupied momentum is $\hbar k_{\rm F}$.

An important consideration is, for our purposes, the connection
between CBF matrix elements, the static structure function, and the
optimization conditions for the ground state.  The static structure
function $S(q)=\frac{1}{N} \bra{\Psi_{\bf o}} \hat\rho_{\vec
  q}\hat\rho_{-\vec q}\ket{\Psi_{\bf o}}$ is routinely obtained in
ground state calculations; for some systems it is also available from
experiments. We can also write $S(q)$ as the weighted average of the
matrix elements (\ref{eq:NWop}),
\begin{eqnarray}
S(q)  &=& \SF{q} +
  \frac{1}{N}\!\sum_{hh'}z_{pp'hh'}N_{pp'hh',0}\,.
\label{eq:SofkfromN}
\end{eqnarray}
where $\SF{q}$ is the static structure function of non-interacting
fermions.

Similarly, the optimization conditions (\ref{eq:Euler}) for the
pair correlation function can, in momentum space, be written
in terms of off-diagonal matrix elements of the Hamiltonian:
\begin{eqnarray}
0 ={\delta E \over \delta \tilde u_2({\bf q},{\bf q}')} &=&
\frac{\left\langle\Phi_{\bf o}\right| F^{\dagger}H' F 
          \left[\hat\rho_{{\bf q}}\hat\rho_{{\bf q}'}
            - \hat\rho_{{\bf q}+{\bf q}'}\right]
          \left|\Phi_{\bf o}\right\rangle}
   {\left\langle\Phi_{\bf o}\vert  F^{\dagger}F \vert\Phi_{\bf o}\right\rangle}
\nonumber\\
&=& \sum_{hh'} 
    \frac{\left\langle\Phi_{\bf o}\right| F^{\dagger}H' F 
          \left|\qerz{p'}\qerz{p}\qver{h}\qver{h'}\Phi_{\bf o}\right\rangle}
   {\left\langle\Phi_{\bf o}\vert  F^{\dagger}F \vert\Phi_{\bf
o}\right\rangle}
= \sum_{hh'}z_{pp'hh'}H'_{pp'hh',0}
\label{eq:zweicoll}
\end{eqnarray}
{\it i.e.\/} the weighted average of the
off-diagonal matrix elements $H'_{0,pp'hh'}$ vanishes for optimized
pair correlations. Both features will provide rules for systematic and
consistent approximation schemes for the operators
${\cal N}(1,2,\ldots,d)$ and ${\cal W}(1,2,\ldots,d)$.

\section{Equations of motion}
\label{sec:EOMs}

\subsection{Excitation operator and action principle}
\label{ssec:StrongAction}

To formulate a theory of excited states for strongly interacting fermions 
we generalize the ansatz (\ref{eq:JastrowWaveFunction}) in analogy to
the pair fluctuations theory for strongly interacting bosons
\cite{JaFe,JaFe2,Jackson71,JacksonSkw,JacksonSumrules,Chuckphonon,eomI}.
We restrict ourselves here to uniform systems.
The system is subjected to a small external perturbation
\begin{equation}
H_{\rm ext}(t) \equiv
\int\!\!d^3r\, h_{\rm ext}({\bf r};t)\, \hat\rho({\bf r})
\label{eq:Hext}
\end{equation}
where $\hat\rho(\rvec)$ is the density operator.  The correlated wave
function for the perturbed state is chosen to be
\begin{eqnarray}
\Big|\Psi(t)\Big\rangle\; &=&
         \exp\bigl[-\I H_{\bf{o},\bf{o}}t/\hbar\bigr]\,
        \Big|\Psi_0(t)\Big\rangle\;,\nonumber\\
\Big|\Psi_0(t)\Big\rangle
&=& {1\over I^{1/2}(t)}\exp\Bigl[\textstyle\frac{1}{2}\displaystyle U(t)\Bigr]\,
\Big|\Psi_{\bf o}\Big\rangle
\label{5.6.1}\\
I(t)&=&\Bigl\langle \Psi_{\bf o}\,\Big|\,
        \exp\Bigl[\textstyle\frac{1}{2}\displaystyle U^\dagger(t)\Bigr]
        \exp\Bigl[\textstyle\frac{1}{2}\displaystyle U(t)\Bigr]
        \,\Big|\,\Psi_{\bf o}\Bigr\rangle \;,\nonumber
\end{eqnarray}
with the excitation operator
\begin{eqnarray}
  U(t) &\equiv&  \sum_{ph}\; \delta u^{(1)}_{ph} (t)\; \perz p \pver h
 + {1\over 2}\;\sum_{pp'hh'}\;
   \delta u^{(2)}_{pp'hh'} (t)\; \perz p \perz {p'}
	\pver{h'} \pver{h}\nonumber\\
&\equiv& U_1(t) + U_2(t)\,.\label{eq:Uop}
\end{eqnarray}
The particle--hole amplitudes $\delta
u^{(1)}_{ph}(t)$ and $\delta u^{(2)}_{pp'hh'}(t)$ are determined
by the stationarity principle for the action
\begin{equation}
{\cal S} \left[
	\delta u^{(1)}_{ph}(t),
	\delta u^{(1)*}_{ph}(t),
	\delta u^{(2)}_{pp'hh'}(t),
	\delta u^{(2)*}_{pp'hh'}(t)
\right] = \int\!\! dt \; {\cal L}(t) \;,
\label{eq:action}
\end{equation}
with the Lagrangian \cite{KermanKoonin,KramerSaraceno,LDavid,rings}
\begin{eqnarray}
  {\cal L}(t) &=& \Big\langle\Psi(t)  \Big| \; H +
  H_{\rm ext}(t)- \I\, \hbar {\partial\over
  \partial t}\; \Big| \Psi(t) \Big\rangle\nonumber\\
	&=& \Big\langle\Psi_0(t)  \Big| \; H ' +
  H_{\rm ext}(t)- \I\, \hbar {\partial\over
  \partial t}\; \Big| \Psi_0(t) \Big\rangle \,.
  \label{eq:Lagrange1}
\end{eqnarray}
A ``boson'' version of the theory is recovered when the particle-hole
amplitudes $\delta u^{(1)}_{ph}(t)$ and $\delta u^{(2)}_{pp'hh'}(t)$
are restricted to {\it local\/} functions that depend only on the
momentum transfers $\qvec^{(\prime)}= \pvec^{(\prime)}
-\hvec^{(\prime)}$.

\subsection{Brillouin conditions}
\label{ssec:Brillouin}

To derive linear equations of motion, the Lagrangian
(\ref{eq:Lagrange1}) must be expanded to second order in the
excitation operator $U(t)$. For the procedure to be meaningful, one
should require that the first order terms vanish.  This is, in
principle, a necessary condition, however, in practice it is not
always possible to satisfy it rigorously.

The first variation of the energy with respect to $\delta
u^{(1)}_{ph}(t)$ and $\delta u^{(1)*}_{ph}(t)$ is
\begin{equation}
 \label{eq:energderiv1}
 \left. \frac{\delta
\left\langle\Psi(t)\right| H' \left|\Psi(t)\right\rangle}
{\delta (\delta u^{(1)}_{ph}(t))}
\right\vert_{\delta u^{(1)}(t) = \delta u^{(2)}(t) = 0}
=  H'_{0,ph} \; \label{eq:eins}
\end{equation}
and its complex conjugate. This term vanishes in the homogeneous
liquid due to momentum conservation.

The variation with respect to $\delta u^{(2)}_{pp'hh'}$ leads to a
similar condition
\begin{equation}
 \left. \frac{\delta \left\langle\Psi(t)\right| H'
\left|\Psi(t)\right\rangle}{\delta(\delta u^{(2)}_{pp'hh'}(t))}
\right\vert_{\delta u^{(1)}(t) = \delta u^{(2)}(t) = 0}
=H'_{0,pp'hh'} = 0  \label{eq:zwei} \ \;
\end{equation}
and its complex conjugate. This condition is not rigorously satisfied
by a Jastrow-Feenberg ground state.  Recall, however, that the
optimization condition (\ref{eq:Euler}) for pair correlations can be
written in terms of off-diagonal matrix elements of $H'$ in the form
(\ref{eq:zweicoll}).  If the correlation operator $F$ is chosen
optimally, {\it i.e.\/} satisfying Eq.~(\ref{eq:Euler}) for all $n$,
the weighted averages of $H_{{\bf o},{\bf n}}$ vanish.
This shows precisely what an optimized ground state does: 
The Jastrow correlation function does not have enough flexibility to 
guarantee the Brillouin condition ~(\ref{eq:zwei}), because $H'_{0,pp'hh'}$
depends non--trivially on four momenta, whereas the two--body 
Jastrow--Feenberg function depends only on the momentum transfer.
Optimization has the effect that the Brillouin conditions are
satisfied in the Fermi-sea average.

To make progress we must assume that in the Lagrangian terms that are
linear in the pair fluctuations are sufficiently small and can be
omitted. Likewise, we also shall assume that the ground state wave
function (\ref{eq:JastrowCorrelations}) is well enough optimized such
that three- and four-body Brillouin conditions are satisfied. In
momentum space, these are
\begin{equation}
\bra{\Psi_0} H' \rho_{\qvec_1},\cdots,\rho_{\qvec_n}\ket{\Psi_0} = 0\,.
\label{eq:Brilln}
\end{equation}

\subsection{Transition density}
\label{ssec:densan}
The quantity of primary interest is the linear density fluctuation
induced by the external field $H_{\rm ext}(t)$.  We regard this
density as a {\it complex\/} quantity; it is understood that the
physical density fluctuation is its real part.  Assuming the
excitation operator (\ref{eq:Uop}), it is
\begin{eqnarray}
  \delta\rho({\bf r};t)
  &=& \sum_{ph}
    \Big\langle\Psi_{\bf o}\Big|\, \hat\rho({\bf r})- \rho
    \,\Big| \Psi_{ph}\Big\rangle \,
    \delta u^{(1)}_{ph}(t)\nonumber\\
 &+& \frac{1}{2}\sum_{pp'hh'}
    \Big\langle\Psi_{\bf o}\Big|\, \hat\rho({\bf r})- \rho
    \,\Big| \Psi_{pp'h'h}\Big\rangle \,
    \delta u^{(2)}_{pp'hh'}(t)\nonumber\\
  &\equiv& \sum_{ph} \rho_{0,ph\!}({\bf r})\,\delta u^{(1)}_{ph}(t)
  + \frac{1}{2}\sum_{pp'hh'\,}\rho_{0,pp'hh'\!}({\bf r})\,
    \delta u^{(2)}_{pp'hh'}(t) \,.
\label{eq:rhoph}
\end{eqnarray}

The matrix elements of the density, $\rho_{0,ph\!}(\rvec)$ and
$\rho_{0,pp'hh'\!}(\rvec)$ with respect to the correlated states can
also be written as linear combinations of the matrix elements
$\rho^{\rm F}_{0,ph}(\rvec)$ with respect to uncorrelated states, and
one-, two-, and three-body matrix elements of the unit operator.  For
the sake of discussion, let us briefly neglect the pair
amplitudes. Since the density operator is local, we can commute
$\hat\rho({\bf r})$ to the right or to the left of the correlation
operator $F$. The form obtained by commuting $\hat\rho({\bf r})$ to
the left is
\begin{equation}
  \rho_{0,ph\!}({\bf r}) =
  \sum_{p'h'}\tilde\rho^{\rm F}_{0,p'h'\!}({\bf r})\,M_{p'h',ph}
  = \tilde\rho^{\rm F}_{0,ph}({\bf r}) +
  \sum_{p'h'}\tilde\rho^{\rm F}_{0,p'h'\!}({\bf r})\,N_{p'h',ph} \;,
\label{eq:rhorep1}
\end{equation}
where $\tilde\rho_{0,ph\!}^{F}(\rvec) \equiv z_{ph}\,\bra{\Phi_{\bf
    o}}\,\hat\rho(\rvec)\!-\!  \rho
\,\ket{\qerz{p}\qver{h}\Phi_{\bf o}} \equiv z_{ph}\, \bra{ h }
\delta\hat\rho({\bf r})\ket{ p}$ are, apart from the normalization
factors $z_{ph}$, the matrix elements of the density operator in a
non-interacting system.

The second form is obtained by commuting  $\hat\rho({\bf r})$ to the right 
of $F$:
\begin{equation}
\rho_{0,ph}({\bf r})
=\frac{1}{z_{ph}^2}\tilde\rho^{\rm F}_{0,ph}({\bf r})
+ \sum_{p'h'}N_{0,pp'hh'}\tilde\rho^{\rm F}_{p'h',0}({\bf r})\,.
\label{eq:rhorep2}
\end{equation}
These two seemingly different expressions are identical, the different
analytic forms appear only because the second quantized formulation
hides the fact that the density operator is local. We will see below
that both forms are useful.

Including pair fluctuations, the fluctuating density (\ref{eq:rhoph}) can
generally be written as
\begin{eqnarray}
\delta\rho({\bf r};t)
&=& \sum_{ph}\tilde\rho^{\rm F}_{0,ph}({\bf r})\left[\sum_{p'h'}
M_{ph,p'h'}\delta u^{(1)}_{p'h'}(t)
+ \frac{1}{2}\sum_{p'p''h'h''}M_{ph,p'p''h'h''}
\delta u^{(2)}_{p'p''h'h''}(t)\right]\,.\;
\end{eqnarray}

A key step that simplifies the structure of the equations of motion
significantly is to introduce a new one-body function.  In analogy to
the boson theory \cite{eomI}, we define new particle-hole amplitudes
$\delta v^{(1)}_{ph}(t)$ through
\begin{equation}
\delta\rho({\bf r};t) \equiv 
\sum_{ph}\rho_{0,ph\!}({\bf r})\,\delta v^{(1)}_{ph}(t)
\label{eq:rhoofv}
\end{equation}
such that
\begin{equation}
\delta\rho({\bf r};t) = \sum_{php'h'}\tilde \rho^{\rm F}_{0,ph}({\bf r})
M_{ph,p'h'}\, \delta v^{(1)}_{p'h'}(t) \,.
\end{equation}
This implies
\begin{equation}
\sum_{p'h'} M_{ph,p'h'}\,\delta v^{(1)}_{p'h'}(t) =
\sum_{p'h'} M_{ph,p'h'}\,\delta u^{(1)}_{p'h'}(t)
+ \frac{1}{2}\sum_{p'p''h'h''}M_{ph,p'p''h'h''}\>
\delta u^{(2)}_{p'p''h'h''}(t)
\;.
\end{equation}
Defining $M^{\rm (I)}_{ph,p'p''h'h''}$ via
\begin{equation}
M^{\phantom{\rm (I)}}_{ph,p'p''h'h''} \equiv  \sum_{p_1h_1} 
  M^{\phantom{\rm (I)}}_{ph,p_1h_1}
  M^{\rm (I)}_{p_1h_1,p'p''h'h''}
\label{eq:M2fact}
\end{equation}
we can formally solve for $\delta v^{(1)}_{ph}(t)$:
\begin{equation}
\delta v^{(1)}_{ph}(t)=
\delta u^{(1)}_{ph}(t)
+ \frac{1}{2}\sum_{p'p''h'h''}M^{\rm (I)}_{ph,p'p''h'h''}\
\delta u^{(2)}_{p'p''h'h''}(t) \,.
\end{equation}
For this operation, the inverse of $M_{ph,p'h'}$ seems to be
needed. As its calculation is not immediately obvious, we hasten to
note that $M^{\rm (I)}_{ph,p'p''h'h''}$ is, in terms of
Jastrow-Feenberg diagrams \cite{CBF2}, a {\em proper subset\/} of the
diagrams contributing to $M_{ph,p'p''h'h''}$. We will discuss the
diagrammatic analysis of $\rho_{0,ph}(\rvec)$ in
App.~\ref{assec:transition_density}.  The diagrammatic construction of
$M^{\rm (I)}_{ph,p'p''h'h''}$ in the spirit of
Eq. (\ref{eq:M2fact}) is carried out in
App.~\ref{assec:MImatrix}.

\subsection{The Lagrangian}
\label{ssec:Lagrangian}

We split the Lagrangian (\ref{eq:Lagrange1}) as
${\cal L}(t) = {\cal L}_{\rm ext}(t)+ {\cal L}_{\rm t}(t)
+ {\cal L}_{\rm int}(t)$,
with
\begin{eqnarray}
{\cal L}_{\rm ext}(t) &=& \Bigl\langle\Psi_0(t)\Bigm|H_{\rm ext} \Bigm| 
   \Psi_0(t) \Bigr\rangle \;,\\
{\cal L}_{\rm t}(t) &=& \biggl\langle\Psi_0(t)  \biggr|  - 
  \I\, \hbar \frac{\partial}{\partial t}\; \biggl| 
   \Psi_0(t) \biggr\rangle \;,\\
{\cal L}_{\rm int}(t) &=& \Bigl\langle\Psi_0(t)\Bigm| \; H'\; \Bigm| 
   \Psi_0(t) \Bigr\rangle \;.
\end{eqnarray}
${\cal L}_{\rm ext}(t)$ is obtained directly from the transition density:
\begin{eqnarray}
    {\cal L}_{\rm ext}(t)
&=&\int\! d^3r\, h_{\rm ext}(\rvec;t)\,\delta\rho(\rvec;t)\nonumber\\
&=&\int\! d^3r\, h_{\rm ext}(\rvec;t)\,{\Re}e\left[
 \sum_{ph}\rho_{0,ph}(\rvec)\, \delta u^{(1)}_{ph}(t)
+\frac{1}{2}\sum_{pp'hh'}\rho_{0,pp'hh'}(\rvec)\, \delta u^{(2)}_{pp'hh'}(t)
\right] \label{eq:Lagrext} \nonumber\\
&=& {\Re}e\sum_{ph}  \int\! d^3r\, h_{\rm ext}(\rvec;t)\,\rho_{0,ph}(\rvec)\>
\delta v^{(1)}_{ph}(t)\,.
\end{eqnarray}
The time-derivative term ${\cal L}_t(t)$ is, to second order in the
fluctuations,
\begin{eqnarray}
{\cal L}_{\rm t}(t) &=&
\frac{\hbar}{2\bigl\langle\Psi_0(t)\ket{\Psi_0(t)}}\;
\Im{\rm m}\;\sum\biggl[
\delta \dot{u}^{(1)}_{ph}(t) \; \bigl\langle\psi(t)
\ket{\perz{p} \pver{h}\psi(t)} \\
&&\qquad\qquad\qquad+ \frac{1}{2}\sum\delta\dot{u}^{(2)}_{pp'hh'}(t) \;
\bigl\langle\Psi_0(t) \ket{ \perz{p} \perz{p'} \pver{h'} \pver{h}\Psi_0(t)}
\biggr] \nonumber\\
&=&\frac{\hbar}{4}\;\Im{\rm m}\;\Biggl[
\sum\delta {u}^{(1)*}_{ph}(t) M_{ph,p'h'} \delta \dot{u}^{(1)}_{p'h'}(t)
 + \frac{1}{2}\sum\delta {u}^{(1)*}_{ph}(t) M_{ph,p'p''h'h''}
\delta \dot{u}^{(2)}_{p'p''h'h''}(t)\nonumber \\
&+& \frac{1}{2}\sum\delta{u}^{(2)*}_{pp'hh'}(t) M_{pp'hh',p''h''}
\delta \dot{u}^{(1)}_{p''h''}(t)
+ \frac{1}{4}\sum\delta {u}^{(2)*}_{pp'hh'}(t) M_{pp'hh',p''p'''h''h'''}
\delta \dot{u}^{(2)}_{p''p'''h''h'''}(t) \Biggr]\,.\nonumber
\end{eqnarray}
Introducing the new amplitudes $\delta v^{(1)}_{ph}(t)$ defined in
Eq.~(\ref{eq:rhoofv}) eliminates the terms that couple the one- and
the two-body amplitudes:
\begin{equation}
{\cal L}_{\rm t}(t)
=\frac{\hbar}{4}\;\Im{\rm m}\;\Biggl[
\sum\delta{v}^{(1)*}_{ph}(t) M_{ph,p'h'} \delta \dot{v}^{(1)}_{p'h'}(t)
+ \frac{1}{4}\sum\delta {u}^{(2)*}_{pp'hh'}(t)
M^{\rm (I)}_{pp'hh',p''p'''h''h'''} \delta \dot{u}^{(2)}_{p''p^{'''}h''h^{'''}}(t)
\Biggr] \;,
\label{eq:Ltime}
\end{equation}
where
\begin{equation}
M^{\rm (I)}_{pp'hh',p''p'''h''h'''}
= M^{\phantom{\rm (I)}}_{pp'hh',p''p'''h''h'''}-
\sum_{p_1p_2h_1h_2} M^{\rm (I)}_{pp'hh',p_1h_1}
M^{\phantom{\rm (I)}}_{p_1h_1,p_2h_2}M^{\rm (I)}_{p_2h_2,p''p'''h''h'''} \,.
\label{eq:Nfour}
\end{equation}
The second term in Eq.~(\ref{eq:Nfour}) cancels, in a diagrammatic
expansion, some terms from the first one ({\it cf}\/.~App.~%
\ref{assec:transition_density}).  From Eqs.~(\ref{eq:Lagrext}) and
(\ref{eq:Ltime}), the advantage of introducing the new particle-hole
amplitudes $\delta v^{(1)}_{ph}(t)$ becomes obvious.

The contributions to the interaction term are classified
according to the involved $n-$body fluctuations $U_n$ as defined in 
(\ref{eq:Uop}),
\begin{equation}
{\cal L}_{\rm int}(t) = 
{\cal L}^{(11)}_{\rm int}(t) +
{\cal L}^{(12)}_{\rm int}(t) +
{\cal L}^{(22)}_{\rm int}(t) \;,
\end{equation}
with
\begin{eqnarray}
{\cal L}_{\rm int}^{(11)}(t) &=& 
\frac{1}{8} \bra{\Psi_{\bf o}}\left[
	U_1^\dagger(t) U_1^\dagger(t) H' + 
	2 U_1^\dagger(t) H' U_1(t) + 
	H' U_1(t) U_1(t)
\right] \ket{\Psi_{\bf o}}\;,\nonumber\\
{\cal L}_{\rm int}^{(12)}(t) &=& 
\frac{1}{4} \bra{\Psi_{\bf o}}\left[
	U_1^\dagger(t) U_2^\dagger(t) H' +
	U_1^\dagger(t) H' U_2(t) +
	U_2^\dagger(t) H' U_1(t) +
	H' U_1(t) U_2(t)
\right] \ket{\Psi_{\bf o}}\;,\nonumber\\
{\cal L}_{\rm int}^{(22)}(t) &=& 
\frac{1}{8} \bra{\Psi_{\bf o}}\left[
	U_2^\dagger(t) U_2^\dagger(t) H' +
	2 U_2^\dagger(t) H' U_2(t) +
	H' U_2(t) U_2(t)
\right]\ket{\Psi_{\bf o}}\,.
\label{eq:Lagrange12}
\end{eqnarray}

If the Brillouin conditions (\ref{eq:eins})--(\ref{eq:zwei}) as well
as their generalizations to higher order fluctuations were satisfied
exactly, all contributions to ${\cal L}_{\rm int}^{(ij)}(t)$
containing $U^\dagger_i(t) U^\dagger_j(t)$ and $U_i(t) U_j(t)$ would
be zero.  For fermions with optimized Jastrow--Feenberg wave functions
it is only true in the averaged sense (\ref{eq:zweicoll}).  These
terms are nevertheless expected to be small in ${\cal L}_{\rm
  int}^{(22)}(t)$ since neglecting these terms is equivalent to
negligible four-body correlations.  Such a simplifying assumption is
not necessary in ${\cal L}_{\rm int}^{(12)}(t)$ and ${\cal L}_{\rm
  int}^{(11)}(t)$ although we will see that the terms containing
$U_1(t)U_2(t)$ and $U_1^\dagger(t)U_2^\dagger(t)$ in ${\cal L}_{\rm
  int}^{(12)}(t)$ are indeed negligible. We keep these terms for the
time being since it will turn out that their omission will suggest,
for consistency reasons, further simplifications.

The next step is to express the interaction term (\ref{eq:Lagrange12})
in terms of the CBF matrix elements introduced on section
\ref{ssec:CBF}. In the following it is understood that we sum over all
quantum numbers when no summation subscripts are spelled out.
\begin{eqnarray}
{\cal L}_{\rm int}^{(11)}(t)
	&=& \frac{1}{ 8}\sum\delta u^{(1)*}_{ph}(t)
        \delta u^{(1)*}_{p'h'}(t) H'_{pp'hh',0} + \mbox{c.c.} 
	 +  \frac{1}{ 4}\sum\delta u^{(1)*}_{ph}(t)
         H'_{ph,p'h'} \delta u^{(1)}_{p'h'}(t) \;,\label{eq:L11}\\
{\cal L}_{\rm int}^{(12)}(t)
	&=& \frac{1}{ 8}\sum\delta u^{(1)*}_{ph}(t)
        \delta u^{(2)*}_{p'p''h'h''}(t) H'_{pp'p''hh'h'',0} + \mbox{c.c.}\nonumber\\
	&+& \frac{1}{ 8}\sum\delta u^{(1)*}_{ph}(t)
        H'_{ph,p'p''h'h''}\delta u^{(2)}_{p'p''h'h''}(t) + \mbox{c.c.}
        \;,\label{eq:L12}\\
{\cal L}_{\rm int}^{(22)}(t)
	&=& \frac{1}{32}\sum\delta u^{(2)*}_{pp'hh'}(t)
        \delta u^{(2)*}_{p''p'''h''h'''}(t) H'_{pp'p''p'''hh'h''h''',0} +
        \mbox{c.c.}\nonumber\\
	&+& \frac{1}{16}\sum\delta u^{(2)*}_{pp'hh'}(t)
        H'_{pp'hh',p''p'''h''h'''} \delta u^{(2)}_{p''p'''h''h'''}(t)\,.\label{eq:L22}
\end{eqnarray}
Substituting $\delta v^{(1)}_{ph}(t)$ for $\delta u^{(1)}_{ph}(t)$
leads to new coefficient functions in the interaction part of the
Lagrangian:
\begin{equation}
{\cal L}_{\rm int}(t)
= {\cal L}^{'(11)}_{\rm int}(t)
+ {\cal L}^{'(12)}_{\rm int}(t)
+ {\cal L}^{'(22)}_{\rm int}(t)
\end{equation}
with
\begin{eqnarray}
{\cal L}_{\rm int}^{'(11)}(t)
	&=& \frac{1}{8}
        \sum\delta v^{(1)*}_{ph}(t)\,\delta v^{(1)*}_{p'h'}(t)\, H'_{pp'hh',0}
        + \mbox{c.c.} 
	+ \frac{1}{4}
        \sum\delta v^{(1)*}_{ph}(t) \, H'_{ph,\,p'h'}\,\delta v^{(1)}_{p'h'}(t)\;, 
	\label{eq:L11p}\\
{\cal L}_{\rm int}^{'(12)}(t)
	&=& \frac{1}{8}
 	\sum\delta v^{(1)*}_{ph}(t) \,\delta u^{(2)*}_{p'p''h'h''}(t)
        \,K^{(ph)}_{p'p''h'h'',0}
        \, + \mbox{c.c.} 
	\nonumber\\
	&+& \frac{1}{ 8}\sum\delta v^{(1)*}_{ph}(t)
        K_{ph,p'p''h'h''} \delta u^{(2)}_{p'p''h'h''}(t) + \mbox{c.c.}
	\label{eq:L12p}\\
{\cal L}_{\rm int}^{'(22)}(t)
	&=& \frac{1}{32}
	\sum\delta u^{(2)*}_{pp'hh'}(t)\, \delta u^{(2)*}_{p''p'''h''h'''}(t)
        \,K^{(pp'hh')}_{p''p'''h''h''',0} 
         \;+ \mbox{c.c.}
	\nonumber\\
	&+&\frac{1}{16}\sum\delta u^{(2)*}_{pp'hh'}(t)\,K_{pp'hh',\,p''p'''h''h'''}\>
        \delta u^{(2)}_{p''p'''h''h'''}(t)\;.
	\label{eq:L22p}
\end{eqnarray}

The new coefficients $K_{{\bf m},{\bf n}}$ are
\begin{eqnarray}
   K_{ph,\,p'p''h'h''} &\equiv& H'_{ph,\,p'p''h'h''}-\sum_{p_1\!h_1}
   H'_{ph,\,p_1\!h_1}\> M^{\rm (I)}_{p_1\!h_1,\,p'p''h'h''}\;,
\label{eq:K12adef}\\
   K^{(ph)}_{p'p''h'h'',0}&\equiv&  H'_{pp'p''hh'h'',0} -
   \sum_{p_1\!h_1}\!H'_{ph\,p_1\!h_1,0}\, M^{\rm (I)}_{p'p''h'h''\!,\,p_1\!h_1}
   \;,
\label{eq:K12bdef}\vspace{0.2cm}\\
  K_{pp'hh',\,p''p'''h''h'''}
  &\equiv& H'_{pp'hh',\,p''p'''h''h'''}
  \nonumber\\
  &-&\sum_{p_1\!h_1}\Bigl(
  M^{\rm (I)}_{pp'hh',\,p_1\!h_1} H'_{p_1\!h_1,\,p''p'''h''h'''} +
  H'_{pp'hh',\,p_1\!h_1}M^{\rm (I)}_{p_1\!h_1,\,p''p'''h''h'''} \Bigr)
 \nonumber\\
 &+& \sum_{p_1\!h_1p_2h_2}M^{\rm (I)}_{pp'hh',\,p_1\!h_1} 
 H'_{p_1\!h_1,\,p_2h_2} M^{\rm (I)}_{p_2h_2,\,p''p'''h''h'''}\,,
\label{eq:K22def}
\end{eqnarray}
and an analogous term for $K^{(pp'hh')}_{p''p'''h''h''',0}$\,. 

\subsection{Equations of motion}
\label{ssec:EOMs}

With the sole approximation to neglect the terms proportional to
$U_2(t)U_2(t)$ and $U_2^\dagger(t)U_2^\dagger(t)$, the Euler equations
become
\begin{eqnarray}
  \sum\Bigr[ \I\hbar M_{ph,p'h'}\frac{\partial}{\partial t} - H'_{ph,p'h'}
      \Bigr] \delta v^{(1)}_{p'h'}(t)
  \;-\;\sum  H'_{pp'hh',0}\,\delta v^{(1)*}_{p'h'}(t)
\label{eq:EOM1}\\
 - \frac{1}{2} \sum\left[ K_{ph,p'p''h'h''}\,\delta u^{(2)}_{p'p''h'h''}(t)
   \;+\; K^{(ph)}_{p'p''h'h'',0}\,\delta u^{(2)*}_{p'p''h'h''}(t) \right]
&=& 2\int\! d^3r\, \rho_{ph,0}({\bf r})\, h_{\rm ext}({\bf r};t) \;,
\vspace{0.2cm}\nonumber\\
  \frac{1}{2}\sum \Bigl[ \I\hbar M^{\rm (I)}_{pp'hh',p''p'''h''h'''}
       \frac{\partial}{\partial t} - K_{pp'hh',p''p'''h''h'''}
       \Bigr]\delta u^{(2)}_{p''p'''h''h'''}(t)
\nonumber\\
 - \sum \Bigr[ K_{pp'hh',p''h''}\,\delta v^{(1)}_{p''h''}(t)
            +  K^{(p''h'')}_{pp'hh',0}\,\delta v^{(1)*}_{p''h''}(t) \Bigr]
&=&   0\;.\
\label{eq:EOM2}
\end{eqnarray}
The time dependence of the external field can be assumed to be harmonic,
with an infini\-te\-simal turn-on component that determines the sign of the
imaginary part
\begin{equation}
h_{\rm ext}({\bf r};t) =
h_{\rm ext}({\bf r};\omega)\left[e^{\I\omega t}+e^{-\I\omega t}\right]
e^{\eta t/\hbar} \, .
\end{equation}
This imposes the time dependence
\begin{eqnarray}
\delta v^{(1)}_{ph}(t) &=&\>
      \delta v^{(1+)}_{ph}(\omega)\;e^{-\I(\omega + \I\eta/\hbar) t} \;+\;
\Bigl[\delta v^{(1-)}_{ph}(\omega)\;e^{-\I(\omega + \I\eta/\hbar) t}\Bigr]^* \, ,
\label{eq:v1omega}\\\nonumber
\delta u^{(2)}_{pp'hh'}(t) &=& 
      \delta u^{(2+)}_{pp'hh'}(\omega)\,e^{-\I(\omega + \I\eta/\hbar) t}+
\Bigl[\delta u^{(2-)}_{pp'hh'}(\omega)\,e^{-\I(\omega + \I\eta/\hbar) t}\Bigr]^* \,.
\end{eqnarray}
Defining
\begin{equation}
E_{pp'hh',p''p'''h''h'''}(\omega) \equiv 
  (\hbar\omega\!+\!\I\eta) M^{\rm (I) }_{pp'hh',p''p'''h''h'''} -
  K_{pp'hh',p''p'''h''h'''}
\label{eq:defEomega}
\end{equation}
the equations of motion for the pair fluctuations are
\begin{eqnarray}
 \frac{1}{2}\sum E_{pp'hh',p''p'''h''h'''}(\phantom{-}\omega) \, 
 \delta u^{(2+)}_{p''p'''h''h'''}(\omega) 
 = \sum \Bigl[ K_{pp'hh',p''h''}\,\delta v^{(1+)}_{p''h''}(\omega)
            +  K^{(p''h'')}_{pp'hh',0}\,\delta v^{(1-)}_{p''h''}(\omega) \Bigr] \,,
\nonumber \\
 \frac{1}{2}\sum E_{pp'hh',p''p'''h''h'''}^*(-\omega) \, 
 \delta u^{(2-)}_{p''p'''h''h'''}(\omega) 
 = \sum \Bigl[ K_{pp'hh',p''h''}^*\,\delta v^{(1-)}_{p''h''}(\omega)
            +  K^{(p''h'')*}_{pp'hh',0}\,\delta v^{(1+)}_{p''h''}(\omega) \Bigr] \,.
\nonumber\\
\label{eq:pair_eq}
\end{eqnarray}

All pair quantities are symmetric under the interchange
of the involved pair variables, {\it e.g.}\/ $(pp',hh')\leftrightarrow
(p'p,h'h)$. We can utilize this feature to replace the
fully symmetric $E_{pp'hh',p''p'''h''h'''}(\omega)$ by an asymmetric
form, e.g.  (\ref{eq:Eomega_appr}) which removes the factor $1/2$
in Eq. (\ref{eq:pair_eq}).

The pair equations (\ref{eq:pair_eq}) are now solved for the $\delta
u^{(2\pm)}_{pp'hh'}(\omega)$ and the solutions are inserted into the
one-body equation.  The latter retains the structure of a TDHF
equation, but with the matrix elements of $H'$ supplemented by
frequency-dependent terms.  We adapt the definition of
$W_{\bf m,n}$ in (\ref{eq:defineW}) by adding these corrections:
\begin{eqnarray}
W_{ph,\,p'h'}(\omega) = W_{ph,\,p'h'}
	&+& \sum K_{ph,\,p_1p_2h_1h_2} \>
        E^{-1}_{p_1p_2h_1h_2,p_1'p_2'h_1'h_2'}(\omega)
        K_{p_1'p_2'h_1'h_2',\,p'h'} \nonumber \\
	&+& \sum K^{(ph)}_{p_1p_2h_1h_2,0}
        E^{*-1}_{p_1p_2h_1h_2,p_1'p_2'h_1'h_2'}(-\omega)
K^{(p'h')*}_{p_1'p_2'h_1'h_2',0} \, , 
	\label{eq:Wpmdef} \\
        W_{pp'hh',\, 0}(\omega) = W_{pp'hh',\, 0}
	&+& \sum K_{ph,\,p_1p_2h_1h_2} \>
        E^{-1}_{p_1p_2h_1h_2,p_1'p_2'h_1'h_2'}(\omega)
        K^{(p'h')}_{p_1'p_2'h_1'h_2',0} \nonumber \\
	&+& \sum K^{(ph)}_{p_1p_2h_1h_2,0}
        E^{*-1}_{p_1p_2h_1h_2,p_1'p_2'h_1'h_2'}(-\omega)
        K^*_{p_1'p_2'h_1'h_2',\,p'h'} \,. 
	\label{eq:Wpmdef_2}
\end{eqnarray}
This TDHF form results also if the terms containing $U_2(t)U_2(t)$ are
retained, but the expressions for the dynamic parts of the $W$-matrices become
lengthier.

The equations of motion for the particle-hole amplitudes are then
\begin{eqnarray}
2\int\! d^3r\> h_{\rm ext}(\rvec;\omega)\,\rho_{0,ph}(\rvec)
&=& \sum_{p'h'}\Bigl[ (\hbar\omega\!+\!\I\eta)\,M_{ph,p'h'} -
    \delta_{p,p'}\delta_{h,h'}\,e_{ph}\Bigr]\, v_{p'h'}^{(1+)}(\omega) \;
\nonumber\\
&-& \sum_{p'h'}\left[W_{ph,\,p'h'}(\omega)\> +
{1\over2}\left(e_{ph}+e_{p'h'}\right)N_{ph,\,p'h'}\> \right]\delta
v_{p'h'}^{(1+)}(\omega)\nonumber\\
&-& \sum_{p'h'}\left[W_{pp'hh',0}(\omega)    +
{1\over2}\left(e_{ph}+e_{p'h'}\right)N_{pp'hh',0}    \right]\delta
v_{p'h'}^{(1-)}(\omega) 
\;.\quad\quad\quad
\label{eq:v1eom_withW(w)} 
\end{eqnarray}

\subsection{Supermatrix representation}
\label{ssec:supermatrix}

We can now carry out exactly the same manipulations as in previous
work \cite{rings} and reduce these equations (\ref{eq:v1eom_withW(w)})
to the form of TDHF equations with energy-dependent effective
interactions.

Equations (\ref{eq:rhorep1}) and (\ref{eq:rhorep2}) express the
density in terms of CBF matrix elements in two different forms.  For
the present purpose, it is convenient to use these two representations
symmetrically,
\begin{equation}
\delta\rho_{0,ph}(\rvec) = \frac{1}{2}
 \left[1+\frac{1}{z_{ph}^2}\right]\tilde \rho^{\rm F}_{0,ph}(\rvec)
+\frac{1}{2}\sum_{p'h'}\left[\tilde\rho^{\rm F}_{0,p'h'} N_{p'h',ph} + 
                   \tilde\rho^{\rm F\,*}_{0,p'h'}(\rvec) N_{0,pp'hh'}\right]
\;.\end{equation}
Using Eqs.~(\ref{eq:rhoofv}) and (\ref{eq:v1omega}),
the density fluctuations can then be written as
\begin{eqnarray}
\delta\rho({\bf r};\omega) 
	&=& \frac{1}{2}\sum_{ph} \left[
          \rho_{0,ph}  ({\bf r})\, \delta v^{(1+)}_{ph}(\omega) +
          \rho_{0,ph}^*({\bf r})\, \delta v^{(1-)}_{ph}(\omega) \right] 
\nonumber \\
	&\equiv& \frac{1}{2}\sum_{ph} \left[
          \tilde \rho_{0,ph}^{\rm F }({\bf r})\,
          \delta c^{(1+)}_{ph}(\omega) +
          \tilde \rho_{0,ph}^{\rm F*}({\bf r})\,
          \delta c^{(1-)}_{ph}(\omega)\right]
\;,
\label{eq:d_rho_r_om}
\end{eqnarray}
({\it cf.}\/ (\ref {eq:rhorep1}) for the definition of
$\tilde\rho_{0,ph}^{\rm F }({\bf r})$).  This defines new amplitudes
$\delta c^{(1\pm)}_{ph}(\omega)$. These relate, apart from the
normalization factors, the observed density to the matrix elements of
the density operator in the non-interacting system.  The equations of
motion can now be simplified by introducing a ``supermatrix''
notation. Particle-hole matrix elements together with their complex
conjugate are combined into vectors, \textit{e.g.}
\begin{equation}
  \tilde{\bm \rho}^{\rm F}
  \equiv \left(\begin{array}{c} \tilde\rho^{\rm F}_{0,ph} \\
  \tilde\rho_{0,ph}^{F*} \end{array}\right)
  \quad;\quad
  \delta{\bf c} 
  \equiv \left(\begin{array}{c} \delta c^{(1+)}_{ph} \\
  \delta c^{(1-)}_{ph}\end{array}\right)
\end{equation}
(and analogously for $\delta v^{(1\pm)}_{ph}$). Equation (\ref{eq:d_rho_r_om})
then simply reads
\begin{equation}
\delta\rho({\bf r};\omega) = \textstyle \frac{1}{2}\, 
\delta{\bf c}(\omega)\cdot \tilde{\bm \rho}^{\rm F}(\rvec) \;.
\label{eq:rhoc} 
\end{equation}
The matrices
\begin{equation}
{\bf N} = \left(\begin{array}{cc}
N_{ph,p'h'} &N_{pp'hh',0}\\
N_{0,pp'hh'} &N_{p'h',ph}\\
\end{array}\right) 
\end{equation}
\label{eq:Nsupermatrix}
and
\begin{equation}
{\bf C} = \frac{1}{2}\left(\begin{array}{cc}
1+\displaystyle{\frac{1}{z_{ph}^2}} & 0 \\
0 & 1+\displaystyle{\frac{1}{z_{ph}^2}}
\end{array}\right) \delta_{p,p'}\delta_{h,h'}+
\frac{1}{2}{\bf N}
\label{eq:Csupermatrix}
\end{equation}
relate the amplitude functions:
\begin{equation}
\delta{\bf c} =  {\bf C}\cdot \delta{\bf v} \; .
\end{equation}

In the driving term on the l.h.s.\ of 
(\ref {eq:v1eom_withW(w)}) we use $\rho_{0,ph} = ({\bf C}\!\cdot\!
\tilde{\bm\rho}^{\rm F})_{0,ph}$ to obtain
\begin{equation}
2\int\!d^3r\> h_{\rm ext}(\rvec;\omega)\,\rho_{0,ph}(\rvec) =
2\,{\bf C}\cdot{\bf h}^{\rm ext}
\;,\end{equation}
where the vector ${\bf h}^{\rm ext}$ is built with the non-interacting states
({\it cf}\/.~$\tilde\rho^{F}_{0,ph}$ in (\ref{eq:rhorep1}))
\begin{equation}
\tilde h_{0,ph}^{F}(\omega) = z_{ph}\,\bra{h}\,h_{\rm ext}({\bf r};\omega)\,\ket{p}\,.
\end{equation}
Defining the $\omega-$dependent matrices
\begin{eqnarray}
\bm\Omega &=&
\left(
\begin{array}{cc}
(\hbar\omega\!+\!\I\eta -e_{ph}) \delta_{p,p'}\delta_{h,h'} & 0 \\
0 &-(\hbar\omega\!+\!\I\eta +e_{ph})\delta_{p,p'}\delta_{h,h'}
\end{array}\right)\;,\nonumber\\
{\bf W} &=&
\left(\begin{array}{cc}
W^{(+)}_{ph,p'h'}(\omega) &W^{(-)}_{pp'hh',0}(\omega)\\
W^{(+)}_{0,pp'hh'}(\omega) & W^{(-)}_{p'h',ph}(\omega)\\
\end{array}\right)\;,
\label{eq:Wmat}
\end{eqnarray}
the equations of motion assume supermatrix form \cite{rings}
\begin{equation}
\left[\bm\Omega + {1\over2}\bm\Omega {\bf N} + {1\over2}{\bf N}\bm\Omega 
      - {\bf W}(\omega)\right]\cdot\delta{\bf v}
\>=\> 2 {\bf C}\cdot{\bf h}^{\rm ext} \,.
\label{eq:eomv1_supmat}
\end{equation}
We now formally define a new, energy--dependent interaction matrix
${\bf V}_{\rm p-h}(\omega)$ by
\begin{equation}
\left[\bm\Omega + {1\over2}\bm\Omega {\bf N}
                + {1\over2}{\bf N}\bm\Omega - {\bf W}\right]
\;\equiv\; {\bf C}\cdot
\biggl[\bm\Omega - {\bf V}_{\!\rm p-h}(\omega)\biggr]\cdot {\bf C} \;.
\label{eq:Vphmatrix}
\end{equation}
Thus the response equations take the simple TDHF form
\begin{equation}
\biggl[\bm\Omega - {\bf V}_{\!\rm p-h}(\omega)\biggr]\cdot\delta{\bf c}
\>=\> 2{\bf h}^{\rm ext} \;.
\label{eq:Vabba}
\end{equation}

With this, we have reformulated the theory for a strongly interacting
system in the TDHF form (\ref{eq:Vabba}) but with an energy dependent
effective interaction.  Our derivation has led to a clear definition
of this effective particle-hole interaction and to a prescription on
how to calculate this from the underlying bare Hamiltonian.

The formal derivation appears to involve the calculation of the inverse of a 
huge matrix. The key point, however, is that the manipulation 
(\ref{eq:Vphmatrix}) can be carried out diagrammatically. 
Then it becomes obvious that many terms occurring in the combination of 
matrices in (\ref{eq:eomv1_supmat}) are {\it not\/} part of
${\bf V}_{\!\rm p-h}(\omega)$. Specifically, these are the chain diagrams
in the direct channel \cite{rings}.

\section{Diagrammatic analysis and local interactions}
\label{sec:localapprox}

\subsection{General strategy}
\label{ssec:genstrategy}

Generally, the non-local operators ${\cal N}(1,2)$ and ${\cal W}(1,2)$
in (\ref{eq:NWop}) consists of up to 4-point functions. Cluster
expansions and resummations have been carried out in
Ref.~\onlinecite{CBF2} and led to reasonably compact representations
in terms of the compound-diagrammmatic quantities of the FHNC
summation method. Nevertheless, due to their non-locality, it is
difficult to deal with these quantities exactly. The simplest
approximation for the operator is to keep just the local terms.  These
are given by the ``direct-direct'' correlation function $\Gamma_{\!\rm
  dd}(|\rvec_1\!-\!\rvec_2|)$ of FHNC theory. This approximation is
adequate but not optimal.

On the other hand, summing $N_{0,pp'hh'}$ over the hole states,
Eq.~(\ref{eq:SofkfromN}), relates ${\cal N}(1,2)$ to the static
structure function. Accurate results are available for $S(q)$, either
from simulations \cite{CeperleyGFMC,BoronatHe3} or from the FHNC-EL
summation technique \cite{annals,polish}. An alternative strategy to
deal with non-local operators is therefore to demand that these
results are reproduced in whatever approximate form one chooses to
use. In this sense, by {\it choosing\/} ${\cal N}(1,2)$ to be local,
{\it naming}\/ the corresponding function $\Gamma_{\!\rm dd}(r)$, and
demanding that this operator in (\ref{eq:SofkfromN}) gives the known
static structure function, we obtain the relationship
\begin{equation}
S(q) = \SF{q}\left[1+\tilde\Gamma_{\!\rm dd}(q)\,\SF{q}\right]
\label{eq:sofk}
\end{equation}
as a {\it definition\/} of $\tilde\Gamma_{\!\rm dd}(q)$ in terms of $S(q)$.  We
adopt this view here and define the ``best'' local
approximation for ${\cal N}(1,2)$ such that it reproduces the best
known $S(q)$. Since the exact $S(q)$ contains a summation of exchange
terms, this implies that their contribution to $S(q)$ is mimicked by a
local contribution to $\tilde\Gamma_{\!\rm dd}(q)$.

An ``optimal'' local approximation for the effective interaction
${\cal W}(1,2)$ can be obtained along similar lines.
From Eqs.~(\ref{eq:CBFph}) and (\ref{eq:defineW}) we have 
\begin{equation}
H'_{0,pp'hh'} = W_{0,pp'hh'} + \frac{1}{2}\left(e_{ph}+e_{p'h'}\right)
N_{0,pp'hh'}
\;.
\label{eq:WNaux}
\end{equation}
The ground state Euler equation for pair correlations
(\ref{eq:zweicoll}) implies that the Fermi sea average of
$H'_{0,pp'hh'}$ vanishes.  Postulating a local ${\cal W}(1,2)\approx
W(r_{12})$, consistency relates this quantity to the local 
approximation of ${\cal N}(1,2)$.  This leads to \cite{polish}
\begin{equation}
\widetilde W(q) = -\frac{t(q)}{\SF{q}}\,\tilde\Gamma_{\!\rm dd}(q) \;.
\label{eq:Wlocal}
\end{equation}

Our procedure of using the relationships (\ref{eq:SofkfromN}) and
(\ref{eq:zweicoll}) to construct local approximations for
$N_{0,pp'hh'}$ and $W_{0,pp'hh'}$ can be generalized to a systematic
definition of optimal local approximations for the matrix elements of
any non-local $d-$body operator: Averaging the matrix elements, which
depend on $d$ particle and $d$ hole momenta, over the Fermi sea,
generates functions of the momentum transfers $\qvec_i\!\equiv
\pvec_i\!-\hvec_i$ only.  Spelling out Fermi occupation functions
$n_\hvec$ and $\bar n_\pvec\!\equiv 1\!-\!n_\pvec$ explicitly, this
reads for a one-body quantity
\begin{equation}
O_\qvec \equiv \frac{\sum_{h} \bar n_\pvec n^{\phantom-}_\hvec \, O_{0,ph}}
                {\sum_{h} \bar n_\pvec n^{\phantom-}_\hvec \, 1      }
  = \frac{1}{N\SF{q}}\sum_{h} 
  \bar n_{\hvec+\qvec}n^{\phantom-}_\hvec\,
  O_{0,\,ph} \;.
\label{eq:FSA}
\end{equation}
The extension to $d$ variables is obvious,
\begin{equation}
   O_{\vec q_1,\ldots,\,\vec q_d} = \sum_{h_1\ldots h_d}\> \prod_{i=1}^d
   \frac{\bar n_{{\bf p}_i} n^{\phantom-}_{{\bf h}_i} }{N\SF{q_i}}\>
   O_{0,\,p_1\ldots p_d\,h_1\ldots h_d} \;,
\label{eq:FSAn}
\end{equation}
as is the extension to matrix elements $O_{{\bf m},{\bf n}\ne{\bf o}}$.  

We emphasize again that the quantities $O_{\vec q_1,\ldots,\,\vec
  q_d}$ contain all exchange and correlation effects in a localized manner. Therefore,
effects related to the $z_{ph}$, as well as CBF corrections to the
$e_{ph}\,$, are already part of $\widetilde W(q)$ and $\tilde
\Gamma_{\!\rm dd}(q)$.  This implies, amongst others,
\begin{equation}
M_{p'h',ph} \approx \delta_{p,p'}\delta_{h,h'} +
\bra{hp'}\Gamma_{\!\rm dd}\ket{ph'}
\;,
\label{eq:MtoGamma}
\end{equation}
and the relationship (\ref{eq:Csupermatrix}) between the supermatrices
${\bf C}$ and ${\bf N}$ simplifies to
\begin{equation}
 {\bf C} = {\bf 1} + \frac{1}{2}{\bf N} \;.
\label{eq:CN}
\end{equation}

\subsection{Matrix elements}
\label{ssec:supermatrel}

The localization procedure discussed above for ${\cal N}(1,2)$ implies
\begin{equation}
{\bf N} =\> \frac{1}{N}\tilde\Gamma_{\!\rm dd}(q)
\left( \begin{array}{cc} \delta_{\vec q,+\vec q'} & \delta_{\vec q,-\vec q'} \\
                         \delta_{\vec q,-\vec q'} & \delta_{\vec q,+\vec q'}
\end{array}\right)\,
\bar n_{\bf p^{\phantom,\!}} \bar n_{\bf p'} 
      n_{\bf h^{\phantom,\!}} n_{\bf h'}
\;.
\label{eq:Nlocal}
\end{equation}
To simplify the notation, the $\delta_{\bf q,\pm\bf q'}$ functions,
together with the Fermi occupation numbers, are understood to be
implicit in all the matrices from now on.  Matrix products, {\it
  i.e.\/}~sums over particle--hole labels, reduce to factors
$\SF{q}$. The inverse of ${\bf C}$ is readily obtained from
(\ref{eq:CN}) as
\begin{equation}
{\bf C}^{-1} =\> {\bf 1} - \,\frac{1}{2N}\tilde X_{\rm dd}(q)
\left(\begin{array}{cc} 1 \,&\, 1 \\
                        1 \,&\, 1 \end{array}\right)\;.
\label{eq:Xlocal}
\end{equation}
with
\begin{equation}
 \tilde X_{\rm dd}(q) = \frac{\tilde\Gamma_{\!\rm dd}(q)}
                        {1 + \SF{q}\,\tilde\Gamma_{\!\rm dd}(q)} \;.
\label{eq:Xdddef}
\end{equation}
In the spirit of the discussion in Sec.~\ref{ssec:genstrategy}, this
is our definition of $\tilde X_{\rm dd}(q)$.  According to (\ref{eq:SFHNC0}),
it can also be identified with the sum of all non-nodal diagrams.

Multiplying ${\bf C}^{-1}$ from both sides to (\ref{eq:Vphmatrix}) yields 
the  $\omega$ dependent effective interactions,
\begin{equation}
{\bf V}_{\!\rm p-h}(\omega) = \frac{1}{N} \left(\begin{array}{cc}
	\tilde V^{\phantom{*}}_{\!_{\rm A}}(q; \phantom{-}\omega) \, &\>
	\tilde V^{\phantom{*}}_{\!_{\rm B}}(q; \phantom{-}\omega) \, \\
	\tilde V^*            _{\!_{\rm B}}(q; -\omega) \, &\>
	\tilde V^*            _{\!_{\rm A}}(q; -\omega) \,.
\end{array}\right)\;.
\label{eq:VABsuper}
\end{equation}

To summarize, the localization of ${\cal N}(1,2)$ in an $S(q)$
conserving manner has {\it uniquely\/} fixed the functions $\tilde
\Gamma_{\!\rm dd}(q)$ and $\tilde X_{\rm dd}(q)$ and, consequently, the
corresponding matrices ${\bf N}$ and ${\bf C}^{-1}$.  Calculating
${\bf V}_{\!\rm p-h}(\omega)$ from (\ref{eq:Vphmatrix}) has thus been
reduced to calculating $V_{\!_{\rm A,B}}(q;\omega)$ from ${\bf W}$.

In order to derive the explicit expressions, we need the optimal local
form of (\ref{eq:Wpmdef}). This involves two steps, calculating the
localized versions of the three-body vertices $K_{ph,p'p''h'h''}$ and
$K^{(ph)}_{p'p''h'h'',0}$, and deriving the inverse of the four-body
energy matrix $\bigl[E(\omega)]^{-1}$.  We expect these quantities to
be sufficiently accurate within the convolution approximation, since
improving on this only marginally changes the results \cite{eomI} for
bosons.

The details of the derivation of the local three-body vertices $\tilde
K_{q,q'q''}$ and $\tilde K^{(q)}_{q'q'',0}$ defined in
(\ref{eq:K12adef})-(\ref{eq:K22def}) can be found in
App.~\ref{assec:Kthree}. These are

\begin{eqnarray}
\tilde K_{q,q'q''}
	&=& \displaystyle \frac{\hbar^2}{2m} \,
		\frac{S(q')S(q'')}{\SF{q}\SF{q'}\SF{q''}}
		\left[
			\qvec\!\cdot\qvec' \, \tilde X_{\rm dd}(q') +
			\qvec\!\cdot\qvec''\, \tilde X_{\rm dd}(q'')
                        - q^2 \tilde{u}_3(q,q',q'')
		\right]
		\nonumber \\ \nonumber \\
	&&\displaystyle +
	\left[1 - \frac{\SF{q'}\SF{q''}}{S(q')S(q'')} \right] ^{-1}
	\tilde K^{(q)}_{q'q'',0} \, ,
	\label{eq:K3loc_1_2} \\
	\nonumber\\
\tilde K^{(q)}_{q'q'',0} &=&\displaystyle
	\frac{\hbar^2}{4m}\,\qG(q)\,\,
        \left[\frac{S(q')S(q'')}{\SF{q'}\SF{q''}}-1\right] 
         \Biggl\{
    \frac{q^2\,\SFthree(q,q',q'')}{\SF{q}\SF{q'}\SF{q''}}
    \;+\; \biggl[
    \frac{\qvec\cdot\qvec'}{\SF{q'}}+\frac{\qvec\cdot\qvec''}{\SF{q''}}\biggr]
    \Biggr\} \nonumber\\
	\label{eq:K3loc}
\end{eqnarray}
Here, $\SFthree(q,q',q'')$ is the three-body static structure function
of non-interacting fermions, defined in Eq.~(\ref{eq:SF3}), and
$\tilde{u}_3(q,q',q'')$ is the ground-state triplet correlation
function \cite{polish}. The implicit momentum conservation functions
$\delta_{\pm \bf q,\bf q'+ \bf q''}$ ensure that both vertices depend
on the magnitudes of the three arguments only.

Going back to the Lagrangian, we realize that the term $\tilde
K^{(q)}_{q'q'',0}$ is the coefficient function of the contributions to
${\cal L}^{'(12)}(t)$ containing $U_1(t)U_2(t)$ which we expect to be
small.  Our numerical applications to be discussed below will support
this expectation. However, the vertex $\tilde K_{q,q'q''}$ contains a
term of the same form. Neglecting $\tilde K^{(q)}_{q'q'',0}$ should,
for consistency, also mean neglecting the same term in $\tilde
K_{q,q'q''}$ which is then given by the very simple first part of
Eq. (\ref{eq:K3loc_1_2}). In this term we recover, apart from $S_F(q)$
factors, also the Bose version of the three-body vertex.

\subsection{Effective interactions}
\label{ssec:VABpm}

Next, the matrix elements (\ref{eq:K3loc_1_2}) and (\ref{eq:K3loc})
are used in (\ref{eq:Wpmdef}) to calculate the dynamic
parts of ${\bf W}$,
\begin{eqnarray}
 W_{ph,p'h'}(\omega) &=&   \,\frac{\delta_{{\bf q},{\bf q}'}}{N}\>
 \Bigl[\widetilde W(q) \>+\> \widetilde W_{\!\rm A}(q;\omega) \Bigr]
 \nonumber\\
 W_{php'h',0}(\omega) &=&   \frac{\delta_{{\bf q},-{\bf q}'}}{N}
 \Bigl[\widetilde W(q) \>+\> \widetilde W_{\!\rm B}(q;\omega) \Bigr]
 \,,
 \label{eq:WABdef}
\end{eqnarray}
where the energy independent part $\widetilde W(q)$ has been defined
in Eq.~(\ref{eq:Wlocal}). Because of the locality of the three-body
matrix elements, we can write for the first dynamic
contribution to (\ref{eq:Wpmdef}),
\begin{eqnarray}
&& \sum_{p_1p_2h_1h_2}\, \sum_{p_1'p_2'h_1'h_2'}  K_{ph,\,p_1p_2h_1h_2}\>
 \Bigl[E(\omega)^{-1}\Bigr]_{p_1p_2h_1h_2,p_1'p_2'h_1'h_2'}\,
 K_{p_1'p_2'h_1'h_2',\,p'h'} 
 \nonumber\\ 
 &=&\;
 \frac{1}{N^2}\!\sum_{q_1q_1'} \tilde K_{q,\,q_1q_2}\,\tilde K_{q_1'q_2',\,q} \>
 \frac{1}{N^2}\!\sum_{h_1h_2h_1'h_2'}  
 \Bigl[E(\omega)^{-1}\Bigr]_{p_1p_2h_1h_2,p_1'p_2'h_1'h_2'}
\nonumber\\
&=& \frac{1}{N^2}\!\sum_{q_1q_2}\tilde K_{q,\,q_1q_2}\,
\tilde E^{-1}(q_1,q_2;\omega)
\tilde K_{q_1q_2,\,q} \> \label{eq:invE_aux}
\end{eqnarray}
with implicit factors $\delta_{{\bf q},{\bf q}_1+{\bf q}_2}\,
\delta_{{\bf q},{\bf q}_1'+{\bf q}_2'}$ for momentum conservation.
The other contributions to (\ref{eq:Wpmdef}) are calculated
analogously.  The inverse four body energy matrix and the pair
propagator
\begin{equation}
 \frac{1}{N^2}\sum_{hh'h''h'''}
 \Bigl[E(\omega)^{-1}\Bigr]_{pp'hh',p''p'''h''h'''}
 \equiv \delta_{q,q''}\delta_{q',q''}\, \tilde E^{-1}(q,q';\omega) \,.
 \label{eq:invE}
\end{equation}
are calculated and discussed in App.\ \ref{asec:ex_pair_eqq}.
Basically, the pair spectrum is built from two particle-hole 
spectra. These are, however, not centered around free particle
spectra but around the Feynman dispersion relation.  Consequently, our
pair propagator also includes {\it two-phonon\/} intermediate states.

The resulting expressions for the energy-dependent 
$\widetilde W_{\!_{\rm {A,B}}}(q; \omega)$ are then
\begin{eqnarray}
 \widetilde W_{\!\rm A}(q;\omega) &=&
  \frac{1}{2N}\sum_{{\bf q}'{\bf q}''} \left[
  |\tilde K_{q,q'q''}|^2 \, \tilde E^{-1}(q',q''; \omega) +
  |\tilde K_{q'q'',0}^{(q)}|^2\, \tilde E^{-1*}(q',q''; -\omega)  \right] \, ,
 \label{eq:resultWA} \\
 \widetilde W_{\!\rm B}(q;\omega) &=&
  \frac{1}{2N}\sum_{{\bf q}'{\bf q}''} \left[
  \tilde K_{q'q'',0}^{(q)}\tilde K_{q,q'q''}\,
  \Bigl(\tilde E^{-1}(q',q''; \omega) 
  +\; \tilde E^{-1*}(q',q''; -\omega)  \Bigr)\right] \, .
  \label{eq:resultWB} 
\end{eqnarray}
Similar to the boson theory, the dynamic parts of the interactions are 
expressed in terms of three-body vertices and an energy denominator, 
the latter now being ``spread'' over the whole width of a 
two-particle-two-hole band.

The last step in our formal derivations is the calculation of 
${\bf V}_{\!\rm p-h}(\omega)$.
Carrying out the operations (\ref{eq:Vphmatrix}) yields the energy-dependent,
but local functions
\begin{eqnarray}
\tilde V_{\!_{\rm A}}(q; \omega) = \tilde V_{\rm p-h}(q)
&+& [\sigma^{+}_q]^2\, \widetilde W  _{\!_{\rm A}}(q; \omega) +
[\sigma^{-}_q]^2\, \widetilde W^*_{\!_{\rm A}}(q;-\omega) \nonumber \\
&+& \sigma^{+}_q\sigma^{-}_q\, \left(
		\widetilde W_{\!_{\rm B}}  (q;  \omega) +
		\widetilde W_{\!_{\rm B}}^*(q; -\omega)
	\right)\;,
\label{eq:VAdef} \\
\tilde V_{\!_{\rm B}}(q; \omega) = \tilde V_{\rm p-h}(q)
&+& [\sigma^{+}_q]^2\, \widetilde W_{\!_{\rm B}}  (q; \omega) +
[\sigma^{-}_q]^2\, \widetilde W_{\!_{\rm B}}^*(q;-\omega) \nonumber \\
&+& \sigma^{+}_q\sigma^{-}_q\, \left(
		\widetilde W_{\!_{\rm A}}  (q; \omega) +
		\widetilde W_{\!_{\rm A}}^*(q;-\omega)
	\right)\;,
\label{eq:VBdef}
\end{eqnarray}
with $\sigma^{\pm}_q \equiv [\SF{q}\pm S(q)]/2S(q)$\,.

\section{Density-density response function}
\label{sec:chi}

\subsection{General form}
\label{ssec:responsefct}

We now derive the density-density response function $\chi(q; \omega)$.
The final result for the dynamic effective interactions,
(\ref{eq:VAdef}), (\ref{eq:VBdef}), 
is inserted into (\ref{eq:Vabba}), which is solved for $\delta{\bf c}$. 
The induced density is then obtained from Eq.~(\ref{eq:d_rho_r_om}).
Using 
$\rho^{\rm F}_{0,ph}({\bf r}) = 
	\frac{\rho}{N} e^{-\I({\bf p}-{\bf h}){\bf r}}$
we obtain
\begin{eqnarray}
\delta\rho(q; \omega)
	&=& \displaystyle\frac{\rho}{2}\displaystyle\sum_{h} \left[
	z_{\vec h+\vec q,\vec h}\, \delta c^{(1+)}_{\vec h+\vec q,\vec h}(\omega) \,
        \bar n_{{\bf h}-{\bf q}}
 	+z_{\vec h-\vec q,\vec h}\, \delta c^{(1-)}_{\vec h-\vec q,\vec h}(\omega) \,
        \bar n_{{\bf h}+{\bf q}}
        \right] \nonumber \\
	&\approx& \displaystyle\frac{N\SF{k}\, \rho}{2}\displaystyle 
        \left[ \delta c^{(1+)}(q; \omega)
          + \delta c^{(1-)}(q; \omega) \right]
        \;,
	\label{eq:rho_from_c}
\end{eqnarray}
where we abbreviate in the second line $\delta c^{(1\pm)}(q;\omega)
\equiv \frac{1}{N}\sum_h \delta c^{(1\pm)}_{ph}(\omega)$.
Spelling out Eqs.~(\ref{eq:Vabba}) explicitly,
\begin{eqnarray}
2\tilde h_{0,ph}^{F}(\omega)
	&=& \left(\pm(\hbar\omega\!+\!\I\eta)-e_{ph}\right) 
            \delta c^{(\pm)}_{ph}(\omega) \nonumber\\
	&-& \tilde V^{\phantom{*}}_{\!_{\rm A}}(q; \omega)\,
		\delta c^{(\pm)}(q; \omega)
	-  \tilde V^*_{\!_{\rm B}}(q;-\omega)\,
		\delta c^{(\mp)}(q; \omega) \;,
	\label{eq:EOMABBA_pm}
\end{eqnarray}
dividing by $\left(\pm(\hbar\omega\!+\!\I\eta)-e_{ph}\right)$ and summing over
$h$ yields
\begin{equation}
\delta c^{(1\pm)}(q; \omega) \>=\>
\left[\frac{2}{N}\,\tilde h_{\rm ext}(q; \omega) 
+ \tilde V^{\phantom{*}\!\!}_{\!_{\rm A}}(q;\omega)\,\delta c^{(1\pm)}(q;\omega)
+ \tilde V^*                _{\!_{\rm B}}(q;-\omega)\,\delta c^{(1\mp)}(q;\omega)
\right]\left\{\!\begin{array}{llll} \kappa_0^{\phantom*}(q;\,\omega)\\
              \kappa_0^*(q;-\omega)\end{array}\right.
\label{eq:EOMABBA_FSA_pm} \\
\end{equation}
with the positive-energy Lindhard function
\begin{equation}
\kappa^{\phantom{*}}_0(q;\omega) \equiv \frac{1}{N}\sum_h
\frac{\bar n_{\bf p} n_{\bf h}}{\hbar\omega - e_{ph} + \I\eta}
\label{eq:chi0pmdef}
\end{equation}
which is related to the full Lindhard function by
\begin{equation}
 \chi_0(q;\omega) = \kappa^{\phantom{*}}_0(q; \omega) + \kappa^*_0(q;-\omega)\,.
\end{equation}

Solving for $\delta c^{(1\pm)}(q;\omega)$ and inserting into 
(\ref{eq:rho_from_c}) we obtain for $\chi(q; \omega)$
\begin{eqnarray}
\chi(q; \omega)
	&=& N(q; \omega) / D(q; \omega) \nonumber\\
N(q; \omega)
	&=& \kappa^{\phantom{*}}_0(q; \omega) + \kappa_0^*(q; -\omega) \nonumber\\
	&-& \kappa^{\phantom{*}}_0(q; \omega)   \kappa_0^*(q; -\omega) \left[
		  \tilde V  _{\!_{\rm A}}(q;  \omega) 
		+ \tilde V^*_{\!_{\rm A}}(q; -\omega)
		- \tilde V  _{\!_{\rm B}}(q;  \omega) 
		- \tilde V^*_{\!_{\rm B}}(q; -\omega)
	\right] \nonumber\\
D(q; \omega) 
	&=& 1
	- \kappa^{\phantom{*}}_0  (q; \omega) \tilde V  _{\!_{\rm A}}(q;  \omega)
	- \kappa_0^*(q;-\omega) \tilde V^*_{\!_{\rm A}}(q; -\omega) \nonumber\\
	&+&   \kappa^{\phantom{*}}_0(q; \omega) \kappa_0^*(q; -\omega)
\left[\tilde V_{\!_{\rm A}}(q; \omega) \tilde V^{*}_{\!_{\rm A}}(q; -\omega)
     -\tilde V_{\!_{\rm B}}(q; \omega) \tilde V^{*}_{\!_{\rm B}}(q; -\omega)
\right]\;.
\label{eq:structurechi}
\end{eqnarray}
Eq.~(\ref{eq:structurechi}) is the TDHF response function for local
and energy dependent interactions. Evidently, the conventional RPA
form (\ref{eq:RPAresponse}) can only be recovered if the interactions
$\tilde V_{\!_{\rm A}}(q; \omega)$ and $\tilde V_{\!_{\rm B}}(q;
\omega)$ are {\em energy independent\/} and {\it equal.\/} Clearly,
our result (\ref{eq:structurechi}) significantly differs from
(\ref{eq:RPAresponse}) with $\Vph{q}$ simply replaced by some energy
dependent $\Vph{q; \omega}$. Such an RPA-like form for the
density-density response function lacks microscopic justification.

\subsection{Long wavelength limit}
\label{ssec:longwaves}

In the limit $q\rightarrow 0$, the spectrum is dominated by collective
excitations, {\it e.g.\/} zero sound or plasmons.  Both vertices
(\ref{eq:K3loc_1_2}) and (\ref{eq:K3loc}) vanish linearly in $q$,
hence $\widetilde W_{\!_{\rm A}}(q;\omega)$ and $\widetilde W_{\!_{\rm
    B}}(q;\omega)$ are quadratic in $q$ as $q\rightarrow 0$.

For neutral systems, the dynamic corrections to the effective
interactions $\tilde V_{\!_{\rm A,B}}(q;\omega)$ in (\ref{eq:VAdef}),
(\ref{eq:VBdef}) are therefore negligible in the long wavelength
limit. The long wavelengths density-density response function is then
given by its RPA form (\ref{eq:RPAresponse}), with the static
particle-hole interaction $\Vph{q}$.  The zero sound speed $c_0$ is
determined by the long wavelength solution of the RPA equation.

For charged quantum fluids, $\sigma_q^\pm\approx \SF{q}/2S(q)$,
hence $\tilde V_{\!_{\rm A}}(q,\omega) = \tilde V_{\!_{\rm B}}(q,\omega)$,
which again implies
the RPA form (\ref{eq:RPAresponse})
\begin{equation}
 \chi(q;\omega) \>=\> \frac{\chi_0(q;\omega)}{1- \chi_0(q;\omega)\,
                      \tilde V_{\!_{\rm A}}(q;\omega)}
\quad \mbox{as}\quad q\!\to\!0
 \,.
 \label{eq:chiCqto0}
\end{equation}
However, now the effective interaction is
\begin{equation}
 \tilde V_{\!_{\rm A}}(q;\omega) \;=\; \Vph{q} +
\frac{S_F^2(q)}{4S^2(q)}
\left[\widetilde W_{\!_{\rm A}}(q;\omega) +\widetilde W_{\!_{\rm A}}(q;-\omega) +
\widetilde W_{\!_{\rm B}}(q;\omega)+\widetilde W_{\!_{\rm B}}(q;-\omega)
\right]\quad \mbox{as}\quad q\!\to\!0\,.
 \label{eq:VAqto0w}
\end{equation}
The static particle-hole interaction
approaches the Coulomb potential
$\tilde v_{\rm c}(q) = 4\pi e^2/q^2$
\begin{equation}
\Vph{q} = \tilde v_{\rm c}(q) + V_0\quad \mbox{as}\quad q\!\to\!0\,.
\label{eq:Vphqto0}
\end{equation}
We can therefore write (\ref{eq:VAqto0w}) as
\begin{equation}
 \tilde V_{\!_{\rm A}}(q;\omega) \;=\;
\tilde V_{\!_{\rm B}}(q;\omega) \;=\;
 \tilde v_{\rm c}(q) + V_0(\omega) \quad \mbox{as}\quad q\!\to\!0\,.
\end{equation} 
As for charged bosons \cite{bosegas}, the two-pair fluctuations modify
the RPA result. The static potential $\Vph{q}$ and $\widetilde W_{\rm
  A,B}(q;\omega)$ contribute for $q\!\to\!0$ at the same level.

\subsection{Static response function}
\label{ssec:staticresponse}

$\tilde E^{-1}(q,q';\omega\!=\!0)$ is real and negative, this is most easily
seen from the representation (\ref{eq:EinvfromImkappa}).  Therefore,
all interactions $\widetilde W_{\!_{\rm A,B}}(q; 0)$ in
(\ref{eq:resultWA})-(\ref{eq:resultWB}) and $\tilde V_{\!_{\rm A,
    B}}(q; 0)$ in (\ref{eq:VAdef})-(\ref{eq:VBdef}) are real.  The
response function (\ref{eq:structurechi}) can again be cast into the
RPA form
\begin{equation}
 \chi(q; 0)
 = \frac{\chi_0(q; 0)}{1 - \tilde V_{\!\rm stat}(q)\, \chi_0(q; 0) }
 \,,
 \label{eq:response-function-static}
\end{equation}
with a static effective interaction
\begin{eqnarray}
	\tilde V_{\!\rm stat}(q) &\equiv& 
	\Vph{q} + \frac{S_{\rm F}^2(q)}{2S^2(q)}\,\Bigl[\widetilde
	 W_{\!_{\rm A}}(q; 0) + \widetilde W_{\!_{\rm B}}(q; 0) \Bigr] 
        \;.
	\label{eq:Vstatic}
\end{eqnarray}
Unlike Eq. (\ref{eq:VAqto0w}), this form holds for all
wavelengths.

For {\it short wavelengths\/} the static response function has the
asymptotic form \cite{HolasReview,VigGiulBook}

\begin{equation}
\chi(q\!\to\!\infty; 0) = -\frac{2}{ t(q)} - \frac{8}{3 t^2(q)}
\frac{\bigl\langle \hat T \bigr\rangle}{N}
+ {\cal O}(q^{-5})\,,
\label{eq:chilarge}
\end{equation}
where $\bigl\langle \hat T \bigr\rangle$ is the kinetic
energy. In the RPA, one obtains in Eq. (\ref{eq:chilarge}) only the
kinetic energy of the non-interacting system. To obtain the correct
asymptotic form, it is therefore necessary to include pair and,
possibly, higher order fluctuations.

Again, we know the result for bosons as a guide: treating pair
fluctuations in the ``convolution'' approximation leads to the correct
asymptotic behavior with $\bigl\langle \hat T \bigr\rangle$ in
(\ref{eq:chilarge}) given in that approximation \cite{eomII}.

We show in App.~\ref{asec:static} that
\begin{equation}
\tilde V_{\rm stat}(q\!\to\!\infty) =
\frac{1}{2} \widetilde W_{\!_{\rm A}}(q\!\to\!\infty;0) \>=\> - \frac{2}{3} 
   \frac{\bigl\langle T \bigr\rangle^{\!\scriptscriptstyle\rm CA} - T_{\rm F}}{N}\, ,
 \label{eq:Vstatqinf}
\end{equation}
where $\langle T \rangle^{\!\scriptscriptstyle\rm CA}$ is the kinetic
energy in ``uniform limit'' or ``convolution'' approximation (\ref{eq:TCA}).
Hence,
inserting the short wavelength expansion of the Lindhard function,
the static response function (\ref{eq:response-function-static}) 
indeed assumes the form (\ref{eq:chilarge})
\begin{eqnarray}
\chi(q; 0)
&=& -\frac{2}{t(q)} - \frac{8}{3 t^2(q)} \frac{\left\langle T
\right\rangle^{\!\scriptscriptstyle\rm CA}}{N}
\quad\mbox{as}\quad q\!\to\!\infty\,,
\label{eq:response-function-static_large_q}
\end{eqnarray}
with the kinetic energy being calculated in the uniform limit approximation 
(\ref{eq:TCA}).

\section{Applications}
\label{sec:applications}

\subsection{Dynamic structure of
\texorpdfstring{$^3$He}{3He}}
\label{ssec:he3}

\subsubsection{Motivation}

The helium fluids are the prime examples of strongly correlated
quantum many-body systems. They have been studied for decades, and
still offer surprises leading to new insight. It is fair to say that
understanding the helium fluids lies at the core of understanding
other strongly correlated systems.  The most important and most
interesting field of application of our theory is therefore liquid
\he3.

Recent developments \cite{eomI,QFS09_He4} have brought manifestly
microscopic theories of \he4 to a level where quantitative predictions
of the excitation spectrum are possible far beyond the roton minimum
without any information other than the underlying microscopic Hamiltonian
(\ref{eq:Hamiltonian}).  \he3 is the more challenging substance for
both, theoretical and experimental investigations. Experimentally, the
dynamic structure function $S(q;\omega)$ of \he3 is mostly determined
by neutron scattering. The results are well documented in a book
\cite{GlydeBook}, the theoretical and experimental understanding a
decade ago has been summarized in Ref.~\onlinecite{GFvDG00}.  Recent
inelastic X-ray scattering experiments have led to a controversy on
the evolution of the zero sound mode at intermediate wave-vectors
\cite{Albergamo,Albergamo_comment,Albergamo_reply}, we will comment on
this issue below.

The RPA (\ref{eq:RPAresponse}) suggests that $S(q;\omega)$ can be
characterized as a superposition of a collective mode similar to the
phonon-maxon-roton in $^4$He, {\it plus\/} an incoherent particle-hole
band which strongly dampens this mode \cite{PinesPhysToday}. The
picture is {\em qualitatively\/} adequate but misses some important
{\em quantitative\/} physics: In \he3 the RPA, when defined through
the form (\ref{eq:RPAresponse}) and such that the sum rules
(\ref{eq:m0})--(\ref{eq:m1}) are satisfied, predicts a zero-sound mode
that is significantly too high. This is consistent with the same
deficiency of the Feynman spectrum (\ref{eq:efeyn}) in \he4.  Drawing
on the analogy to \he4 \cite{PinesPhysToday}, the cure for the problem
is, as pointed out above, to include pair fluctuations $\delta
u^{(2)}_{pp'hh'} (t)$ in the excitation operator.

An alternative, namely to lower the collective mode's energy by
introduction of an effective mass in the Lindhard function, leads to
various difficulties: First, one violates the sum rules
(\ref{eq:m0})--(\ref{eq:m1}), {\it i.e.\/} one disregards well
established information on the system. Second, the effective mass is
far from constant; it has a strong peak around the Fermi momentum
\cite{PethickMass,ZaringhalamMass,Bengt,he3mass}, a secondary maximum
around $2 \qkf$, and then quickly falls off to the value of the bare
mass. In fact, it is not even clear if the notion of a ``single
(quasi-)particle spectrum'' that is characterized by a momentum
is adequate at these wave numbers.
 
The localization procedure of Sec.~\ref{sec:localapprox} implies that
the only input needed for the application of our theory is the static
structure function $S(q)$, whereas the single-particle spectrum is
that of a free particle.  We hasten to state that we do {\it not\/}
claim that the precise location of the single-particle spectrum is
completely irrelevant for the energetics of the zero sound; we only
claim that the {\it dominant mechanism\/} in Bose and Fermi fluids is
the same, namely pair-fluctuations.  In order to maintain the sum
rules (\ref{eq:m0})--(\ref{eq:m1}), any modification of the
particle-hole spectrum must go along with an inclusion of exchange
effects. At the level of single-particle fluctuations
\cite{LDavid,rings}, such a calculation is quite feasible
\cite{NaiThesis,QFS09_exch}. However, to describe the dynamics of \he3
correctly, it is insufficient to include only the CBF single particle
energies (\ref{eq:CBFph}).  These suggest a smooth spectrum with an
effective mass slightly less than the bare mass, in contradiction to
the highly structured spectrum mentioned already above.

\subsubsection{Collective mode}

\begin{figure}
   \includegraphics[width=0.48\textwidth]{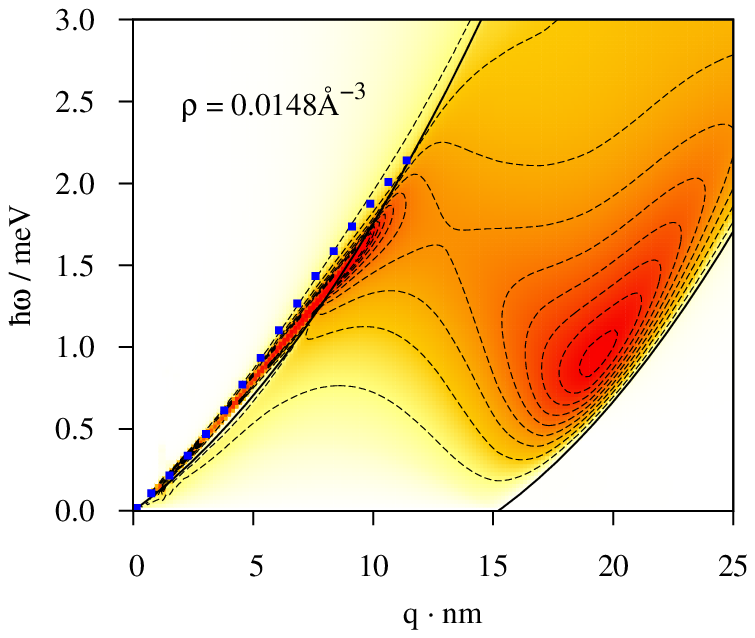}
\hfill
   \includegraphics[width=0.48\textwidth]{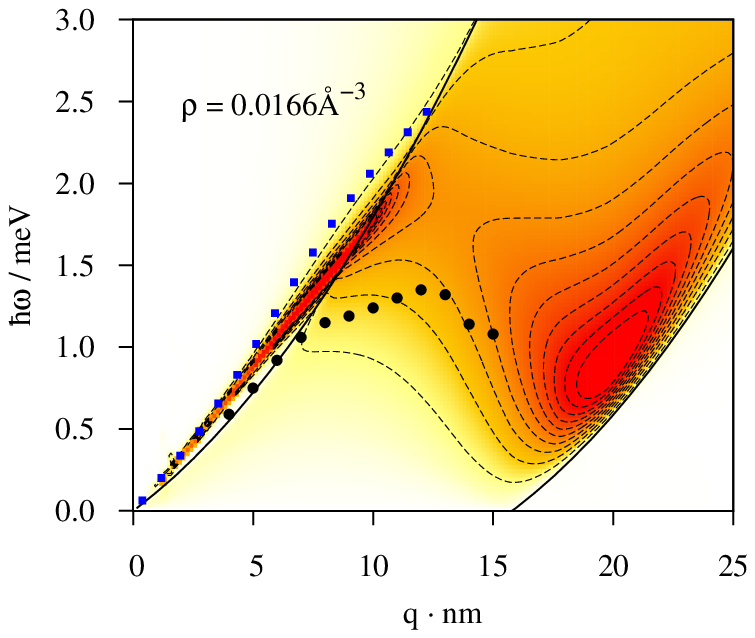}\\
   \includegraphics[width=0.48\textwidth]{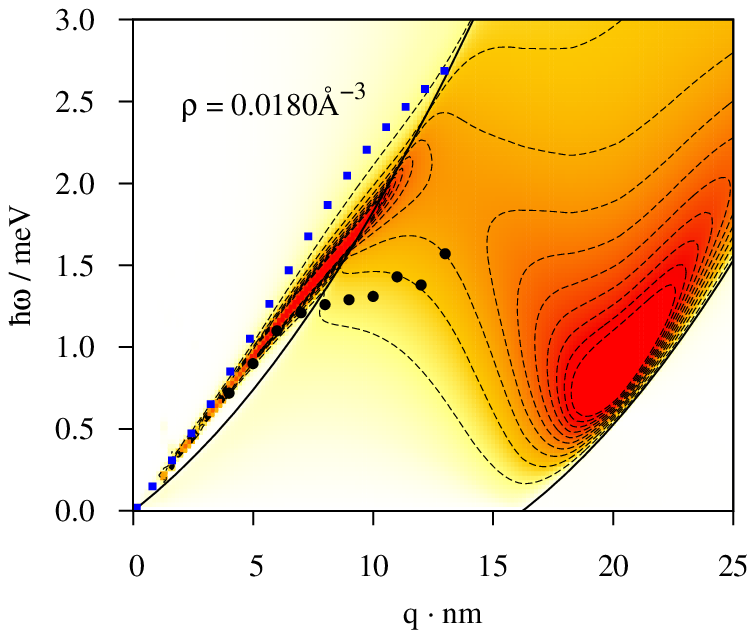}
\hfill
   \includegraphics[width=0.48\textwidth]{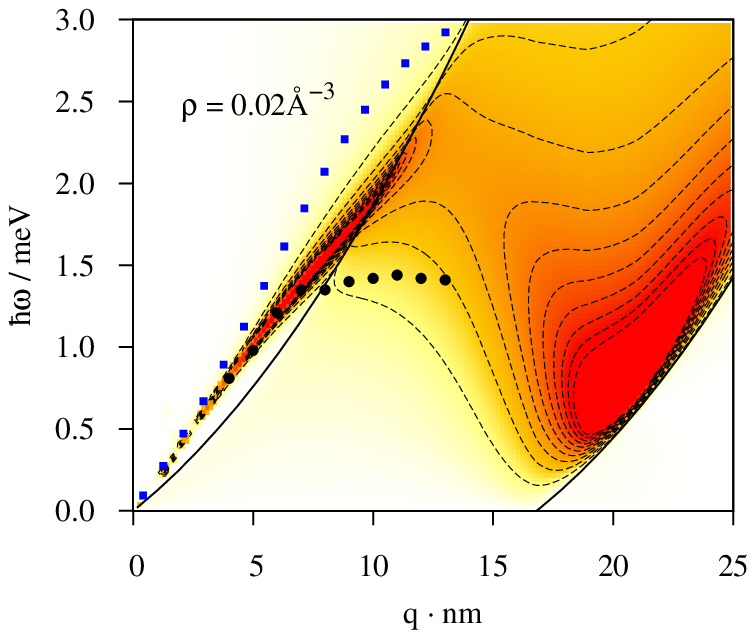}\\
   \caption{(Color online) $S(q;\omega)$ of $^3$He, for the densities
     $\rho= 0.0148,\; 0.0166,\; 0.018,\; 0.02 $\AA$^{-3}$. The
     experimental results for the collective mode (dots) are from
     inelastic neutron scattering experiments at the ILL
     (Ref. \onlinecite{Fak94}). The densities 0.0166, 0.0180 and 0.0200
     \AA$^{-3}$ correspond in good approximation to the pressures
     $p=0,\;5,\;10$ bar \cite{GRE86,OuY}.  Dashed lines are
     equidistant contours marking the same absolute value in all
     plots. Solid lines are the boundaries of the particle-hole
     continuum for $m^*=m$.  The blue boxes show the RPA
     result for the collective mode.
     \label{fig:He3_contour}}
\end{figure}

For our calculations we have used input from the FHNC-EL calculations
of Ref. \onlinecite{polish} that utilizes the Aziz-II potential
\cite{AzizII} and includes optimized triplet correlations as well as
four- and five-body elementary diagrams.  An overview of our
results for bulk \he3 and a comparison with both the RPA and
experimental data is shown in Figs.~\ref{fig:He3_contour} for four
different densities.  The most prominent consequence of pair
fluctuations is a change in energy and strength of the collective
mode and its continuation into the particle-hole band. Pair
fluctuations also contribute a continuum background outside the
particle-hole continuum.

At long wavelengths, the collective mode is sharp and well defined
above the particle-hole band, exhausting most of the sum rules
(\ref{eq:m0}) and (\ref{eq:m1}). In this regime, the RPA provides a
faithful description of the physics. This is in accordance with the
observation that the dynamic correction to the effective interactions
vanish, for neutral systems, in the long-wavelength limit.  With
increasing density, the speed of sound increases and the phonon
becomes farther separated from the particle-hole band.

\begin{figure}
\includegraphics[width=0.48\textwidth]{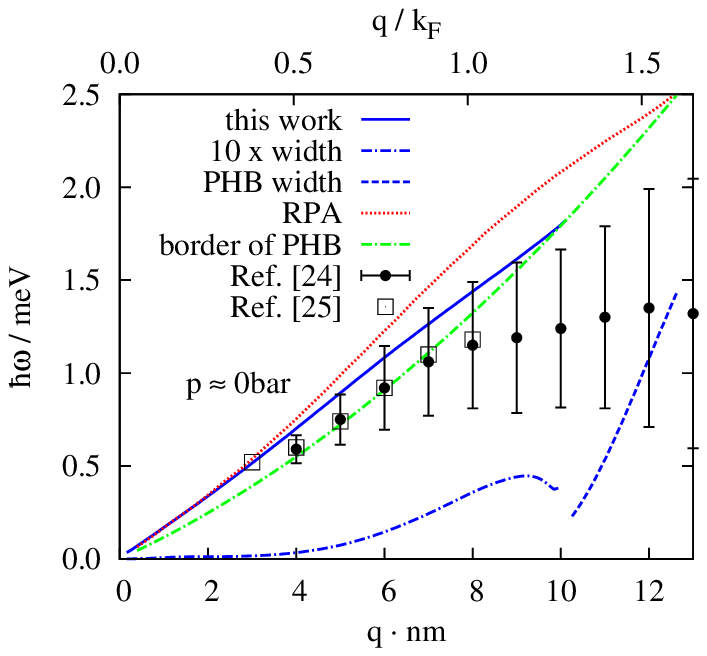}\hfill
\includegraphics[width=0.48\textwidth]{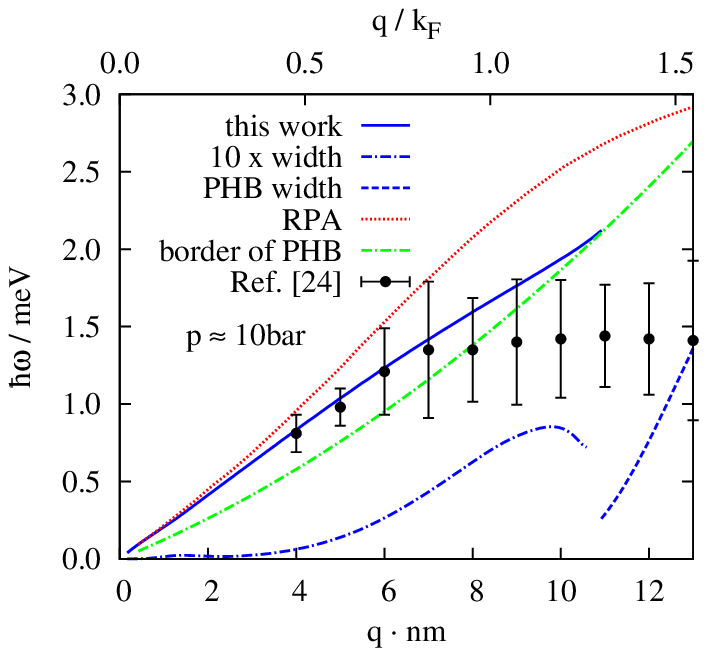}
\caption{Zero sound mode calculated within the pair fluctuation theory
(full blue line), 
RPA (red chained line) and experimental data by the ILL group 
\cite{GFvDG00} (square symbols) and \cite{Fak94} 
(circles). The bars indicate the width of the fit to the data,
the line at the bottom of the figure gives the width due to pair fluctuations 
enhanced by a factor of 10 to make it visible. The dashed blue line gives the 
FWHM of the mode within the particle hole continuum. Left part: $\rho=0.0166
$\AA$^{-3}$, right part: $\rho=0.02 $\AA$^{-3}$.
\label{fig:zero_sound_166}}
\end{figure}

Further details are shown in Fig.~\ref{fig:zero_sound_166}.  At
intermediate wavelengths the collective mode bends down due to the
attractiveness of the effective interaction. This is where the dynamic
theory starts to deviate visibly from the RPA.  Evidently, pair
fluctuations are the major cause for lowering the energy of the
collective mode, although they do not completely bridge the
discrepancy between the RPA and experiments \cite{Fak94,GFvDG00}. This
is expected because, for bosons, pair fluctuations bridge only about
two thirds of the gap between the Feynman and the experimental roton
energy \cite{Chuckphonon,eomI}.  Three-body and higher-order
fluctuations are also important \cite{eomII}. We expect that these
corrections are smaller in \he3 due to its lower density, yet not
negligible.

When the collective mode enters the particle-hole band, a slight kink
in the position of the maximum in $S(q;\omega)$ is expected, as well
as an abrupt broadening of the mode.  At saturated vacuum pressure,
shown in the left part of Fig.~\ref{fig:zero_sound_166}, these effects
are difficult to identify in the experiments \cite{GFvDG00}.  A
possible reason is that the observed mode stays always very close to
the particle-hole band. The measured mode width in
Fig.~\ref{fig:zero_sound_166} gives no clear indication of the upper
boundary of the particle-hole band other than that it is determined by
a spectrum with an average effective mass of $m^* \lessapprox m$.

The situation is much clearer at higher pressure: With increasing
density, the speed of sound increases, separating the collective mode
farther from the particle-hole band. For $\rho\!=0.02\,$\AA$^{-3}$ a
clear kink is identified at $q \!\approx5\,\mathrm{nm}^{-1}$
(Fig.~\ref{fig:zero_sound_166} right part).  The broadening is also
more abrupt and, in particular, does not increase for larger values of
$q$. Similar to SVP, explaining these data requires a boundary of the
particle-hole band that is even above that of the non-interacting
Fermi fluid. Damping due to multiparticle excitations is, on the other
hand, for both densities far too small to account for the
experimentally seen broadening of the zero sound mode.

\subsubsection{\texorpdfstring
	{Frequency dependence of $S(q;\omega)$}
	{Frequency dependence of S(q;omega)}
}

\begin{figure}
\includegraphics[width=0.48\textwidth]{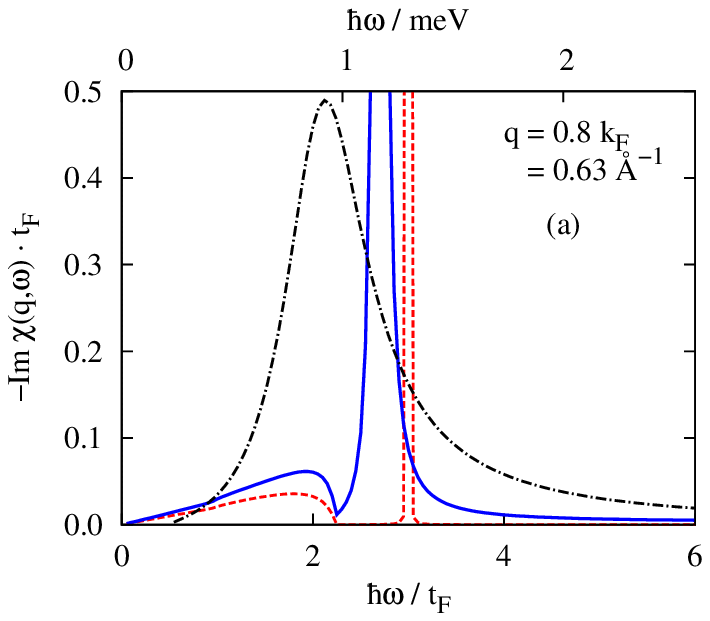}\hfill
\includegraphics[width=0.48\textwidth]{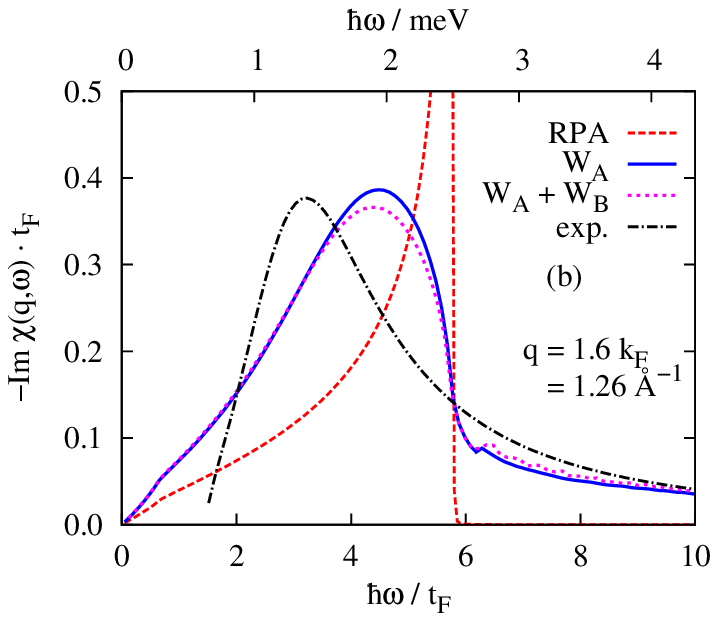}\\

\includegraphics[width=0.48\textwidth]{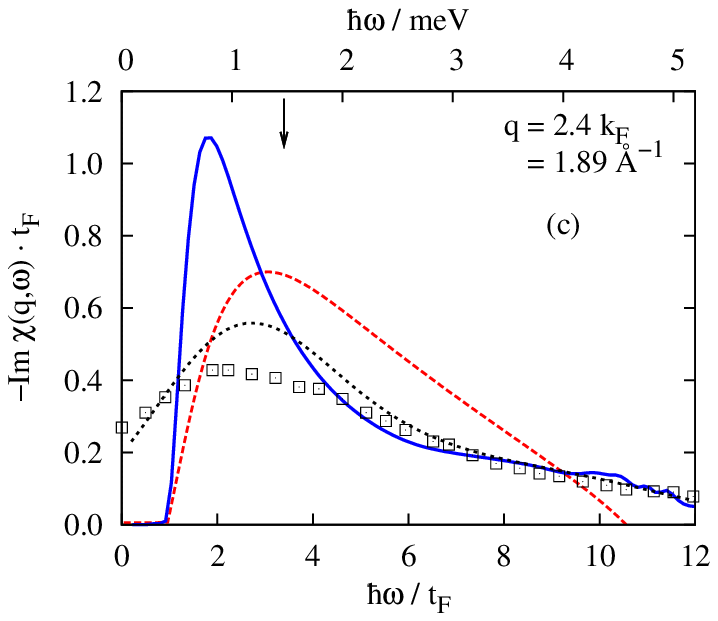}\hfill
\includegraphics[width=0.48\textwidth]{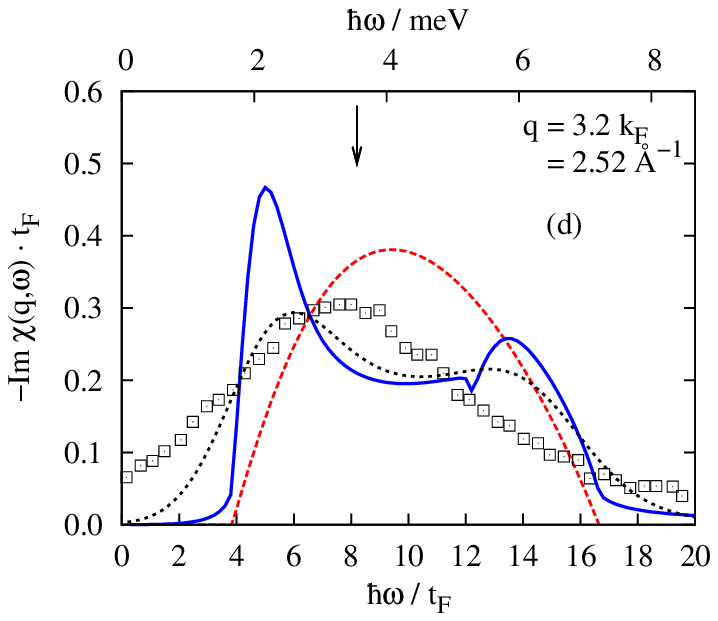}

\caption{(Color online) $S(q;\omega)$ for \he3 as a function of energy at
  $\rho=0.0166$\AA$^{-3}$ for a sequence of momentum transfers
  $q=0.8,\; 1.6,\; 2.4,\; 3.2 \;\qkf$ (a)-(d). Also shown is the RPA
  (dashed, red). The solid blue line is the result of this work with
  the simplified $\widetilde W_{\!_{\rm A}}(q;\omega)$ and $\widetilde
  W_{\!_{\rm B}}(q,\omega)=0$ as discussed in the text. In pane (b),
  we also show the results when the full $\widetilde W_{\!_{\rm
      A}}(q;\omega)$ and $\widetilde W_{\!_{\rm B}}(q;\omega)$ of
  Eqs. (\ref{eq:resultWA}) and (\ref{eq:resultWB}) are retained (short
  dashed magenta line).  The results from the different approximations
  are almost indistinguishable in panes (a),(c) and (d) and therefore
  not shown. The black dash-dotted line in panes (a) and (b) are fits
  to the experimental results of Ref. \onlinecite{Fak94}. In panes (c)
  and (d) we indicate the maximum of the experimentally observed
  dynamic structure function by an arrow. We also plot in panes (c)
  and (d) recent inelastic X-ray diffraction data obtained by
  Albergamo \textit{et al.\/} \cite{Albergamo} (boxes) as well as our
  theoretical results folded with the experimental resolution (dashed
  line).
  \label{fig:He3_cut}}
\end{figure}

For a quantitative discussion we show in Fig.~\ref{fig:He3_cut} the
dynamic structure factor as a function of frequency at a sequence of
wave vectors. We conclude that the RPA quantitatively and even
qualitatively differs from our theory and the experiment. Including
pair fluctuations improves the agreement with experiment
significantly. The arrows in panes (c) and (d) indicate the maximum of
the experimentally observed dynamic structure function.

In Fig.~\ref{fig:He3_cut}(b) we also show the consequence of the
plausible simplification of our theory discussed already in connection
with Eqs. (\ref{eq:K3loc_1_2}) and (\ref{eq:K3loc}): We neglect all
terms that vanish for bosons as well as for large momentum
transfers $q,q',q''\ge 2k_F$. This is $\tilde K^{(q)}_{q'q'',0}$ and,
consequently, the second term in $K_{q, q' q''}$,
Eq. (\ref{eq:K3loc_1_2}). The three-body vertex is then given by
the first term in Eq.  (\ref{eq:K3loc_1_2}), see also
(\ref{eq:K3loc_1_2_large_q}). This simplifies the effective
interactions significantly: Only the first term of
Eq. (\ref{eq:resultWA}) for $\widetilde W_{\!\rm A}(q;\omega)$
contributes, and $\widetilde W_{\!\rm B}(q;\omega)$ is neglected.
Fig.~\ref{fig:He3_cut}(b) shows that these simplifications modify our
results only marginally, the form (\ref{eq:K3loc_1_2_large_q}) can
therefore be considered a practical and useful simplification of our
theory.

Figs.~\ref{fig:He3_cut}(c) and \ref{fig:He3_cut}(d) show our results
for the two momentum transfers $q = 2.4\,\qkf = 1.89\,$\AA$^{-1}$ and
$q = 3.2\,\qkf = 2.52\,$\AA$^{-1}$.  Recent X-ray scattering
experiments in that momentum range
\cite{Albergamo,Albergamo_comment,Albergamo_reply} appeared to support
the notion of a high-momentum collective mode without visible damping
by incoherent particle-hole excitations. Figs.~\ref{fig:He3_cut}(c)
and \ref{fig:He3_cut}(d) show that pair fluctuations lead to a
narrowing of the strength of $S(q;\omega)$ compared to the RPA. To
facilitate the comparison with experiments, we have convoluted our
result with the instrumental resolution of $1.58\,{\rm meV}$, the
results are also shown in Figs.~\ref{fig:He3_cut}(c) and
\ref{fig:He3_cut}(d).  After this, our results agree quite well with
the experimental spectrum. Also, the location of the observed peak
intensity for $q = 2.4\,\qkf$ appears to be consistent with our
calculation. The RPA is, on the other hand, too broad to explain the
data. We also point out that a value of the effective mass close to
$m^*\approx m$ is consistent with our theoretical calculations
\cite{he3mass}. We have to conclude therefore that the observed width
of the X-ray data are also consistent with our picture.

After a regime of strong damping we see in Figs.~\ref{fig:He3_contour}
an intensity peak at momentum transfer of $q\approx 2.5 \qkf$. With
increasing density, this peak moves towards the lower edge of the
particle-hole band and becomes sharper. Such a peak should be
identified with the remnant of the roton excitation in \he4, broadened
by the particle-hole continuum. The overall agreement with the
experiment is quite good, see Fig.~1 of Ref. \onlinecite{Fak94}. Our theory
predicts a ``roton minimum'' that is slightly above the observed energy;
this is expected because for bosons a similar effect is observed.  To
obtain a higher accuracy, triplet- and higher order fluctuations must
be included \cite{eomII}.

\subsubsection{Static response}

\begin{figure}
\centerline{\includegraphics[width=8.6cm]{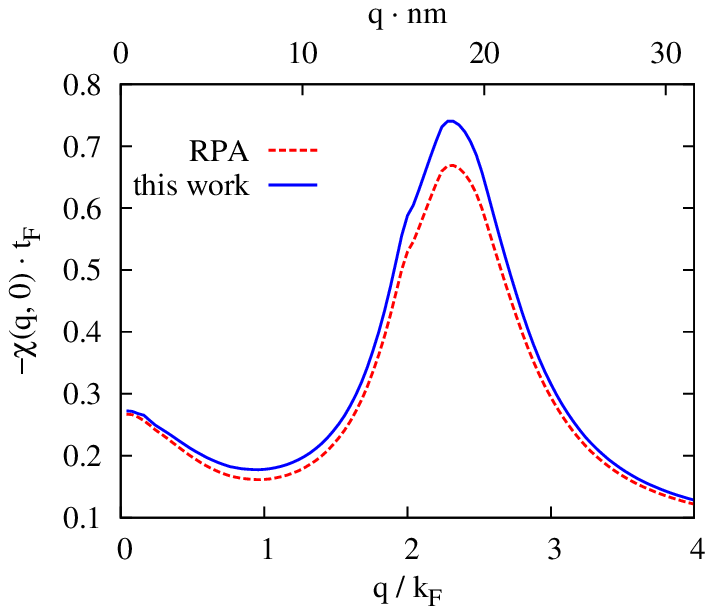}}
\caption{(color online) Static response of \he3 at $\rho=0.0166$\AA$^{-3}$.
The red curve shows the RPA result whereas the blue line is the result
of this work. 
\label{fig:static_resp}}
\end{figure}
\begin{figure}
\centerline{\includegraphics[width=8.6cm]{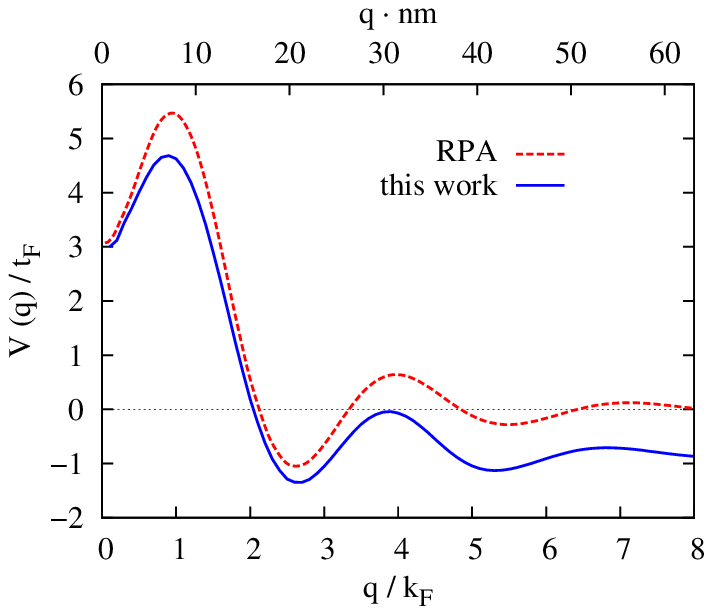}}
\caption{(color online) Effective interaction of \he3 at $\rho=0.0166$\AA
$^{-3}$. 
The red curve shows the static effective interaction $\Vph{q}$ whereas
the blue line is $\tilde V_{\rm stat}(q)$.
\label{fig:static_pot}}
\end{figure}

For completeness, and because the quantity should be obtainable by
experiments and simulations similar to those for \he4
\cite{MCS92a,CaupinJordi} and on bulk jellium \cite{MCS95}, we show in
Fig.~\ref{fig:static_resp} the static response function $\chi(q,0)$
of \he3 at $\rho=0.0166$\AA$^{-3}$. The main peak, which is a result
of the local symmetry in the fluid, is visibly raised compared to the
RPA result. We suspect, form experience with the boson theory, that
this peak is still a bit underestimated.

The comparison also lets us assess the validity of an energy
independent particle hole interaction. Fig.~\ref{fig:static_pot} shows
a comparison between the FHNC $\Vph{q}$ and the static effective
interaction (\ref{eq:Vstatic}). Evidently, the qualitative structure
is very similar, in particular $\Vph{q\!\to\!0} = \tilde V_{\rm
  stat}(q\!\to\!0)$ as discussed in Sec. \ref{ssec:longwaves}.  The
most visible difference is that $\tilde V_{\rm stat}(q)$ approaches a
constant for large $q$, see Eqs. (\ref{eq:chilarge}) and
(\ref{eq:Vstatqinf}).

\subsection{Electron liquid}
\label{ssec:egas}

The second typical area of application of microscopic many-body
methods is the electron liquid \cite{VigGiulBook,SiT81}. It provides
the basic under\-standing of valence electron correlations in simple
metals. In its two-component version it has proved useful for
describing the electron-hole liquid in semiconductors.

Compared to the helium fluids, the soft repulsion of the Coulomb
interaction induces substantially weaker correlations. Therefore,
electrons are much less challenging than \he3 and the RPA (or slightly
modified versions) contain much of the relevant physics.

Correlations are somewhat more pronounced in layered realizations of
the electron liquid, such as Si- and GaAs-AlGaAs
hetero-structures. For electrons on He surfaces preliminary results
show \cite{Camerino2d} that at very low densities, again, a roton-like
structure evolves for intermediate wave vectors.

We have seen that pair fluctuations contribute, already at long wave
lengths, to the static response function, see our discussion in
Secs.~\ref{ssec:longwaves}-\ref{ssec:staticresponse}.  Most important
are, of course, those effects that are {\em qualitatively\/} new
consequences of multiparticle fluctuations. These are the
short--wavelength behavior of the static response function and the
appearance of a new feature in the dynamics structure function, namely
the ``double plasmon'' excitation. The latter has raised new interest
\cite{SHV05,HSS08} in studying the dynamics of electrons at metallic
densities in this $(q;\omega)$ region.

\subsubsection{Double Plasmon}
\label{sssec:doubleplas}

\begin{figure}[h]
	\includegraphics[width=0.48\textwidth]%
        {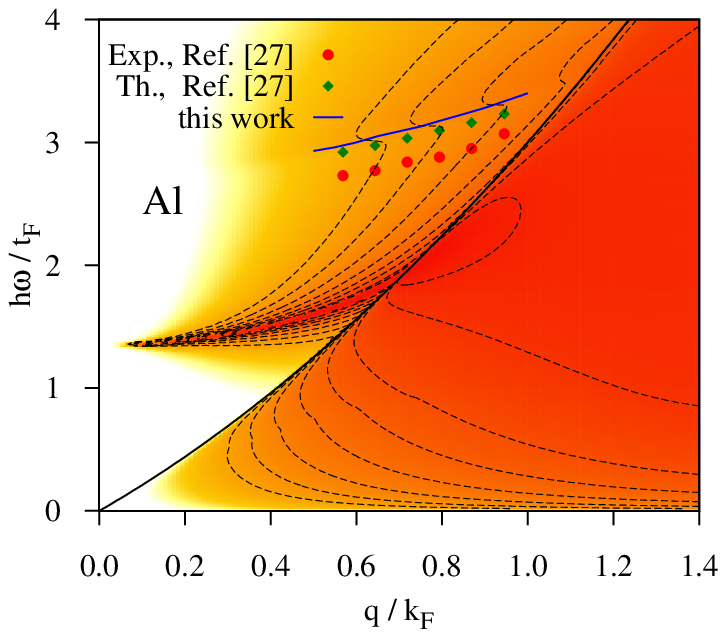}
	\includegraphics[width=0.48\textwidth]%
        {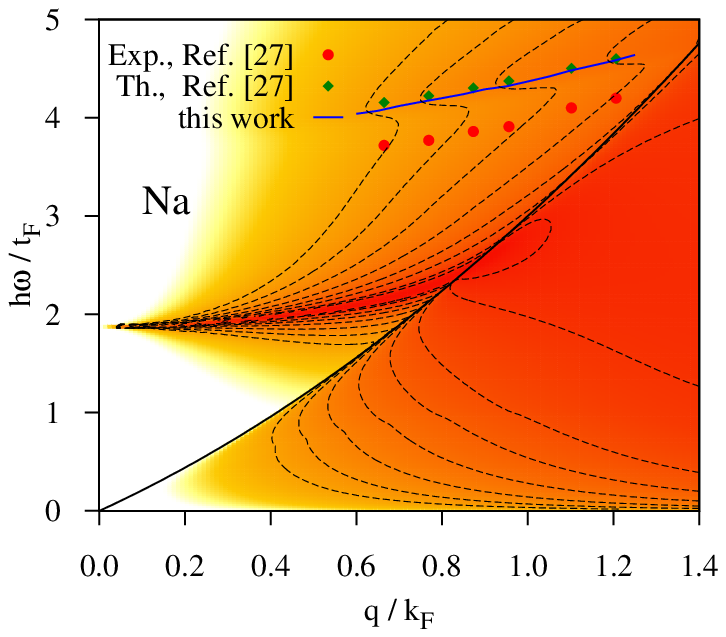}
\caption{(Color online) The figure shows $S(q;\omega)$ of an electron liquid
with density parameters $r_s = 2.06$ and $r_s = 3.99$ appropriate for Al and
Na, respectively. As in Figs.~\ref{fig:He3_contour}, dark red regions
correspond to high intensity (logarithmic scale).
 The blue line is the position of the
  double-plasmon peak obtained in the present work, red dots are
  experimental results \cite{HSS08} from inelastic X-ray scattering
  and green diamonds results from Green's functions calculations
  \cite{StG00,HSS08}.
  \label{fig:egas_contour}}
\end{figure}

Figure~\ref{fig:egas_contour} shows the dynamic structure factor
$S(q;\omega)$ obtained from the pair fluctuation theory.  We have
chosen two different densities $\rho\equiv 3/(4\pi r_s^3a_{\rm B}^3)$,
corresponding to Al, $r_s \!=\! 2.06$, and Na, $r_s \!=\!
3.99$. Immediately obvious are the finite width ({\it i.e.\/}\
lifetime) of the plasmon above the particle-hole band, and a second
peak-like structure around twice the plasma frequency $\omP$.

\begin{figure}[h]
\includegraphics[width=0.48\textwidth]%
{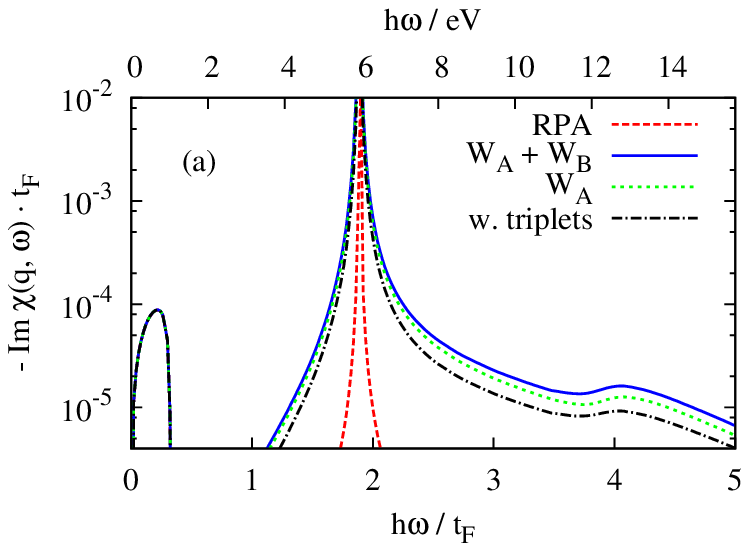}
\hfill
\includegraphics[width=0.48\textwidth]%
{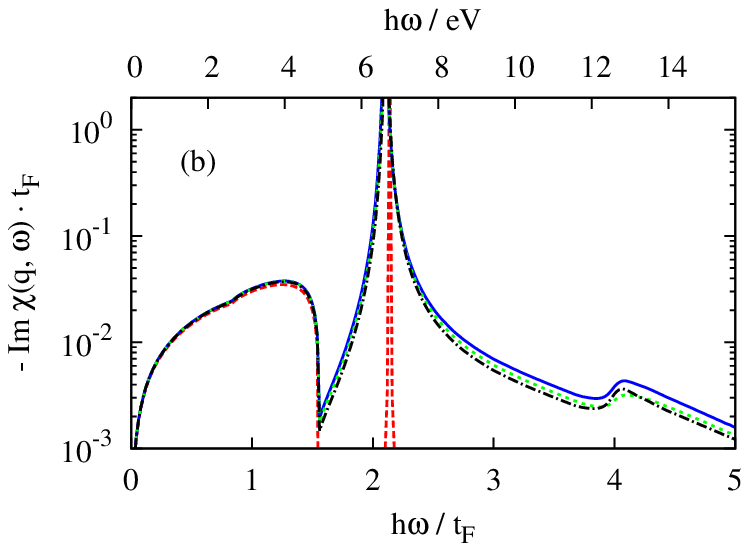}\\
\includegraphics[width=0.48\textwidth]%
{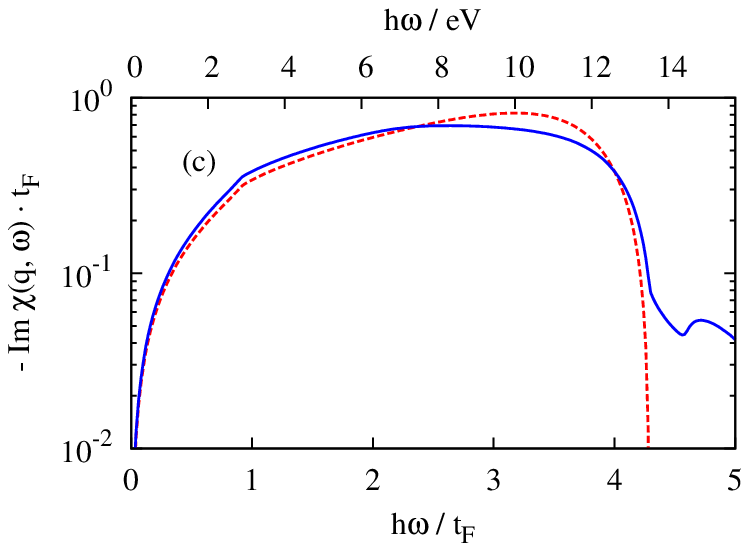}\hfill
\includegraphics[width=0.48\textwidth]%
{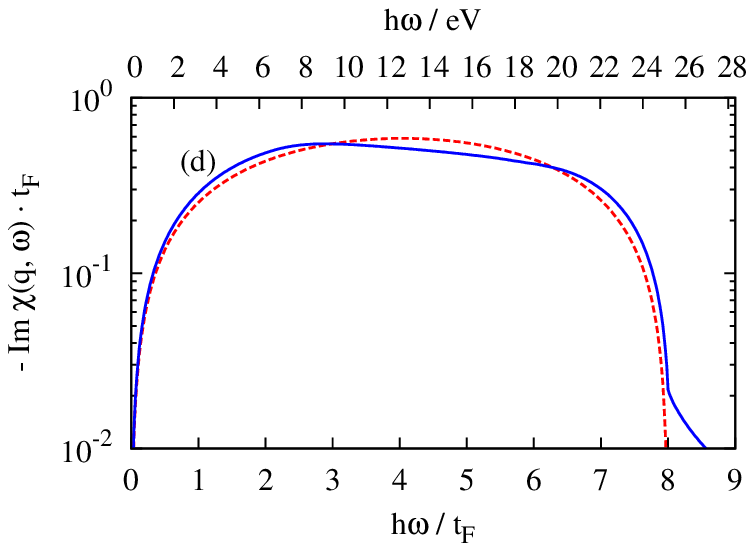}
\caption{%
(Color online) $S(q_0;\omega)$ for Na ($r_s\! = 3.99$), at wave vectors $q_0$
(a) $0.15\,k_\mathrm{F}$,
(b) $0.6 \,k_\mathrm{F}$,
(c) $1.3 \,k_\mathrm{F}$, and
(d) $2.0 \,k_\mathrm{F}$.
The full (blue) lines are our pair fluctuation theory,
dashed (red) lines are the RPA results using
$\widetilde V_{\!\scriptscriptstyle\rm p_{\bar{\ }\!}h}(q)$. To make the plasmon
visible, the RPA data have been broadened artificially by adding an imaginary 
frequency of $10^{-5}$eV$/\hbar$.
The dotted (green) lines in (a) and (b) refer to neglecting
$K^{(q)}_{q'q'',0}$ in Eqs.~(\ref{eq:K3loc_1_2})-(\ref{eq:K3loc}),
and the dash-dotted (black) lines include ground state triplet
correlations.  At larger momentum transfers these effects are too small 
to be visible.
\label{fig:egas_cuts}
}
\end{figure}

Characteristic cuts at constant wave vectors $q$ are shown in
Fig.~\ref{fig:egas_cuts} for Na. In parts (a) and (b) the plasmon is
outside the particle-hole band and rather sharp; the second peak
slightly above $2\hbar\omega_{\rm p} \!= 4.5\,t_{\rm F}$ is clearly
visible. We identify this feature, which has also been observed
experimentally \cite{HSS08}, with the ``double-plasmon''.

The ``double-plasmon'' excitation is due to the emergence of an
imaginary part in $\tilde V_{\!_{\rm A}}(a,\omega)$ at $\omega =
2\omP$, caused by the appearance of an imaginary part of the pair
propagator $\tilde E^{-1}(q',q'';\omega)$. It is therefore a genuine
multipair effect. The properties of the pair propagator are discussed
in in App.~\ref{assec:PairProp}. From (\ref{eq:Einvdoublpole}) we
obtain for the double-pole part of the dynamic interaction
(\ref{eq:VAqto0w})
\begin{eqnarray}
 {\Im}m\,\tilde V_{\!_{\rm A}}(q\!\to\!0;\omega)  &=&
 \frac{9\hbar^2\omP^2}{16t_{\rm F}^2}\,\frac{\pi}{8N}\!\sum_{{\bf q}'} 
 \left[\displaystyle\frac{\qkf}{q}K_{q,q'q''}\right]^2\times
 \nonumber\\&&\hfill 
 z^2(q')\, \left[ \delta(2\hbar\omega_{\rm c}(q')-\hbar\omega)  +
 \delta(2\hbar\omega_{\rm c}(q')+\hbar\omega) \right]\,.
\end{eqnarray}

In Fig.~\ref{fig:egas_cuts}(c), the plasmon is broad and
Landau-damped, while the double-plasmon still shows a clear structure,
even at the brink of entering the particle-hole continuum. Some
structure in the spectrum persists to even higher momentum transfers:
At $q\! = 2.0\,k_{\rm F}$ in Fig.~\ref{fig:egas_cuts}(d), traces of
the ordinary as well as the double plasmon show up as a faint
double-peak structure, with its minimum where the RPA yields a
single maximum.

\begin{figure}
  \centerline{\includegraphics{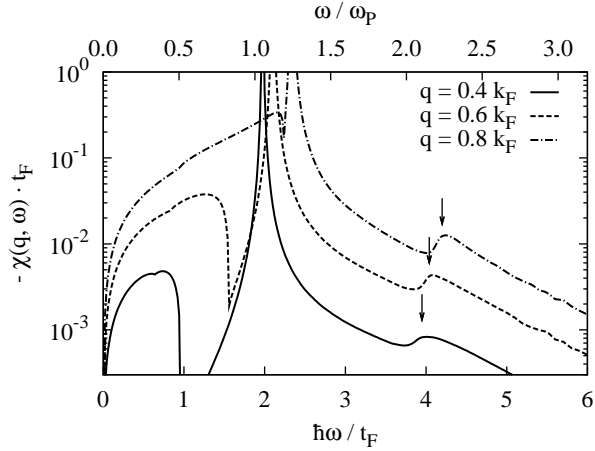}}
\caption{Cuts of the density-density response function at Na-density
($r_s = 3.99$), for constant momentum transfer $q = 0.4\,k_\mathrm{F}$
(solid line), $q = 0.6\,k_\mathrm{F}$ (dashed line) and $q = 0.4\,k_\mathrm{F}$
(dash-dotted line). The arrows mark the position of the
double-plasmon.
\label{fig:double_plasmon_multi_cut}}
\end{figure}

We now investigate the nature of the slight but measurable \cite{HSS08} peak 
in the loss function at approximately twice the plasmon frequency $\omP$.
 Fig.~\ref{fig:double_plasmon_multi_cut} shows $S(q,\omega)$ for $r_s = 3.99$
for three different momentum transfers, the position of the
double plasmon is marked with arrows.

We have already shown in Figs.~\ref{fig:egas_contour} the location of
the double plasmon excitation and a comparison with the experimental
inelastic X-ray scattering data \cite{SHV05,HSS08}. The double-plasmon
is also accessible by Green's function methods \cite{StG00}. These
results are very close to those of our pair fluctuation theory. This
can be understood from the fact that the leading terms of the
long-wavelength part of the pair propagator actually contain no
correlation effects, see Eq. \ref{eq:Einv_dpl_expansion}. Hence,
theories that are less well suited than CBF for the description of
strong correlations should, similar to the single plasmon, give the
right answer.  The remaining discrepancy with experiments must
therefore be attributed to lattice effects.  Fig.
\ref{fig:double_plasmon_multi_cut} shows more details of $S(q,\omega)$
at a sequence of three different momentum transfers for $r_s = 3.99$
(the position of the double-plasmon is marked with arrows), in
particular in order to assess the relative strength of the
double-plasmon excitation compared to the underlying continuum.

\subsubsection{Static Response}

\begin{figure}[h]
\includegraphics[width=0.48\textwidth]{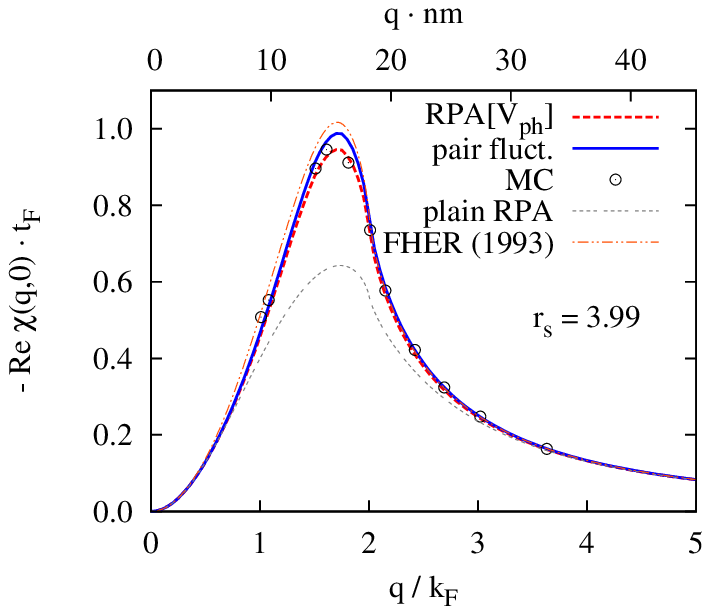}\hfill
\includegraphics[width=0.48\textwidth]{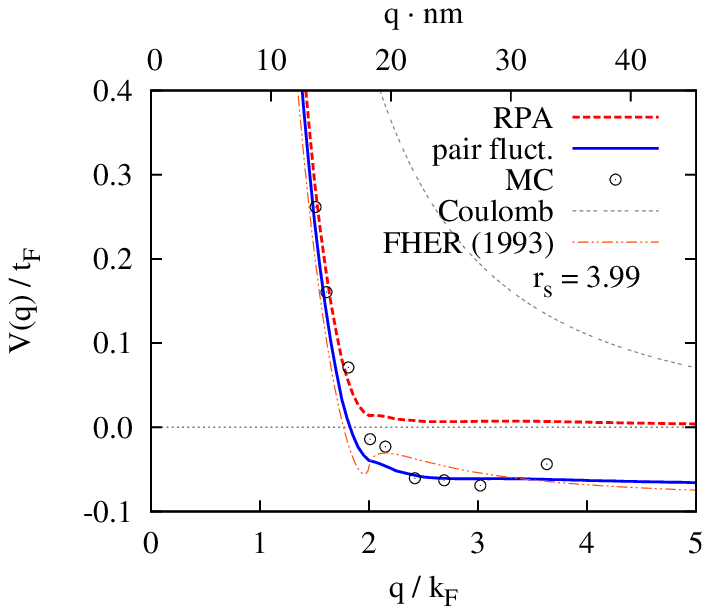}
\caption{%
Static response function (left), and static effective interaction (right)
of the electron liquid at $r_s\! = 3.99$.
Full blue lines are our results, black dash-dotted lines a fit based
on the simulations \cite{CdSOP98,MCS95}. 
Dotted red and thin broken lines show the RPA with $\Vph{q}$ and 
$\tilde v_{\rm c}(q)$, respectively.
\label{fig:Chistategas}
}
\end{figure}
Monte Carlo studies of the static response function $\chi(q;0)$ were
performed for two- and three-dimensional $^4$He
\cite{MCS92a,CaupinJordi} and on bulk jellium \cite{MCS95} for
$r_s\!=2,\,5$ and 10.  While $\chi(q;\omega)$ is accessible
experimentally, for electron liquids it is popular to define a {\it
  static local field correction\/} to the Coulomb interaction $\tilde
v_{\rm c}(q)$ via \cite{VigGiulBook}
\begin{equation}
	\tilde V_{\!\rm stat}(q) \>\equiv\>
	\tilde v_{\rm c}(q)\,(1-G(q)) \;.
	\label{eq:LFF_Def}
\end{equation}
From our analysis it is clear that a response function in the RPA form
can be defined only for $q\!\to\!0$ and at $\omega\!=\!0$. 
Therefore, only in these two cases such a function is a physically meaningful 
quantity.

In the $q\!\to\!\infty$ limit, our theory yields a finite value for
$V_{\!\rm stat}(q)$, resulting in $G(q) \propto q^2$, whereas
$\Vph{q}$ falls off like the bare potential. This correct
$q-$dependence arises {\it solely}\/ from multiparticle fluctuations.
In Fig.~\ref{fig:Chistategas} we compare our results with the Monte
Carlo data, and with curves calculated from an analytic analytic fit
for $-v_{\rm c}(q)\,G(q)$ obtained from the latter \cite{CdSOP98}.
The agreement is remarkably good.

No trace of a possible ``hump'' in $G(q)$ around $2k_{\rm F}$ as a
remnant of some charge- or spin-density wave instability was found in
the simulations, but it also was not fully conclusively ruled out. Our
results, clearly, do not yield any such peak structure at $2k_{\rm F}$
either.

\section{Summary}
\label{sec:Summary}

We have presented the fermion version of theories of the
dynamic response of Bose fluids that have been developed in the past
successfully by Jackson, Feenberg, and Campbell. These methods form
the basis of our present understanding of the dynamics of Bose
fluids. Our derivations were admittedly lengthy but eventually led to
a reasonably compact formulation of the dynamic response of correlated
Fermi fluids. Our final result could be formulated as a set of TDHF
equations in terms of dynamic and non-local effective interactions.

For the first applications we have reduced the theory to a practical
level capturing the relevant physics, while avoiding many of the
technical complications.  In particular the version of the equations
of motion spelled out in Appendix \ref{asec:recipe} has proved to
be adequate for systems as different as \he3 and homogeneous
electrons.  It is hardly more complicated than TDHF.  The sole
required input is the static structure function $S(q)$ which can, in
principle, also be obtained from simulations. Our developments have
led to {\it quantitative\/} improvements of our understanding of \he3
and electrons as well as to the description of {\it qualitatively
  new\/} effects like mode-mode coupling, multiparticle spectra, and
damping.

We have, at various places, commented on the role of the particle-hole
spectrum. In the homogeneous electron liquid, the interaction
corrections to the single-particle spectrum are relatively small
\cite{annals,IKP84}, the theory formulated here should therefore
suffice for many purposes. The situation is more difficult in \he3: As
is seen from our results, good agreement with experiments can be
reached by assuming a spectrum of non-interacting fermions.  In
particular looking at the zero-sound damping suggests that, at
$q\approx k_F$, the boundary of the single-particle continuum should
be close (perhaps even above) to the one given by a non-interacting
spectrum, {\it cf.\/} Fig. \ref{fig:zero_sound_166}.  This is not in
contradiction to experiments \cite{GRE83,GRE86} suggesting an
effective mass ratio $m^*/m\approx 3$ at the Fermi surface. One reason
is that the effective mass ratio drops rapidly with distance from the
Fermi surface. The more fundamental reason however, is that the concept of
describing the particle-hole excitations by a spectrum that depends on
momentum only is questionable at elevated wave numbers. More
precisely, the single-particle motion is described by a non-local,
energy dependent self-energy. Upon closer examination it becomes clear
that exchange effects are intimately related to self-energy
corrections and exchange effects must
therefore be included simultaneously.

In independent work, we have used the ideas of CBF theory as well as
the Aldrich-Pines pseudopotential theory to calculate the
single-particle propagator in \he3.  In both three and two dimensions,
we found good agreement between the theoretical effective mass near
the Fermi surface, and that obtained experimentally from specific heat
measurements \cite{Bengt,he3mass,2dmass}.  However, the somewhat {\it
  ad-hoc\/} use of the effective interactions in that work is still
awaiting rigorous justification. This is the subject of future work.

\begin{acknowledgments}
  A part of this work was done while one of us (EK) visited the
  Physics Department at the University at Buffalo, SUNY. Discussions with
  C. E. Campbell, H. Godfrin and R. E. Zillich are gratefully
  acknowledged. This work was supported, in part, by the Austrian
  Science Fund FWF under project P21264.
\end{acknowledgments}

\vfill\eject
\appendix
\section{Ground state theory}
\label{asec:ground_state}

\subsection{The essence of FHNC-EL}
\label{assec:fhnc0}

For the sake of the discussions of this work we here briefly review
the essence of variational FHNC theory.  The diagram expansion and
summation procedure that is used to derive, for the variational wave
function (\ref{eq:JastrowWaveFunction}) a set of equation for the
calculation and optimization of physical observables has been
described at length in review articles \cite{Johnreview} and
pedagogical literature \cite{KroTriesteBook}. Details on the specific
implementation for \he3 are given in Ref. \onlinecite{polish}.

Here, we spell out a reduced set of equations. These do not provide the
quantitatively best implementation \cite{polish} of the FHNC-EL theory,
but they contain the relevant physics:  They provide, in the language of 
perturbation theory, a self-consistent approximate summation of ring-- and 
ladder diagrams \cite{parquet1}, thereby capturing both, long- as well as
short-ranged features.

In the simplest approximation \cite{Mistig}, which contains, as we
shall see momentarily, the ``RPA'' expression (\ref{eq:RPAresponse}),
the Euler equation (\ref{eq:Euler}) can be written in the form \cite{polish}
\begin{equation}
        S(q) = {S_{\rm F}(q)\over \sqrt{1 +
                2{\displaystyle S_{\rm F}^2(q)\over\displaystyle t(q)}
        \Vph{q}}} \,,
\label{eq:FermiPPA0}
\end{equation}
where $t(q) = \hbar^2 q^2/2m$ is the kinetic energy of a free particle,
and
\begin{equation}
        V_{\!\scriptscriptstyle\rm p_{\bar{\ }\!}h\!}(r)=\>
        \left[1+ \Gamma_{\!\rm dd}(r)\right]v(r)
        + {\hbar^2\over m}\left|\nabla\sqrt{1+\Gamma_{\!\rm dd}(r)}\right|^2
+ \Gamma_{\!\rm dd}(r)w_{\rm I}(r)
\label{eq:VddFermi0}
\end{equation}
is what we call the ``particle-hole interaction''. Auxiliary quantities
are the ``induced interaction''
\begin{equation}
        \tilde w_{\rm I}(q)=-t(q)
        \left[{1\over S_{\rm F}(q)}-{1\over S(q)}\right]^2
        \left[{S(q)\over S_{\rm F}(q)}+\frac{1}{2}\right].
\label{eq:inducedFermi0}
\end{equation}
and the ``direct-direct correlation function'' 
\begin{equation}
   \tilde \Gamma_{\!\rm dd}(q) = \bigl(S(k)-S_{\rm F}(q)\bigr)/S^2_{\rm F}(q)
   \label{eq:GFHNC}
\end{equation}
(see also
Eq. (\ref{eq:sofk})). Eqs.~(\ref{eq:FermiPPA0})--(\ref{eq:GFHNC})
form a closed set which can be solved by iteration.  Note that the
Jastrow correlation function has been eliminated entirely.

The relationship (\ref{eq:FermiPPA0}) between the static structure function
$S(q)$ and the particle-hole interaction $\Vph{q}$ can also be
derived from Eq.~(\ref{eq:RPAresponse}), if the Lindhard function is replaced
with its ``mean spherical'' or ``collective'' approximation (CA),
\begin{equation}
        \chi_0^{\scriptscriptstyle\rm CA}(q; \omega) =
        {\displaystyle {2 t(q)}
        \over(\hbar\omega+\I\eta)^2- t^2(q)/S^2_{\rm F}(q)}  \; .
\label{eq:msa0}
\end{equation}
The essence of this approximation is to replace the branch cut in
$\chi_0(q; \omega)$ by a single pole; its strength chosen such that
the first two sum rules
agree when evaluated with the full Lindhard function $\chi_0(q; \omega)$ or
in the collective approximation $\chi_0^{\scriptscriptstyle\rm CA}(q; \omega)$,
{\it i.e.}
\begin{eqnarray}
\Im m\int\! d\omega\> \chi_0^{\scriptscriptstyle\rm CA}(q; \omega) &&=
\Im m\int\! d\omega\> \chi_0(q; \omega)\nonumber\\
\Im m\int\! d\omega\> \omega\,\chi_0^{\scriptscriptstyle\rm CA}(q; \omega) &&=
\Im m\int\! d\omega\> \omega\,\chi_0(q; \omega) \;.
\label{eq:MSA}
\end{eqnarray}
In fact,
(\ref{eq:RPAresponse}) together with (\ref{eq:msa0}) or, alternatively,
\begin{equation}
 \Vph{q} = \frac{t(q)}{2}
\left(\frac{1}{S^2(q)}-\frac{1}{S_{\rm F}^2(q)}\right)
\label{eq:Vph}
\end{equation}
can be used \cite{polish} to {\it define\/} the particle-hole interaction
from an accurately known $S(q)$.

The energy, consisting of kinetic and potential energy $\langle T\rangle +
\langle V \rangle$, is \cite{polish}
\begin{eqnarray}
E &=& \frac{3}{5}Nt_{\rm F}  + E_{\rm R} +  E_{\rm Q}\,,\label{eq:EJF} \\
 E_{\rm R} &=& \frac{\rho N}{2}\int\! d^3r\>
 \left[g(r)\, v(r) + \frac{\hbar^2}{m}\bigl(1+ C(r)\bigr)
 \left|\nabla\sqrt{1+\Gamma_{\!\rm dd}(r)}\right|^2\right]\,,
 \label{eq:ER}\\
 E_{\rm Q} &=& \frac{N}{4}\int\!\frac{d^3q}{(2\pi)^2\rho}\>
 t(q)\left[S^2_{\rm F}(q)-1 - S^2(q)+S(q)\,\right]\,
 \tilde\Gamma_{\!\rm dd}^2(q)\label{eq:EQ}\,.
\end{eqnarray}
Here, $t_{\rm F}$ is the Fermi energy, and, in this approximation,
\begin{equation}
\tilde C(q) = S_{\rm F}(q)-1+(S_{\rm F}^2(q)-1)\tilde\Gamma_{\!\rm dd}(q)\,.
\label{eq:C0ofk}
\end{equation}

To make the connection with the limiting behavior of $\chi(q,0)$
in Sec.~\ref{ssec:staticresponse}, we next spell out what is known as the
``uniform limit'' or ``collective'' approximation (CA). Products of 
functions which {\it in coordinate space\/} vanish for $r\rightarrow\infty$ 
are considered small. This implies to expand 
$\nabla\sqrt{1+\Gamma_{\!\rm dd}(r)} 
\approx \frac{1}{2}\nabla\Gamma_{\!\rm dd}(r)$ and to neglect $C(r)$.
The kinetic energy then is
\begin{equation}
 \left\langle T \right\rangle^{\!\scriptscriptstyle\rm CA}
 \;=\; T_\mathrm{F} + \frac{1}{4}\sum_{\vec q}
 t(q) \,S(q)\, \tilde X^2_{\rm dd}(q)\,.
\label{eq:TCA}
\end{equation}
Here, $T_\mathrm{F}=3Nt_{\rm F}/5$,
and $\tilde X_{\rm dd}(q)$ is the ``non-nodal'' function. In our
reduced FHNC approximation, $\tilde X_{\rm dd}(q)$ is related to the
static structure factor by
\begin{equation}
        \tilde X_{\rm dd}(q) = \frac{1}{\SF {q}} - \frac{1}{S(q)} \;.
\label{eq:SFHNC0}
\end{equation}

\section{Diagrammatic analysis}
\label{asec:diagrams}

\subsection{Transition density}
\label{assec:transition_density}

\begin{figure}
\centerline{\includegraphics[width=0.8\textwidth]{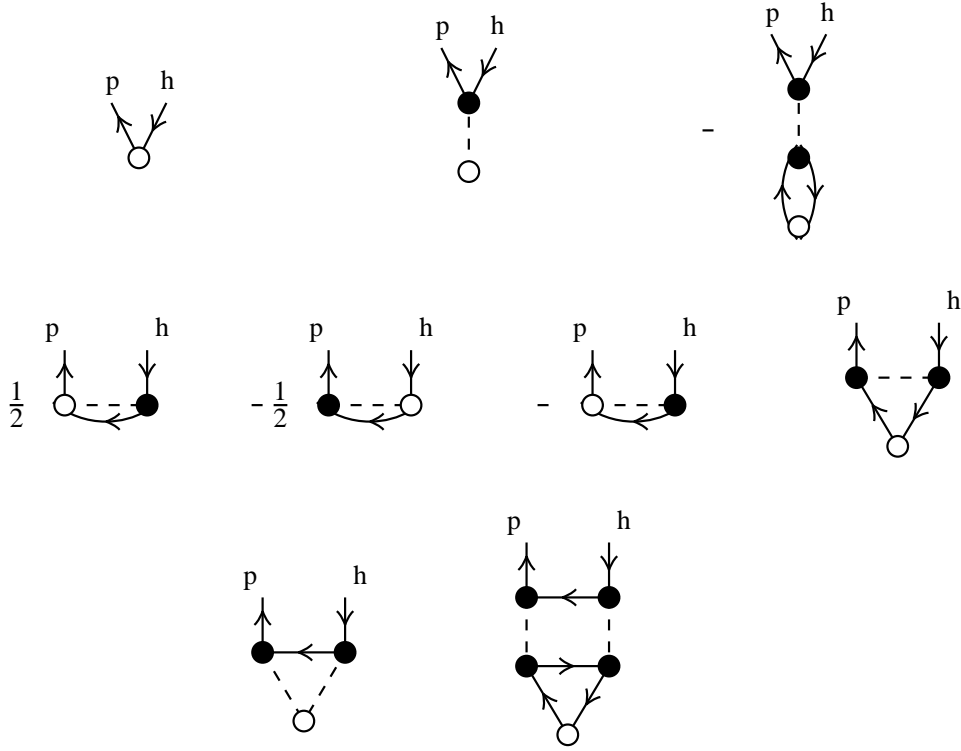}}
\caption{Diagrammatic representation of some contributions to
$\rho_{0,ph}({\bf r})$. The upper row shows the diagrams defining the 
local approximation. The second row are the leading exchange diagrams and 
the third row shows two corrections due to the non-locality of ${\cal N}(1,2)$.
\label{fig:M1mat}}
\end{figure}

We first examine the diagrammatic structure of CBF matrix elements
$\rho^{\phantom{\rm F}}_{0,ph}(\rvec)$ of the density operator,
~(\ref{eq:rhorep1},\,\ref{eq:rhorep2}). The simplest approximation for
$M_{ph,p'h'}$ has been spelled out in Eq.~(\ref{eq:MtoGamma}), the
corresponding approximation for $\rho^{\phantom{\rm F}}_{0,ph}({\bf
  r})$ is
\begin{equation}
   \rho^{\phantom{\rm F}}_{0,ph}({\bf r}) \>=\>
   \rho^{\rm F}_{0,ph }({\bf r})+ \rho\!\int\!\!d^3r'\!\int\!\!d^3r''\>
   \Bigl[\delta(\rvec\!-\!\rvec') -
   \displaystyle\frac{\rho}{\nu}\displaystyle
         \ell^2(|\rvec\!-\!\rvec'|\qkf) \Bigr]\,
   \Gamma_{\!\rm dd}(\rvec'\!-\!\rvec'')\, \rho^{\rm F}_{0,ph}({\rvec''})
   \,.
\label{eq:rhophSFGamma}
\end{equation}

The diagrammatic representation of some leading diagrams contributing
to $\rho^{\phantom{\rm F}}_{0,ph}(\rvec)$ is shown in Fig.
  \ref{fig:M1mat}.  As usual, open points represent particle
  coordinates $\rvec_i$, while filled points indicate an integration
  over the associate coordinate space and a density factor.  Dashed
  lines connecting points $\rvec_i$ and $\rvec_j$ represent a function
  $\Gamma_{\!\rm dd}(r_{ij})$, and oriented solid lines an exchange
  function $\ell(r_{ij}\qkf)$. New elements are particle- and hole-states, depicted as upward (particles) or downward (holes) lines
  entering or leaving the diagram.

  The three leading terms (\ref{eq:rhophSFGamma}) are shown in the
  upper row of Fig.~\ref{fig:M1mat}.  In the second row of
  Fig.~\ref{fig:M1mat} we show the leading exchange diagrams. In the
  representation (\ref{eq:rhorep1}), these originate from the factors
  $z_{ph}$ in the definition of the $\tilde\rho_{0,ph}(\rvec)$, these
  are shown as the first two diagrams. Exchange terms also originate
  from the matrix element $\left\langle ph'\vert \Gamma_{\!\rm dd} \vert
    hp' \right\rangle_a$, these are shown as third and fourth diagram
  in that row. Evidently there is a partial cancellation. The diagrams
  shown in that row also serve as an example for how the
  representations (\ref{eq:rhorep1}) and (\ref{eq:rhorep2}) are equal:
  Starting from the form (\ref{eq:rhorep2}), the diagrams originating
  from the $z_{ph}$-factors ({\em i.e.\/}\ the first two diagrams in
  the second row), have opposite signs; and the exchange term of
  $\left\langle pp'\vert \Gamma_{\!\rm dd} \vert hh' \right\rangle_a$ 
  yields the third diagram with interchanged particle- and
  hole labels. The sum of all three diagrams is the same.

\subsection{\texorpdfstring{The $M^{\rm (I)}$ matrix}
{The M I matrix}}
\label{assec:MImatrix}

\begin{figure}
\centerline{\includegraphics[scale=0.8]{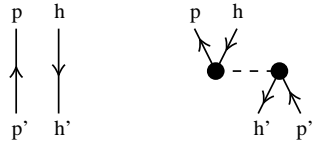}}
\caption{Diagrammatic representation of the local approximation
for $M_{ph,p'h'}$. 
\label{fig:diagram_M11}}
\end{figure}

Our next task is to show that the diagrams representing $M^{\rm
  (I)}_{ph,p'p''h'h''}$ are a proper subset of those contributing to
$M_{ph,p'p''h'h''}$. We restrict ourselves here to the simplest
case, which is the numerically implemented version.
We start with the two-body matrix $M_{ph, p'h'}$\,.  As spelled out 
in Eq.~(\ref{eq:MtoGamma}), besides the $\delta$-function, the leading 
contribution is the {\it local\/} term in the two-body operator
\begin{equation}
{\cal N}_{\rm loc}(1,2) = \Gamma_{\!\rm dd}(r_{12})\,.
\label{eq:Nloc}
\end{equation}
The diagrammatic representation of this approximation for $M_{ph,
  p'h'}$ is shown in Fig.~\ref{fig:diagram_M11}.

\begin{figure}
\centerline{\includegraphics[scale=0.7]{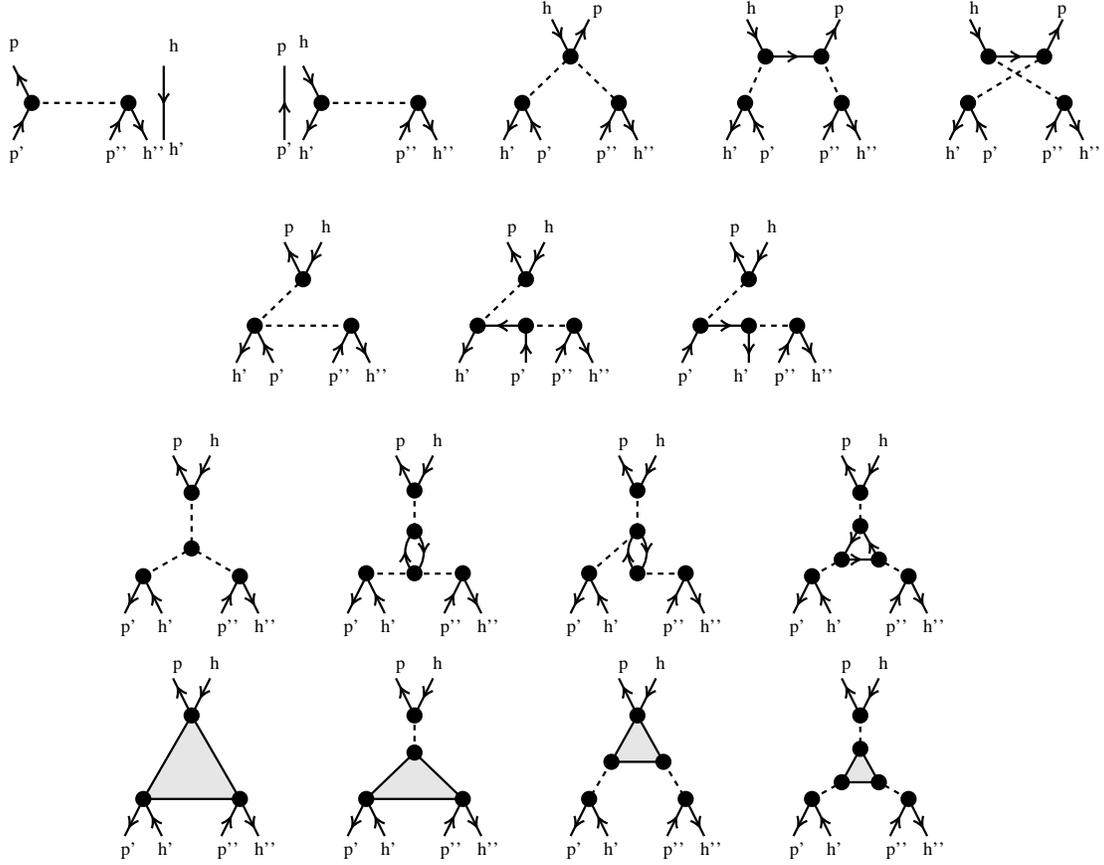}}
\caption{Diagrams of $M_{ph,p'p''h'h''}$ in the convolution approximation
(\ref{eq:N12_most_general}). Graphs obtained by exchanging the pairs $(p'h')$ 
and $(p''h'')$ are to be added. The last row shows some diagrams
containing ground state triplet correlations (shaded triangle), all of these
contribute to $M^{(I)}_{ph,p'p''h'h''}$. 
\label{fig:diagram_M12}}
\end{figure}

A diagrammatic expansion of the matrix elements $M_{ph,p'p''h'h''}$
can be derived in exactly the same way as the corresponding expansions
of the two-body matrix elements \cite{CBF2}. Generally, the
$M_{ph,p'p''h'h''}$ are matrix elements of a non-local three-body
operator, which can be expressed in terms of FHNC diagrams.
Restricting ourselves again to the numerically implemented level, we
need these quantities in an approximation equivalent to the ``uniform
limit approximation'' \cite{Chuckphonon} for bosons.  We generalize
this approach to fermions by keeping all diagrams contained in the
Bose case plus those, where the end points of the correlation
functions are linked by exchange paths (the bosonic $g(r_{ij})-\!1$ is
identified with the direct-direct correlation function $\Gamma_{\!\rm
  dd}(r_{ij})\,$). This procedure has already been used for deriving
the optimal triplet correlations for the fermion ground state
\cite{polish}. The diagrammatic representation of this approximation
is shown in Fig.~\ref{fig:diagram_M12}, the analytic form is
\def\bracketops#1#2#3{\ensuremath{ \langle #1 | #2 | #3 \rangle}}
\begin{eqnarray}
M^{\rm\scriptscriptstyle CA}_{ph,p'p''h'h''}
&=& \delta_{h,h'}\,\bracketops{p h''}
{ \Gamma_{\!\rm dd}(1,2) }{p' p''}
  - \delta_{p,p'}\,\bracketops{h'h''}
{ \Gamma_{\!\rm dd}(1,2) }{h  p''} \nonumber \\
&&+ \frac{1}{2}\>\bracketops{ph'h''}{\Gamma_{\!\rm dd}(3,1)
  \, \Gamma_{\!\rm dd}(1,2)}{hp'p''} \nonumber \\		
&&- \frac{1}{2}\>\sum_{h_1} \bracketops{ph''}{ \Gamma_{\!\rm dd} }{h_1p''} \,
 \bracketops{h'h_1 }{ \Gamma_{\!\rm dd} }{p'h} 
  - \frac{1}{2}\>\sum_{h_1} \bracketops{ph'}{ \Gamma_{\!\rm dd} }{h_1p'}   \,
\bracketops{h''h_1}{ \Gamma_{\!\rm dd} }{p''h} \nonumber \\	
&&+\> \bracketops{ph'h''}{\Gamma_{\!\rm dd}(1,2) \, \Gamma_{\!\rm dd}(2,3)}{hp'p''}
\nonumber \\							
&&-\> \sum_{h_1} \bracketops{ph'} { \Gamma_{\!\rm dd} }{hh_1} \,
\bracketops{h''h_1}{ \Gamma_{\!\rm dd} }{p''p'} 
- \sum_{h_1} \bracketops{ph_1}{ \Gamma_{\!\rm dd} }{hp'}  \, \bracketops{h'h''}
{\Gamma_{\!\rm dd} }{h_1p''} \nonumber \\                        
&&+\> \bracketops{p h' h''}{ \Gamma^{\scriptscriptstyle\rm CA}_{\!\rm ddd}(1,2,3) }{h p' p''} \nonumber \\
&&+\> \left\lbrace (p'h') \leftrightarrow (p''h'')\right\rbrace
\label{eq:N12_most_general} \,.
\end{eqnarray}
Here, in convolution approximation,
\begin{eqnarray}
\Gamma^{\scriptscriptstyle\rm CA}_{\!\rm ddd}(\rvec_1, \rvec_2, \rvec_3)
=& & \frac{\rho}{2} \int \!  d^3r_4 \,
\Gamma_{\!\rm dd}(\vec r_1 - \vec r_4) \Gamma_{\rm dd}(\vec r_2 - \vec r_4)
\Gamma_{\!\rm dd}(\vec r_3 - \vec r_4) \nonumber \\
&+& \frac{\rho^2}{2\nu} \int\! d^3r_4\,d^3r_5 \,
\ell^2(|\vec r_4-\vec r_5|\qkf) \Gamma_{\!\rm dd}(\vec r_1 - \vec r_4)
\Gamma_{\!\rm dd}(\vec r_2 - \vec r_5) \Gamma_{\rm dd}(\vec r_3 - \vec r_5)
\nonumber \\
&+& \frac{\rho^2}{\nu} \int\! d^3r_4\,d^3r_5 \,
\ell^2(|\vec r_4 - \vec r_5|\qkf) \Gamma_{\!\rm dd}(\vec r_1 - \vec r_4)
\Gamma_{\!\rm dd}(\vec r_3 - \vec r_4) \Gamma_{\rm dd}(\vec r_2 - \vec r_5)
\nonumber \\
&+& \frac{\rho^3}{\nu^2} \int\! d^3r_4\,d^3r_5\,d^3r_6 \,
\ell(|\vec r_4 - \vec r_5|\qkf) \ell(|\vec r_5 - \vec r_6|\qkf)
\ell(|\vec r_6 - \vec r_4|\qkf) \nonumber \\
& & \times \Gamma_{\!\rm dd}(\vec r_1 - \vec r_4)
\Gamma_{\!\rm dd}(\vec r_2 - \vec r_5) \Gamma_{\rm dd}(\vec r_3 - \vec r_6)
\label{eq:Gamma_ddd_def} \, .
\end{eqnarray}
The first two lines are invariant under exchanging
$\rvec_2 \leftrightarrow \rvec_3$, equivalent to exchanging 
$(p'h') \leftrightarrow (p''h'')$ in (\ref{eq:N12_most_general}).

Optimized triplet correlations improve the description of the
ground-state structure, in particular in the area of the peak of the
static structure function and also improve, for bosons, the density
dependence of the spectrum \cite{Chuckphonon}.  These correlations add
another term to the three-body function $\Gamma^{\scriptscriptstyle\rm
  CA}_{\!\rm ddd}(\rvec_1, \rvec_2, \rvec_3)$. The expressions are
lengthy \cite{polish}, we refrain from spelling them out here and just
show the diagrammatic representation of some typical terms in the last
row of Fig. \ref{fig:diagram_M12}.

Per definition in (\ref{eq:M2fact}), $M^{\rm (I)}_{ph,p'p''h'h''}$ is to be 
constructed such that its matrix product with $M_{ph,p'h'}$ reproduces
$M_{ph,p'p''h'h''}$.  A low-order manifestation of this is easily verified 
with choosing for $M^{\rm (I)}_{ph,p'p''h'h''}$ the uniform limit diagrams
shown in the first row of Fig.~\ref{fig:diagram_M12},
\begin{eqnarray}
 M^{\rm (I)\,\scriptscriptstyle CA}_{ph,p'p''h'h''}&=& \Bigr\{
  \delta_{h,h'} \bra{ph''}\Gamma_{\!\rm dd}\ket{p'p''}
 -\delta_{p,p'} \bra{h'h''}\Gamma_{\!\rm dd}\ket{hp''}
 +(p'h') \leftrightarrow (p''h'') \Bigr\} \nonumber\\
&+& \sum_{p_1} \bra{ph''}  \Gamma_{\!\rm dd} \ket{p_1p''}
               \bra{p_1h'} \Gamma_{\!\rm dd} \ket{hp'}
 -  \sum_{h_1} \bra{ph'}   \Gamma_{\!\rm dd} \ket{h_1p'}
               \bra{h_1h''}\Gamma_{\!\rm dd} \ket{hp''}
 \label{eq:MI_spelled_out}\\
&=& \frac{1}{N}\delta_{\bf q , \bf q' + \bf q''} \,
 \bar n_{\bf p} \bar n_{\bf p'} \bar n_{\bf p''} 
 n_{\bf h} n_{\bf h'} n_{\bf h''} \times \nonumber \\
&& \biggl[\Bigl\{\tilde\Gamma_{\!\rm dd}(q'')\,(
  \delta_{h,h'}\!-\!\delta_{p,p'} )
 \>+\>(p'h') \leftrightarrow (p''h'') \Bigr\}\nonumber\\
&&\phantom{\biggl[}
 + \frac{1}{N}\tilde\Gamma_{\!\rm dd}(q'') \tilde\Gamma_{\!\rm dd}(q')\,
  (\bar n_{{\bf h}+{\bf q}'} - n_{{\bf h}+{\bf q}''}) \,\biggr] 
 \label{eq:MI_spelled_out_loc}
\end{eqnarray}
where the term originating from triplet correlations has not been spelled out.

Generally, $M^{\rm (I)}_{ph,p'p''h'h''}$ is represented by the subset
of $M_{ph,p'p''h'h''}$ diagrams that can {\it not}\/ be cut into two
pieces, one connected to the labels $ph$ and the other to
$p'p''h'h''$, by cutting either two exchange lines, or cutting the
diagram in an internal point.  The third row of
Fig.~\ref{fig:diagram_M12} shows such contributions.

$M^{\rm (I)\,\scriptscriptstyle CA}_{ph,p'p''h'h''}$ depends
non-trivially on three particle and three hole quantum numbers.  We
define the localized version as its Fermi sea average,
Eq.~(\ref{eq:FSAn}),
\begin{eqnarray}
\tilde M^{\rm (I)\,\scriptscriptstyle CA}_{q,\,q'q''} 
&\equiv& \frac{1}{\SF{q}\SF{q'}\SF{q''}}\,\frac{1}{N}\!\sum_{hh'h''} M^{\rm (I)\,\scriptscriptstyle CA}_{ph,p'p''h'h''} \nonumber\\
&=& 
\delta_{\bf q , \bf q' + \bf q''}
\left[\left[\frac{S(q')S(q'')}{\SF{q'}\SF{q''}} -1\right]
\frac{\SFthree(q,q',q'')}{\SF{q}\SF{q'}\SF{q''}}
+\frac{S(q')S(q'')}{\SF{q'}\SF{q''}} \tilde u_3(q,q',q'')\right] .
\qquad
\label{eq:MI3av}
\end{eqnarray}
Here, the relationship (\ref{eq:GFHNC}) was used for the connection
between $\tilde \Gamma_{\!\rm dd}(q)$ and $S(q)$, and
\begin{equation}
\SFthree(q,q',q'')\equiv
\frac{1}{N}\sum_h n_{\vec h}\bar{n}_{\vec h-\vec q}
\left[\bar{n}_{\vec h+\vec q'}-n_{\vec h+\vec q''} \right]
\label{eq:SF3}
\end{equation} 
is the three-body static structure function of non-interacting
fermions. 

\subsection{Three-body vertices}
\label{assec:Kthree}

We now apply the localization procedure (\ref{eq:FSAn}) to the
three-body vertices.  Starting with (\ref{eq:K12bdef}), we have
\begin{equation}
 \tilde K^{(q)}_{q'q'', 0} \equiv
 N^2\,K^{(q)}_{q'q'', 0} \>=\> \frac{1}{N\,\SF{q}\SF{q'}\SF{q''}}\,
 \sum_{hh'h''}\left[H'_{pp'p''hh'h'',0} -
   \sum_{p_1\!h_1}\!H'_{ph\,p_1\!h_1,0}\, M^{\rm (I)}_{p'p''h'h''\!,\,p_1\!h_1}\right]\,.
 \label{eq:defKatilde}
\end{equation}
As discussed in Sec.~\ref{ssec:Brillouin},
the Euler equations (\ref{eq:Euler}) for the ground state
optimizations ensure that the Fermi sea average (\ref{eq:zwei}) of
$H'_{pp'p''hh'h'',0}$ vanishes.  For the matrix elements $H'_{php'h',0}$
Eqs.\ (\ref{eq:sofk})-(\ref{eq:Wlocal}) yield
\begin{equation}
H'_{php'h',0} = \frac{1}{2N}\delta_{\bf q +\bf q', \bf 0}\!
\biggl[e_{ph}+e_{p'h'}  -2\frac{t(q)}{\SF{q}} \biggr]\,
\tilde \Gamma_{\!\rm dd}(q) .
\end{equation}
Therefore, using (\ref{eq:MI_spelled_out_loc}) for $M^{\rm(I)}_{ph,p'p''h'h''}$
\begin{eqnarray}
\frac{1}{N^3}\!\sum_{hh'h''} K^{(ph)}_{p'p''h'h'',0} &=&
-\frac{1}{N^3}\!\sum_{hh'h''}\sum_{p_1h_1} H'_{ph\,p_1h_1,0}\,
 M^{\rm (I)}_{p'p''h'h'',p_1h_1} \nonumber\\
 &=& -\frac{1}{2N^3}\,\tilde \Gamma_{\!\rm dd}(q)\,\SF{q}\,
 \sum_{h'h''h_1} \Bigl(e_{h_1-q,h_1} - \frac{t(q)}{\SF{q}}\Bigr)\,
M^{\rm (I)}_{p'p''h'h'',(h_1-q)h_1} \nonumber\\
&=& \frac{\delta_{{\bf q} + {\bf q'} + {\bf q''}, {\bf 0}}}{N^2} \,
	\frac{\hbar^2}{4m} \, \qG(q)
	\left[\frac{S(q')S(q'')}{\SF{q'}\SF{q''}}-1\right] \nonumber \\
& &\times
	\biggl[q^2\, \SFthree(q,q',q'') + {\bf q} \cdot
	\left[{\bf q}''\,\SF{q'} +{\bf q}'\,\SF{q''} \right] \SF{q}  \biggr] \, .
\label{eq:H3_aux0}
\end{eqnarray}
This term vanishes when $q$ and $q'$ are larger than $2\,\qkf$. It is
also zero if the matrix element $H'_{ph\,p_1h_1,0}$ in
Eq.~(\ref{eq:H3_aux0}) is replaced by its Fermi sea average.  We
therefore expect this term to be small, in particular since it has no
analog in the Bose limit. Note also that triplet ground state
correlations do not contribute to this term.  Dividing by the
normalization factors $\SF{q}\SF{q'}\SF{q''}$ leads to the result
(\ref{eq:K3loc}).

To calculate a localized version of
the vertex $K_{ph,p'p''h'h''}$, Eq. (\ref{eq:K12adef}), we need
\begin{equation}
 \tilde K_{q,q'q''} \equiv
 N^2\,K_{q,q'q''} \>=\> \frac{1}{N\,\SF{q}\SF{q'}\SF{q''}}\,
 \sum_{hh'h''}\left[ H'_{ph,\,p'p''h'h''}-\sum_{p_1\!h_1}
   H'_{ph,\,p_1\!h_1}\> M^{\rm (I)}_{p_1\!h_1,\,p'p''h'h''}\right]
 \label{eq:defKbtilde}
\end{equation}
with
\begin{equation} 
H'_{ph,p'h'} = \;\delta_{\bf q\,,\,\bf q'} \biggl\{ 
 \delta_{h,h'}\,e_{ph} + \displaystyle
 \frac{1}{2N}\biggl[e_{ph}+e_{p'h'}  -2\frac{t(q)}{\SF{q}} \biggr]\,
 \tilde \Gamma_{\!\rm dd}(q)\biggr\} \,.
 \label{eq:H12aloc}
\end{equation}

We first separate the contribution that survives in the boson
limit. Starting with the identity
\begin{equation}
 \sum_{h'h''} \vert \Psi_{p'p''h'h''} \rangle \>=\>
 F\, \Bigl[ \hat\rho_{\vec q'}\hat\rho_{\vec q''} -
 \sum_{h'} a^{\dagger}_{h'+q'+q''} a^{\phantom\dagger}_{h'}\, 
  (\bar{n}_{\vec h'+\vec q''}- n_{\vec h'+\vec q'}) \,
  \Bigr]\,\vert\Phi_{\bf o}\rangle 
 \label{eq:H3_aux2}
\end{equation}
we have
\begin{equation}
 \sum_{hh'h"} H'_{ph,\,p'p''h'h''} \>=\> 
 \left\langle\Psi_{\bf o}\left\vert\,\hat\rho_{\vec q}H'
 \hat\rho_{\vec q'}\hat\rho_{\vec q''}\right\vert\Psi_{\bf o}\right\rangle
 \>-\>
\sum_{hh'}(\bar{n}_{\vec h'+\vec q''}- n_{\vec h'+\vec q'})\, H_{ph,\,h'+q\,h'}
 \,.
 \label{eq:H3_aux1}
\end{equation}
Postulating that three-body correlations have been optimized we can
simplify the first term
\begin{equation}
 \frac{1}{2N}\,
 \left\langle\Psi_{\bf o}\left\vert\,\Bigl[\bigl[\hat\rho_{\vec q},H'\bigr],
 \hat\rho_{\vec q'}\hat\rho_{\vec q''}\Big]\, 
 \right\vert\Psi_{\bf o}\right\rangle
 \>=\> -\frac{\hbar^2}{2m}\qvec\cdot\Bigl[\qvec''\,S(q')+
                                \qvec' \,S(q'')\Bigr] \,.
\end{equation}
For the form (\ref{eq:H12aloc}), the second term in (\ref{eq:H3_aux1}) is
\begin{eqnarray}
&-&\frac{1}{N}\sum_{hh'}H_{ph,\,h'+q\,h'}\,
 (\bar{n}_{\vec h'+\vec q''} -n_{\vec h'+\vec q'})=
 \frac{\hbar^2}{2m}\qvec\cdot\Bigl[\qvec''\,\SF{q'}+\qvec'\,\SF{q''}\Bigr]
 \nonumber\\
&+&\;\frac{\hbar^2}{4m}\qG(q) \biggl[q^2\, \SFthree(q,q',q'')
        +  {\bf q}\cdot\Bigl[{\bf q}''\,\SF{q'} + 
                             {\bf q}'\,\SF{q''}\Bigr]\SF{q}\biggr]\,.
\end{eqnarray}

The remaining term of $\tilde K_{q,q'q''}$ in (\ref{eq:K12bdef}),
$-\sum_{p_1h_1} H'_{ph,\,p_1h_1}\, M^{\rm (I)}_{p_1h_1,p'p''h'h''}$,
contains contributions originating from the diagonal and the
off-diagonal parts of $H'_{ph,p_1h_1}$, Eq. (\ref{eq:H12aloc}).
The off-diagonal part is identical to the
expression (\ref{eq:H3_aux0}),
whereas the contribution from the diagonal term gives
\begin{eqnarray}
-\frac{1}{N}\sum_{h,h',h''} e_{ph} \,M^{\rm (I)}_{ph,p'p''h'h''}&=&
\frac{\hbar^2}{2m}\qvec\cdot\Bigl[\qvec''\,\SF{q'}+\qvec'\,\SF{q''} 
 \Bigr]
\left[\frac{S(q')S(q'')}{\SF{q'}\SF{q''}}-1\right] \nonumber\\
&-&\frac{\hbar^2q^2}{2m}S(q')S(q'')
\tilde u_3(q,q',q'').
\end{eqnarray}
Collecting the individual contributions we obtain  Eq.~(\ref{eq:K3loc_1_2}).

\subsection{Four-body coupling matrix element}
\label{assec:Kfour}

In Eq.~(\ref{eq:Nfour}) we have defined the irreducible four-body
coupling matrix element $M^{\rm (I)}_{pp'hh',p''p'''h''h'''}$.  Again,
``irreducible'' means that in the diagrammatic representation left
and right arguments can not be separated by cutting a particle and a
hole line.  In analogy to the Bose case the ``convolution'' (``uniform
limit'') approximation is obtained by retaining the leading order
diagrams
\begin{eqnarray}
 M^{\rm (I)\,\scriptscriptstyle CA}_{pp'hh',p''p'''h''h'''} \>\equiv\>
 M_{ph,p''h''}\, M_{p'h',p'''h'''} \>+\>
M_{ph,p'''h'''}\, M_{p'h',p''h''} \,.
\label{eq:MIfour}
\end{eqnarray}
This contains all diagrams with up to two correlations.  A consistent
improvement of the convolution approximation involves an infinite
resummation.  For bosons \cite{eomI} this had only a marginal effect.
We expect a similarly small improvement for fermions.

The approximation for $K_{pp'hh',p''p'''h''h'''}$ consistent with 
(\ref{eq:MIfour}) is to keep all diagrams containing only one correlation 
function $\Gamma_{\!\rm dd}(r)$, 
\begin{eqnarray}
K^{\rm \scriptscriptstyle CA}_{pp'hh',p''p'''h''h'''} &\equiv&
 \delta_{p,p''}\delta_{h,h''}\,e_{ph}\,M_{p'h',p'''h'''} \>+\>
 \delta_{p',p'''}\delta_{h',h'''}\,e_{p'h'}\,M_{ph,p''h''}
 \nonumber\\
&+& \{p''h''\leftrightarrow p'''h'''\} \,.
\label{eq:Kfour}
\end{eqnarray}
Note that both $M^{\rm(I)\, \scriptscriptstyle CA}_{pp'hh',p''p'''h''h'''}$ and
$K^{\rm\scriptscriptstyle CA}_{pp'hh',p''p'''h''h'''}$ contain explicit
particle- and hole-labels.
Again, we no longer spell out the superscript ``CA'' in the following.

A word is in order about the symmetry of both
quantities. Eqs. (\ref{eq:MIfour}) and (\ref{eq:Kfour}) show that both
operators are the sum of two term that differ from each other merely by
the interchanging $\{p''h''\leftrightarrow p'''h'''\}$. We have discussed
in connection with Eq. (\ref{eq:pair_eq}) that it is legitimate to replace
$M^{\rm (I)}_{pp'hh',p''p'''h''h'''}$ and $K_{pp'hh',p''p'''h''h'''}$
by their asymmetric form.

\section{Pair propagator}
\label{asec:ex_pair_eqq}

\subsection{Pair energy matrix}
\label{assec:EpairConvApp}

{\it A priori\/}, $E_{pp'hh',p''p'''h''h'''}(\omega)$ is a function of
four hole and four particle momenta as well as the energy. In the
uniform limit approximation we can, however, express the inverse in
terms of two-body quantities.  From (\ref{eq:MIfour}) and
(\ref{eq:Kfour}) we obtain the pair energy matrix
\begin{eqnarray}
 E_{pp'hh',p''p'''h''h'''}(\omega)  &=&
  (\hbar\omega\!+\!\I\eta) M_{ph,p''h''}\, M_{p'h',p'''h'''} 
 \nonumber\\&&
 \>-\> (\delta_{p,p''}\delta_{h,h''}\,e_{ph})\,M_{p'h',p'''h'''} 
 \>-\> M_{ph,p''h''}\,(\delta_{p',p'''}\delta_{h',h'''}\,e_{p'h'}) \, .
 \label{eq:Eomega_appr}
\end{eqnarray}
To calculate its inverse,
write (\ref{eq:Eomega_appr}) as
\begin{eqnarray}
\sum_{p_1h_1p_2h_2} M^{-1}_{ph,p_1h_1}M^{-1}_{p'h',p_2h_2}
 E_{p_1p_2h_1h_2,p''p'''h''h'''}(\omega)  \;=\;
  (\hbar\omega\!+\!\I\eta) \delta_{p,p''}\delta_{h,h''}\,
                           \delta_{p',p'''}\delta_{h,h'''} \nonumber
 \\
 \>-\> (M^{-1}_{ph,p''h''}\,e_{p''h''})\> \delta_{p',p'''}\delta_{h,h'''}
 \>-\> \delta_{p,p''}\delta_{h,h''}\> (M^{-1}_{p'h',p'''h'''}\,e_{p'''h'''})
\end{eqnarray}
Use now, for two commuting operators ${\rm A,B}$
\begin{equation}
 \Bigl[(\hbar\omega\!+\!\I\eta) - {\rm A} - {\rm B} \Bigr]^{-1}
 \;=\;
 -\!\int\limits_{-\infty}^{\infty}\!\frac{d\hbar\omega'}{2\pi\I}\>
 \Bigl[(\hbar\omega'\!+\!\I\eta)- {\rm A} \Bigr]^{-1}\>
 \Bigl[\hbar(\omega\!-\!\omega'\!+\!\I\eta)- {\rm B} \Bigr]^{-1}
 \,,
\end{equation} 
which can be proved by series expansion.
Consequently, we have
\begin{eqnarray}
 E^{-1}_{pp'hh',p''p'''h''h'''}(\omega)\, 
 \;=\;
 -\!\int\limits_{-\infty}^{\infty}\!\frac{d\hbar\omega'}{2\pi\I}\>
 \kappa_{ph,p''h''}(\omega')\>
 \kappa_{p'h',p'''h'''}(\omega\!-\!\omega')
 \label{eq:Einvphfromkappa}
\end{eqnarray}
with
\begin{equation}
 \kappa_{ph,p'h'}(\omega) \>\equiv\> \left[(\hbar\omega\!+\!\I\eta)
 M_{ph,p'h'} -\> \delta_{pp'}\delta_{hh'}\,e_{ph}\right]^{-1}
 \,.
 \label{eq:kappadef}
\end{equation}
For our choice (\ref{eq:MtoGamma}) of $M_{p'h',ph}$,
we can calculate $\kappa_{ph,p'h'}(\omega)$ analytically,
\begin{eqnarray}
 \kappa_{ph,p'h'}(\omega) &=&
 \frac{\delta_{p,p'}\delta_{h,h'}}{\hbar\omega - e_{ph} +\I\eta}
 \\\nonumber &-&
 \frac{1}{\hbar\omega - e_{ph} +\I\eta}\>
\frac{\hbar\omega\,\tilde\Gamma_{\rm dd}(q)/N}
{1+\,\hbar\omega\,\tilde\Gamma_{\rm dd}(q)\>\kappa^0(q;\omega)}\,\>
 \frac{1}{\hbar\omega - e_{p'h'} +\I\eta}
 \,,
 \label{eq:kappalocal}
\end{eqnarray}
where $\kappa^0(q;\omega)$ has been defined in Eq. (\ref{eq:chi0pmdef}).

According to Eqs.~(\ref{eq:Wpmdef}) and (\ref{eq:WABdef}), the dynamic parts
of the interactions are obtained from matrix products of
$E^{-1}_{pp'hh',p''p'''h''h'''}(\omega)$ as given in (\ref{eq:Einvphfromkappa})
with the three-body vertices (\ref{eq:K3loc_1_2}) and (\ref{eq:K3loc}).
The latter being local functions, only sums over the hole states
enter $V_{\!_{\rm A,B}}(q;\omega)$.
\begin{equation}
\tilde E^{-1}(q_1,q_2; \omega)
\equiv \frac{1}{N^2}\!\sum_{h_1h_2h_1'h_2'}  
 E^{-1}_{p_1p_2h_1h_2,p_1'p_2'h_1'h_2'} (\omega)\
 \;=\;
 -\!\int\limits_{-\infty}^{\infty}\!\frac{d\hbar\omega'}{2\pi\I}\>
 \kappa(q_1;\omega')\> \kappa(q_2;\omega\!-\!\omega')
 \label{eq:Einvqfromkappa}
\end{equation}
with
\begin{equation}
 \kappa(q;\omega) \>\equiv\> \frac{1}{N}\sum_{hh'} \kappa_{ph,p'h'}(\omega)
 \>=\> \frac {\kappa_0(q;\omega)}{1+
              \hbar\omega\tilde\Gamma_{\rm dd}(q)\kappa_0(q;\omega)}\,.
\label{eq:kappa_qomega}
\end{equation}
Using Kramers-Kronig relations, we obtain the useful alternative representation
\begin{equation}
\tilde E^{-1}(q_1,q_2;\omega) \>=\>
\int\limits_{-\infty}^\infty\! \frac{d(\hbar\omega_1)d(\hbar\omega_2)}{\pi^2}\>
\frac{\Im m \kappa(q_1;\omega_1) \,\Im m \kappa(q_2;\omega_2)}
{\hbar\omega_1+\hbar\omega_2-\hbar\omega-\I\eta}\,.
\label{eq:EinvfromImkappa}
\end{equation}

\subsection{Properties of the pair propagator}
\label{assec:PairProp}

\subsubsection{Properties of $\kappa(q;\omega)$}
\label{assec:propkappa}

The structure of $\kappa(q;\omega)$ resembles that of $\chi(q;\omega)$ in 
the RPA.
It features a particle-hole continuum $\kappa_{\rm cont}(q;\omega)$,
and, possibly, a ``collective mode'' with a dispersion relation given by
\begin{equation}
 1 + \kappa_0\bigl(q;\omega_{\rm c}(q)\bigr)\, \hbar\omega_{\rm c}(q)
 \, \tilde \Gamma_{\!\rm dd}(q)=0\,.
\label{eq:KappaColl}
\end{equation} 
We can therefore write
\begin{eqnarray}
 {\Im}m\, \kappa(q,\omega) &=& z(q)\pi\,\delta(\hbar\omega-\hbar\omega_c(q)) + 
 {\Im}m\, \kappa_{\rm cont}(q;\omega)\,,
 \nonumber\\
 z(q) &=& \frac{\kappa_0(q;\omega)}{\tilde\Gamma_{\!dd}(q)\,
 \frac{d}{d\omega}\,\omega\kappa_0(q;\omega)} \,\bigg|_{\omega_{\rm c}(q)}
 \,.
 \label{eq:kappacoll}
\end{eqnarray}


$\kappa(q,\omega)$ satisfies the following sum rules which we write in the
suggestive way
\begin{eqnarray}
\frac{S^2(q)}{S_{\rm\scriptscriptstyle F}^2(q)}
\int\limits_0^{\infty}\! \frac{d(\hbar\omega)}{\pi}\> 
\Im m \kappa(q;\omega)
 &=& - S(q)
\label{eq:m0_kappa}
\\
\frac{S^2(q)}{S_{\rm\scriptscriptstyle F}^2(q)}
 \int\limits_0^{\infty}\! \frac{d(\hbar\omega)}{\pi}\> \hbar\omega\>
 \Im m \kappa(q;\omega)
 &=& - t(q)\,.
\label{eq:m1_kappa}
\end{eqnarray}
Eq.~(\ref{eq:m0_kappa}) is proved by extending the integration to
$-\infty$, noting that $\kappa_0(q;\omega)$ is real on the negative
$\omega$ axis. Since $\kappa_0(q;\omega)$ has no poles in the upper
complex plane, we can evaluate the integral along a circle, using the
asymptotic expansion
\begin{equation}
\kappa_0(q;\omega\!\to\!\infty) \>=\> \frac{\SF{q}}{\hbar\omega} + 
 \frac{t(q)}{\hbar^2\omega^2}
+ {\cal O}(\hbar\omega)^{-3}\,. 
\label{eq:kappa_asympt}
\end{equation}
The proof of Eq.~(\ref{eq:m1_kappa}) proceeds along the same line,
subtracting the asymptotic expansion of $\kappa(q;\omega)$ beforehand.
From Eqs. (\ref{eq:m0_kappa}), (\ref{eq:m1_kappa}) it is clear that
the analytic properties of $S^2(q)\,
\kappa(q;\omega)/S^2_{\rm\scriptscriptstyle F}(q)$ are similar to those of
the density-density response function
$\chi^{\rm RPA}(q;\omega)$.  For bosons, the two functions coincide 
exactly: Identifying $\tilde\Gamma_{\!\rm dd}(q) = S(q)\!-\!1$ and
$\SF{q}\!=\!1$, $\kappa^0(q;\omega)$ consists of a single mode, so that
\begin{equation}
\kappa_0(q;\omega) = \frac{1}{\hbar\omega\!+\!\I\eta -t(q)} \,, \qquad
\kappa(q;\omega) = \frac{1}{S(q)}
\frac{1}{\hbar\omega\!+\!\I\eta -\varepsilon(q)} \,.
\label{eq:kappaRPA}
\end{equation}

\begin{figure}
\centerline{\includegraphics[height=7.4cm]{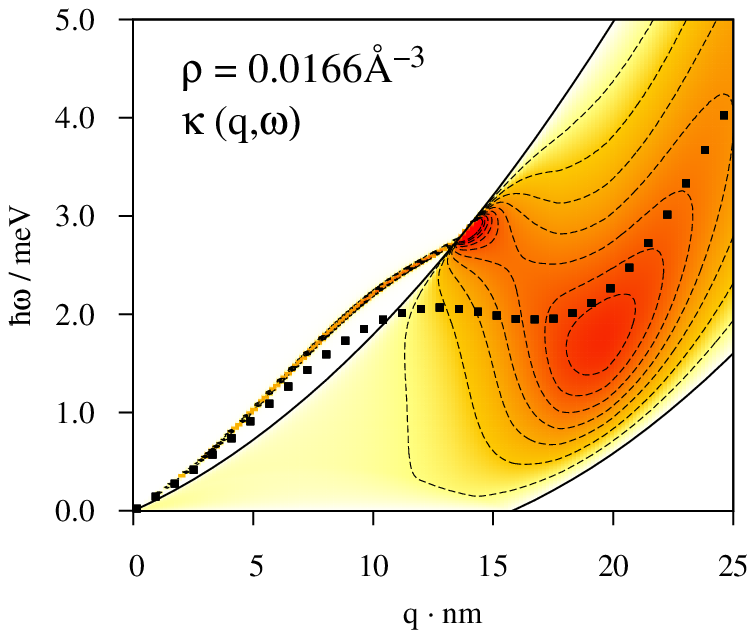}
            \includegraphics[height=7.4cm]{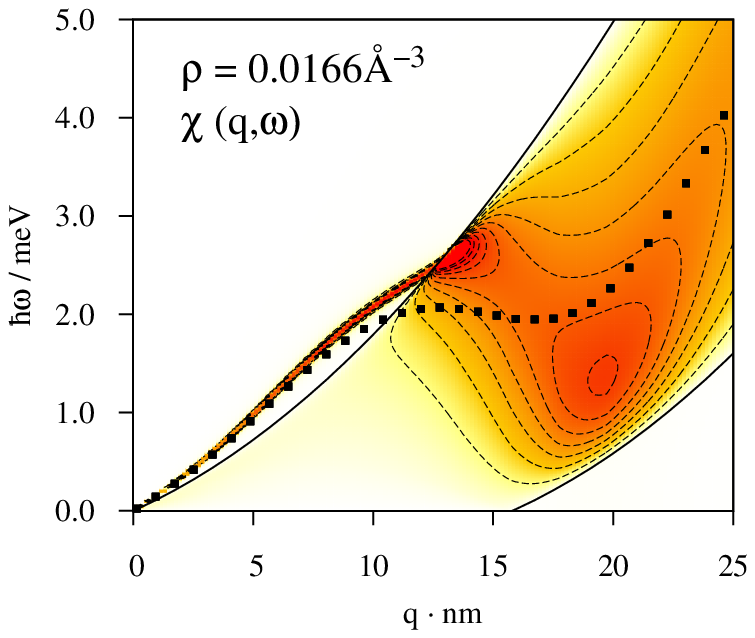}}
\caption{Imaginary part of the scaled propagator
$S^2(q)\,\kappa(q,\omega)/S_F^2(q)$ (left) and of $\chi^{\rm RPA}(q,\omega)$
(right) at the density $\rho = 0.0166\,$\AA$^{-3}$. The black squares 
show, for reference, the Feynman dispersion relation $\varepsilon(q)$.
\label{fig:E_q13}}
\end{figure}

Figure \ref{fig:E_q13} further confirms this similarity for \he{3} at
saturation density.  Expectedly, a solution of
Eq.~(\ref{eq:KappaColl}) is found to lie within a few percent of the
RPA zero sound mode.

\subsubsection{Properties of $\tilde E^{-1}(q,q';\omega)$}
\label{assec:propinvE}

Equations (\ref{eq:m0_kappa}) and (\ref{eq:m1_kappa})
lead to the sum rules for the pair propagator,
\begin{eqnarray}
 \int\limits_{-\infty}^{\infty}\! \frac{d(\hbar\omega)}{\pi}\> 
 \Im m E^{-1}(q,q';\omega) &=&
 -\frac{S_F^2(q)}{S(q)}\frac{S_F^2(q')}{S(q')}\,.
\label{eq:m0pair}
 \\
 \int\limits_{-\infty}^{\infty}\! \frac{d(\hbar\omega)}{\pi}\> \hbar\omega\>
 \Im m E^{-1}(q,q';\omega) &=&
 -\frac{S_F^2(q)}{S(q)}\frac{S_F^2(q')}{S(q')}
 \left(\varepsilon(q)+\varepsilon(q')\right) \,.
\label{eq:m1pair}
\end{eqnarray}
The proof of (\ref{eq:m0pair}) is best carried out starting from the
representation (\ref{eq:EinvfromImkappa}),
\begin{eqnarray}
\int\limits_0^{\infty} \frac{d(\hbar\omega)}{\pi} \Im m E^{-1}(q_1,q_2;\omega)&=&
-\int \limits_{0}^{\infty}\! \frac{d\hbar\omega_1}{\pi}\>
\Im m \kappa(q_1;\omega_1)
\int\limits_0^{\infty} \frac{d(\hbar\omega)}{\pi}\>
\Im m \kappa(q_2;\omega-\omega_1)\,.
\end{eqnarray}
The $\hbar\omega$ integral in the last term can be extended to $-\infty$
since $\Im m \kappa(q;\omega)$ is real on the negative $\omega$-axis.

If Eq.~(\ref{eq:KappaColl}) has a solution, the pair propagator has 
a collective mode. From (\ref{eq:kappacoll}) we obtain 
\begin{equation}
{\Im}m\,\tilde E^{-1}(q_1,q_2;\omega) =  
 \pi\,z(q_1)\,z(q_2)\>
 \delta(\hbar\omega_{\rm c}(q_1)+\hbar\omega_{\rm c}(q_2)-\hbar\omega) \,.
\label{eq:Einvdoublpole}
\end{equation}
This is the origin of two-phonon excitations, or the double-plasmon
in charged systems.

The two-particle-two-hole band consists of three parts which may
overlap. The first one is the continuum--continuum (c-c) coupling,
where the contribution of each $\kappa(q,\omega)$ in
(\ref{eq:Einvqfromkappa}) comes from its particle hole band.  This
defines the two-particle-two-hole ``tube'' in $(q,q';\omega)$ space.
Its boundaries are
\begin{equation}
 e_{\rm min}(q)+e_{\rm min}(q') \>\leq\>\hbar\omega\>\leq\>
 e_{\rm max}(q)+e_{\rm max}(q') \,,
\label{eq:cccoupling}
\end{equation}
where $e_{\rm min}$ and $e_{\rm max}$ denote the upper and lower border of 
each single-particle-hole band, respectively.  

The other two parts of $E^{-1}(q,q';\omega)$ arise from
continuum--mode (c-m) coupling, they are identical apart from
interchanging $q$ and $q'$. Their boundaries are
\begin{equation}
 e_{\rm min}(q)+\hbar\omega_{\rm cm}(q') \>\leq\>\hbar\omega\>\leq\>
 e_{\rm max}(q)+\hbar\omega_{\rm cm}(q') \,.
\label{eq:cmcoupling}
\end{equation}

Finally, we consider three limits of the pair propagator.
First, in the non-interacting case, $\tilde \Gamma_{\!\rm dd}(q) = 0$, 
we simply obtain a sum over two-pair energy denominators
\begin{equation}
\tilde E^{-1}_{\rm F}(q,q';\omega)
=\> -\int\frac{ d\hbar\omega'}{2\pi\I}
\kappa_0(q';\omega-\omega')\kappa_0(q;\omega')
=\> \frac{1}{N^2}\sum_{hh'}\frac{1}{\hbar\omega\!+\!\I\eta -e_{ph}-e_{p'h'}} \, ,
\end{equation}
{\it i.e.\/} the two-particle energy denominator appropriate for
perturbation theory in a weakly interacting Fermi system.

Second, (\ref{eq:kappaRPA}) reproduces the energy
denominator appearing in the boson theory,
\begin{equation}
\tilde E^{-1}_{\rm bos}(q,q';\omega)
= \frac{1}{S(q)S(q')}\>
\frac{1}{ \hbar\omega\!+\!\I\eta-\varepsilon(q)-\varepsilon(q')  } \,.
\label{eq:E_pair_bose}
\end{equation}

Finally, we consider the ``collective'' or ``uniform limit'' approximation.
Following (\ref{eq:MSA}) we replace $\kappa_0(q;\omega)$ by
that single-pole approximation which ensures its correct $\omega^0$ and
$\omega^1$ sum rules. This gives
\begin{eqnarray}
 \kappa^{\rm\scriptscriptstyle CA}_0(q;\omega) \>=\>
 \frac{\SF{q}}{\hbar\omega\!+\!\I\eta -t(q)/\SF{q}} 
 \,,
 \\
 \kappa^{\rm\scriptscriptstyle CA}(q;\omega)=
 \frac{S_{\rm\scriptscriptstyle F}^2(q)}{S(q)}\,
 \frac{1}{\hbar\omega\!+\!\I\eta -\varepsilon(q)}
 \,,
\end{eqnarray}
and
\begin{equation}
E^{-1}_{\rm\scriptscriptstyle CA}(q,q';\omega)
= \frac{S_F^2(q)}{S(q)}\frac{S_F^2(q')}{S(q')}
\frac{1}{\hbar\omega\!+\!\I\eta-\varepsilon(q)-\varepsilon(q')  }
 \,.
\end{equation}

The boson limit as well as the collective approximation demonstrate
the effect of correlations: The single-particle energies get shifted
and form a band around the ``Feynman-spectrum''. Note that the collective
approximation satisfies the sum rules (\ref{eq:m0pair})-(\ref{eq:m1pair})
exactly.

\subsubsection{Pair propagator for charged systems}
\label{asssec:PairCharged}

For charged systems, the dispersion of the solution of Eq.~(\ref{eq:KappaColl}) 
has, unlike the plasmon, a term that is linear in the wave number:
\begin{equation}
 \hbar\omega_{\rm c}(q) = \omP +
 \frac{t_{\rm F}}{6}\,\frac{q}{k_{\rm F}} -
 \frac{9t_{\rm F}^2}{4 \hbar\omP} \left( \frac{q}{\qkf} \right)^2
 + {\cal O}(q^3)
 \,.
\end{equation}
For the strength of this mode we obtain
\begin{equation}
 z(q,\omega_{\rm c}(q)) \>=\>
 \frac{9\hbar\omP}{16t_{\rm F}} - \frac{3}{32}\frac{q}{\qkf} \,.
\end{equation}
Hence, to leading order, for the pole of $E^{-1}(q_1,q_2;\omega)$ in
(\ref{eq:Einvdoublpole}) we obtain
\begin{equation}
\Im m \, \tilde E^{-1}(q',q';\omega)
= - \pi \, \left( \frac{9}{16}  \frac{\hbar\omP}{t_{\rm F}}
- \frac{3}{32}\frac{q'}{\qkf} \right)^2 \,
\delta\left(\hbar\omega - 2\hbar\omP -
\frac{t_{\rm F}}{3} \frac{q'}{\qkf} \right) \ \mbox{as}\ q'\rightarrow 0.
\label{eq:Einv_dpl_expansion}
\end{equation}
Note that the location of double-plasmon pole contains, in leading order
in the momentum transfer, no information on many-body correlations.

\section{Large momentum limit}
\label{asec:static}

For large momenta, $S(q)\!-\!1$ falls off at least as $q^{-4}$.  The
vertices (\ref{eq:K3loc_1_2}) and (\ref{eq:K3loc}) fall off as $q^{-1}$
and as $q^{-2}$, respectively, hence we have
\begin{eqnarray}
	\tilde K_{q, q' q''}      &\approx& 
	\frac{S(q')\,S(q'')}{\SF{q'}\,\SF{q''}}\, 
	\frac{\hbar^2}{2m}\, \bigg[ \vec q \cdot \vec q' \, \qX(q') +
        \vec q \cdot \vec q'' \, \qX(q'') \bigg] \, ,
	\label{eq:K3loc_1_2_large_q}
	\\\nonumber
        \tilde K^{(q)}_{q'q'',0} &\approx& 0 \, .
\end{eqnarray}
As a consequence, $\widetilde W_{\!_{\rm B}}(q; 0)$ is negligible for
large momenta, and only the first term in Eq.~(\ref{eq:resultWA}))
contributes to $\widetilde W_{\!_{\rm A}}(q; 0)$.

For large $q$ either $q'$ or $q''$ (or both) must be large.
(let $q''\!\ge\!q'$, the symmetry in ${\bf q'} \leftrightarrow {\bf q''}$ 
just yielding a factor of two).
Since $\qX(q)$ falls off for large $q$, the dominant contribution of 
(\ref{eq:K3loc_1_2_large_q}) then arises from small $q'$ and we can write
\begin{eqnarray}
\tilde W_A(q\rightarrow\infty,0)
&=& \left(\frac{\hbar^2}{2m}\right)^2\frac{1}{N}\sum_{{\bf q}'}
\left(\frac{S(q')}{\SF{q'}}\right)^2
\left[ \vec q \cdot \vec q' \, \qX(q')\right]^2 \tilde E^{-1}(q',q'';0)\nonumber\\
&=& \frac{t(q)}{3} \frac{1}{N} \sum_{{\bf q}'}t(q')
\left[\frac{S(q')}{\SF{q'}}\qX(q')\right]^2 \tilde E^{-1}(q',q;0)\,.
\label{eq:WA_static_1}
\end{eqnarray}

We now use the representation (\ref{eq:Einvqfromkappa}) for the pair propagator 
\begin{equation}
\tilde E^{-1}(q',q;0) = -\int\limits_{-\infty}^\infty\! \frac{d\hbar\omega'}{\pi}\>
 {\Re}e\, \kappa(q',\omega')\, {\Im}m\, \kappa(q,-\omega') \,.
\end{equation}
Since $\kappa^0(q\!\gg\!\qkf;\omega) = 1/(\hbar\omega -t(q) +\I\eta)$ we have
\begin{equation}
 \kappa(q\!\to\!\infty;\omega) \>=\> 
 \frac{1}{S(q)}\> \frac{1}{\hbar\omega - \varepsilon(q) +\I\eta} \,.
\end{equation}
Consequently,
\begin{eqnarray}
\tilde  E^{-1}(q',q\!\to\!\infty;0) &=&   \frac{1}{S(q)}\>
 {\Re}e\, \kappa(q',-\textstyle\frac{1}{\hbar}\displaystyle\varepsilon(q))
 \nonumber\\ &=&
 -\frac{1}{t(q)} \frac{S{\rm _F}^2(q')}{S(q')} \,,
\label{eq:Einv_static}
\end{eqnarray}
where the last equality follows from the high-frequency limit 
$\kappa^0(q';\omega)\to \SF{q'}/\omega$.
Insertion into (\ref{eq:WA_static_1}) yields
\begin{equation}
\tilde W_A(q\rightarrow\infty,0) \>=\>
 -\frac{1}{3N} \sum_{{\bf q}'} 
 t(q') S(q') \left[\qX(q')\right]^2  \,,
\end{equation}
which together with Eq.~(\ref{eq:TCA}) gives the result (\ref{eq:Vstatqinf}).

\section{Sum rules}
\label{asec:sum_rules}

For bosons, the $\omega^0$ and $\omega^1$ sum rules (\ref{eq:m0}) and
(\ref{eq:m1}) are satisfied exactly \cite{JacksonSumrules} in the
sense that the result of the frequency integration is independent of
the level at which pair fluctuations are treated. This feature
provides an unambiguous method to determine the static particle-hole
interaction $\Vph{q}$ through the sum rule (\ref{eq:m0}) from the
static structure function.

The proof of the $m_1$ sum rule is identical to the one for
bosons: Due to the symmetry
\[\chi(q;\omega) = \chi^*(q,-\omega)\]
we can write
\begin{equation}
m_1 = -\frac{1}{2\pi}\Im m \int_{-\infty}^\infty d(\hbar\omega)\,
\hbar\omega\,\chi(q;\omega)\,.
\end{equation}
All poles of $\chi(q; \omega)$ are in the lower half plane, allowing to
close the integral in the upper half plane. For large
$\omega$ we have, however,
\begin{equation}
\chi_0(q;\omega) - \chi^{\rm RPA}(q;\omega)\propto \omega^{-4}
\qquad
\chi_0(q;\omega) - \chi(q;\omega)\propto \omega^{-4}
\label{eq:chiasympt}
\end{equation}
since
\begin{equation}
{\widetilde{V}_{A,B}(q; \omega)} = \tilde V_{\rm ph}(q) +
\frac{\mathrm{const.}}{\omega}\qquad
{\mathrm as}\qquad\omega\rightarrow\infty.
\end{equation} 
We have therefore
\begin{equation}
\Im m \int_{-\infty}^\infty d(\hbar\omega) \,\hbar\omega\,\chi(q;\omega)
= \Im m \int_{-\infty}^\infty d(\hbar\omega)\,
\hbar\omega\,\chi^{\rm RPA}(q;\omega)
= \Im m \int_{-\infty}^\infty d(\hbar\omega)\, 
\,\hbar\omega\,\chi(q;\omega)\,.
\end{equation}

For fermions, the frequency integration in (\ref{eq:m0}) must be
carried out numerically, which is best done by Wick rotation along the
imaginary axis.  The result of the integration is no longer rigorously
independent of the approximation used for the response function.


\begin{figure}
\centerline{\includegraphics[width=8.6cm]{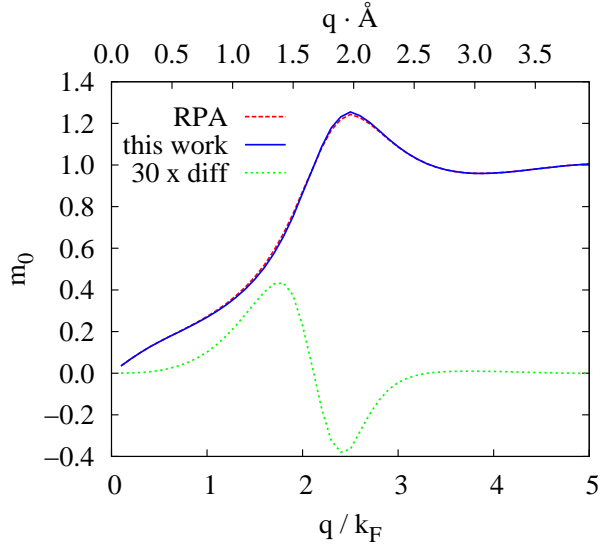}}
\caption{Result of the $m_0$ sum rule for $^3$He at saturated vapor pressure.
The purple dashed line shows the FHNC $S(q)$ and the blue short
dashed line the result from the pair fluctuation theory; the
dashed green line shows the difference, magnified by a factor of 30 to make it
visible.
\label{fig:sumrule_166}}
\end{figure}

Fig. \ref{fig:sumrule_166} compares the $m_0$ sum rule calculated
within the RPA and the pair excitation theory.  Evidently, the
discrepancy is very small. One can understand by comparing with the
boson theory: If we restricted the fluctuation operators $\delta
u^{(1)}_{ph} (t)$ and $\delta u^{(2)}_{pp'hh'} (t)$ to be functions of
momentum transfers $\qvec = \pvec - \hvec$ and $\qvec' = \pvec' -
\hvec'$, we would end up with a density-density response function that
is formally identical to that of bosons and would, hence, lead to an
$S(q)$ that is independent of the treatment of the pair
fluctuations. The expectation that the inclusion of the particle-hole
structure of the two-pair energy denominator makes only a small
difference is verified in Fig.~\ref{fig:sumrule_166}.  Thus, it is
also legitimate in the pair-excitation theory to obtain the static
particle-hole interaction $\tilde V_{\rm p-h}(q)$ from the static
structure function $S(q)$ through Eqs. (\ref{eq:m0}) and
(\ref{eq:RPAresponse}).

\section{Implementation Recipe}
\label{asec:recipe}

This section provides, for the convenience of the reader and easy
further reference, a compilation of all necessary ingredients to
implement the theory.  Mostly a summary of sections
\ref{sec:localapprox} and \ref{ssec:responsefct}, we deliberately
refrain from any explanation to avoid redundancy and keep it as
compact as possible.

We have shown in our applications to \he3 and the electron liquid that
for practical purposes, only one of the local three-body vertices is
necessary:
\begin{eqnarray}
\tilde K_{q,q'q''}
	&=& \displaystyle \frac{\hbar^2}{2m} \,
		\frac{S(q')S(q'')}{\SF{q}\SF{q'}\SF{q''}}
		\left[
			\qvec\!\cdot\qvec' \, \tilde X_{\rm dd}(q') +
			\qvec\!\cdot\qvec''\, \tilde X_{\rm dd}(q'') - q^2 \tilde{u}_3(q,q',q'')
		\right]\,,
\end{eqnarray}
where ${u}_3(q,q',q'')$ is the three-body ground state correlation
\cite{polish}. The effective interaction
$\widetilde W_{\!_{\rm A}}(q,\omega)$ is then
\begin{equation}
 \widetilde W_{\!_{\rm A}}(q;\omega) =
  \frac{1}{2N}\sum_{{\bf q}'}
  |\tilde K_{q,q'q''}|^2 \, \tilde E^{-1}(q',q''; \omega)
\end{equation}
whereas  $\widetilde W_{\!_{\rm B}}(q,\omega)$ vanishes.
Consequently, the
components of the (energy--dependent) interaction matrix
${\bf V}_{\!\rm p-h}(\omega)$ are
\begin{eqnarray}
\tilde V_{\!_{\rm A}}(q; \omega) = \tilde V_{\rm p-h}(q)
&+& [\sigma^{+}_q]^2\, \widetilde W  _{\!_{\rm A}}(q; \omega) +
[\sigma^{-}_q]^2\, \widetilde W^*_{\!_{\rm A}}(q;-\omega) \, ,
\\
\tilde V_{\!_{\rm B}}(q; \omega) = \tilde V_{\rm p-h}(q)
&+& \sigma^{+}_q\sigma^{-}_q\, \left(
		\widetilde W_{\!_{\rm A}}  (q; \omega) +
		\widetilde W_{\!_{\rm A}}^*(q;-\omega)
	\right)\;,
\end{eqnarray}
with $\sigma^{\pm}_q \equiv [\SF{q}\pm S(q)]/2S(q)$.

Finally we need the pair propagator:
\begin{eqnarray}
\tilde E^{-1}(q_1,q_2; \omega)
 &=&
 -\!\int\limits_{-\infty}^{\infty}\!\frac{d\hbar\omega'}{2\pi\I}\>
 \kappa(q_1;\omega')\> \kappa(q_2;\omega\!-\!\omega')
\\
 \kappa(q;\omega)
 &=&\frac {\kappa_0(q;\omega)}{1+
              \hbar\omega\tilde\Gamma_{\rm dd}(q)\kappa_0(q;\omega)}
\end{eqnarray}
with the partial Lindhard functions:
\begin{equation}
\kappa^{\phantom{*}}_0(q;\omega) \equiv \frac{1}{N}\sum_h
\frac{\bar n_{\bf p} n_{\bf h}}{\hbar\omega - e_{ph} + \I\eta}
\end{equation}
The simplifications of the interactions do not
significantly simplify the form (\ref{eq:structurechi}) of
the density-density response function.


\begin{thebibliography}{67}
\expandafter\ifx\csname natexlab\endcsname\relax\def\natexlab#1{#1}\fi
\expandafter\ifx\csname bibnamefont\endcsname\relax
  \def\bibnamefont#1{#1}\fi
\expandafter\ifx\csname bibfnamefont\endcsname\relax
  \def\bibfnamefont#1{#1}\fi
\expandafter\ifx\csname citenamefont\endcsname\relax
  \def\citenamefont#1{#1}\fi
\expandafter\ifx\csname url\endcsname\relax
  \def\url#1{\texttt{#1}}\fi
\expandafter\ifx\csname urlprefix\endcsname\relax\def\urlprefix{URL }\fi
\providecommand{\bibinfo}[2]{#2}
\providecommand{\eprint}[2][]{\url{#2}}

\bibitem[{\citenamefont{Kerman and Koonin}(1976)}]{KermanKoonin}
\bibinfo{author}{\bibfnamefont{A.~K.} \bibnamefont{Kerman}} \bibnamefont{and}
  \bibinfo{author}{\bibfnamefont{S.~E.} \bibnamefont{Koonin}},
  \bibinfo{journal}{Ann. Phys. (NY)} \textbf{\bibinfo{volume}{100}},
  \bibinfo{pages}{332} (\bibinfo{year}{1976}).

\bibitem[{\citenamefont{Kramer and Saraceno}(1981)}]{KramerSaraceno}
\bibinfo{author}{\bibfnamefont{P.}~\bibnamefont{Kramer}} \bibnamefont{and}
  \bibinfo{author}{\bibfnamefont{M.}~\bibnamefont{Saraceno}},
  \emph{\bibinfo{title}{Geometry of the time-dependent variational principle in
  quantum mechanics}}, vol. \bibinfo{volume}{140} of
  \emph{\bibinfo{series}{Lecture Notes in Physics}}
  (\bibinfo{publisher}{Springer}, \bibinfo{address}{Berlin, Heidelberg, and New
  York}, \bibinfo{year}{1981}).

\bibitem[{\citenamefont{Thouless}(1972)}]{ThoulessBook}
\bibinfo{author}{\bibfnamefont{D.~J.} \bibnamefont{Thouless}},
  \emph{\bibinfo{title}{The quantum mechanics of many-body systems}}
  (\bibinfo{publisher}{Academic Press}, \bibinfo{address}{New York},
  \bibinfo{year}{1972}), \bibinfo{edition}{2nd} ed.

\bibitem[{\citenamefont{Chen et~al.}(1982)\citenamefont{Chen, Clark, and
  Sandler}}]{LDavid}
\bibinfo{author}{\bibfnamefont{J.~M.~C.} \bibnamefont{Chen}},
  \bibinfo{author}{\bibfnamefont{J.~W.} \bibnamefont{Clark}}, \bibnamefont{and}
  \bibinfo{author}{\bibfnamefont{D.~G.} \bibnamefont{Sandler}},
  \bibinfo{journal}{Z. Physik A} \textbf{\bibinfo{volume}{305}},
  \bibinfo{pages}{223} (\bibinfo{year}{1982}).

\bibitem[{\citenamefont{Krotscheck}(1982)}]{rings}
\bibinfo{author}{\bibfnamefont{E.}~\bibnamefont{Krotscheck}},
  \bibinfo{journal}{Phys. Rev. A} \textbf{\bibinfo{volume}{26}},
  \bibinfo{pages}{3536} (\bibinfo{year}{1982}).

\bibitem[{\citenamefont{Jackson and Feenberg}(1961)}]{JaFe}
\bibinfo{author}{\bibfnamefont{H.~W.} \bibnamefont{Jackson}} \bibnamefont{and}
  \bibinfo{author}{\bibfnamefont{E.}~\bibnamefont{Feenberg}},
  \bibinfo{journal}{Ann. Phys. (NY)} \textbf{\bibinfo{volume}{15}},
  \bibinfo{pages}{266} (\bibinfo{year}{1961}).

\bibitem[{\citenamefont{Campbell and Krotscheck}(2009)}]{eomI}
\bibinfo{author}{\bibfnamefont{C.~E.} \bibnamefont{Campbell}} \bibnamefont{and}
  \bibinfo{author}{\bibfnamefont{E.}~\bibnamefont{Krotscheck}},
  \bibinfo{journal}{Phys. Rev. B} \textbf{\bibinfo{volume}{80}},
  \bibinfo{pages}{174501/1} (\bibinfo{year}{2009}).

\bibitem[{\citenamefont{Landau}(1957{\natexlab{a}})}]{LandauFLP1}
\bibinfo{author}{\bibfnamefont{L.~D.} \bibnamefont{Landau}},
  \bibinfo{journal}{Sov. Phys. JETP} \textbf{\bibinfo{volume}{3}},
  \bibinfo{pages}{920} (\bibinfo{year}{1957}{\natexlab{a}}).

\bibitem[{\citenamefont{Landau}(1957{\natexlab{b}})}]{LandauFLP2}
\bibinfo{author}{\bibfnamefont{L.~D.} \bibnamefont{Landau}},
  \bibinfo{journal}{Sov. Phys. JETP} \textbf{\bibinfo{volume}{5}},
  \bibinfo{pages}{101} (\bibinfo{year}{1957}{\natexlab{b}}).

\bibitem[{\citenamefont{Feynman}(1954)}]{Feynman}
\bibinfo{author}{\bibfnamefont{R.~P.} \bibnamefont{Feynman}},
  \bibinfo{journal}{Phys. Rev.} \textbf{\bibinfo{volume}{94}},
  \bibinfo{pages}{262} (\bibinfo{year}{1954}).

\bibitem[{\citenamefont{Feynman and Cohen}(1956)}]{FeynmanBackflow}
\bibinfo{author}{\bibfnamefont{R.~P.} \bibnamefont{Feynman}} \bibnamefont{and}
  \bibinfo{author}{\bibfnamefont{M.}~\bibnamefont{Cohen}},
  \bibinfo{journal}{Phys. Rev.} \textbf{\bibinfo{volume}{102}},
  \bibinfo{pages}{1189} (\bibinfo{year}{1956}).

\bibitem[{\citenamefont{Jackson and Feenberg}(1962)}]{JaFe2}
\bibinfo{author}{\bibfnamefont{H.~W.} \bibnamefont{Jackson}} \bibnamefont{and}
  \bibinfo{author}{\bibfnamefont{E.}~\bibnamefont{Feenberg}},
  \bibinfo{journal}{Rev. Mod. Phys.} \textbf{\bibinfo{volume}{34}},
  \bibinfo{pages}{686} (\bibinfo{year}{1962}).

\bibitem[{\citenamefont{Feenberg}(1969)}]{FeenbergBook}
\bibinfo{author}{\bibfnamefont{E.}~\bibnamefont{Feenberg}},
  \emph{\bibinfo{title}{Theory of Quantum Fluids}}
  (\bibinfo{publisher}{Academic}, \bibinfo{address}{New York},
  \bibinfo{year}{1969}).

\bibitem[{\citenamefont{Jackson}(1971)}]{Jackson71}
\bibinfo{author}{\bibfnamefont{H.~W.} \bibnamefont{Jackson}},
  \bibinfo{journal}{Phys. Rev. A} \textbf{\bibinfo{volume}{4}},
  \bibinfo{pages}{2386} (\bibinfo{year}{1971}).

\bibitem[{\citenamefont{Jackson}(1973)}]{JacksonSkw}
\bibinfo{author}{\bibfnamefont{H.~W.} \bibnamefont{Jackson}},
  \bibinfo{journal}{Phys. Rev. A} \textbf{\bibinfo{volume}{8}},
  \bibinfo{pages}{1529} (\bibinfo{year}{1973}).

\bibitem[{\citenamefont{Jackson}(1974)}]{JacksonSumrules}
\bibinfo{author}{\bibfnamefont{H.~W.} \bibnamefont{Jackson}},
  \bibinfo{journal}{Phys. Rev. A} \textbf{\bibinfo{volume}{9}},
  \bibinfo{pages}{964} (\bibinfo{year}{1974}).

\bibitem[{\citenamefont{Chang and Campbell}(1976)}]{Chuckphonon}
\bibinfo{author}{\bibfnamefont{C.~C.} \bibnamefont{Chang}} \bibnamefont{and}
  \bibinfo{author}{\bibfnamefont{C.~E.} \bibnamefont{Campbell}},
  \bibinfo{journal}{Phys. Rev. B} \textbf{\bibinfo{volume}{13}},
  \bibinfo{pages}{3779} (\bibinfo{year}{1976}).

\bibitem[{\citenamefont{Campbell and Krotscheck}(2010{\natexlab{a}})}]{eomII}
\bibinfo{author}{\bibfnamefont{C.~E.} \bibnamefont{Campbell}} \bibnamefont{and}
  \bibinfo{author}{\bibfnamefont{E.}~\bibnamefont{Krotscheck}},
  \emph{\bibinfo{title}{Dymanic many body theory {III}: Multi-particle
  fluctuations in bulk {$^4$He}}} (\bibinfo{year}{2010}{\natexlab{a}}),
  \bibinfo{note}{in preparation}.

\bibitem[{\citenamefont{{Aldrich III} and Pines}(1978)}]{ALP78}
\bibinfo{author}{\bibfnamefont{C.~H.} \bibnamefont{{Aldrich III}}}
  \bibnamefont{and} \bibinfo{author}{\bibfnamefont{D.}~\bibnamefont{Pines}},
  \bibinfo{journal}{J. Low Temp. Phys.} \textbf{\bibinfo{volume}{31}},
  \bibinfo{pages}{689} (\bibinfo{year}{1978}).

\bibitem[{\citenamefont{Iwamoto and Pines}(1984)}]{IP84}
\bibinfo{author}{\bibfnamefont{N.}~\bibnamefont{Iwamoto}} \bibnamefont{and}
  \bibinfo{author}{\bibfnamefont{D.}~\bibnamefont{Pines}},
  \bibinfo{journal}{Phys. Rev. B} \textbf{\bibinfo{volume}{29}},
  \bibinfo{pages}{3924} (\bibinfo{year}{1984}).

\bibitem[{\citenamefont{Clark}(1979)}]{Johnreview}
\bibinfo{author}{\bibfnamefont{J.~W.} \bibnamefont{Clark}}, in
  \emph{\bibinfo{booktitle}{Progress in Particle and Nuclear Physics}}, edited
  by \bibinfo{editor}{\bibfnamefont{D.~H.} \bibnamefont{Wilkinson}}
  (\bibinfo{publisher}{Pergamon Press Ltd.}, \bibinfo{address}{Oxford},
  \bibinfo{year}{1979}), vol.~\bibinfo{volume}{2}, pp.
  \bibinfo{pages}{89--199}.

\bibitem[{\citenamefont{Fabrocini et~al.}(2002)\citenamefont{Fabrocini,
  Fantoni, and Krotscheck}}]{KroTriesteBook}
\bibinfo{author}{\bibfnamefont{A.}~\bibnamefont{Fabrocini}},
  \bibinfo{author}{\bibfnamefont{S.}~\bibnamefont{Fantoni}}, \bibnamefont{and}
  \bibinfo{author}{\bibfnamefont{E.}~\bibnamefont{Krotscheck}},
  \emph{\bibinfo{title}{Introduction to Modern Methods of Quantum Many--Body
  Theory and their Applications}}, vol.~\bibinfo{volume}{7} of
  \emph{\bibinfo{series}{Advances in Quantum Many--Body Theory}}
  (\bibinfo{publisher}{World Scientific}, \bibinfo{address}{Singapore},
  \bibinfo{year}{2002}).

\bibitem[{\citenamefont{Scherm et~al.}(1987)\citenamefont{Scherm,
  Guckelsberger, F\aa{}k, Sk{\"o}ld, Dianoux, Godfrin, and Stirling}}]{SGFS87}
\bibinfo{author}{\bibfnamefont{R.}~\bibnamefont{Scherm}},
  \bibinfo{author}{\bibfnamefont{K.}~\bibnamefont{Guckelsberger}},
  \bibinfo{author}{\bibfnamefont{B.}~\bibnamefont{F\aa{}k}},
  \bibinfo{author}{\bibfnamefont{K.}~\bibnamefont{Sk{\"o}ld}},
  \bibinfo{author}{\bibfnamefont{A.~J.} \bibnamefont{Dianoux}},
  \bibinfo{author}{\bibfnamefont{H.}~\bibnamefont{Godfrin}}, \bibnamefont{and}
  \bibinfo{author}{\bibfnamefont{W.~G.} \bibnamefont{Stirling}},
  \bibinfo{journal}{Phys. Rev. Lett.} \textbf{\bibinfo{volume}{59}},
  \bibinfo{pages}{217} (\bibinfo{year}{1987}).

\bibitem[{\citenamefont{F{\aa}k et~al.}(1994)\citenamefont{F{\aa}k,
  Guckelsberger, Scherm, and Stunault}}]{Fak94}
\bibinfo{author}{\bibfnamefont{B.}~\bibnamefont{F{\aa}k}},
  \bibinfo{author}{\bibfnamefont{K.}~\bibnamefont{Guckelsberger}},
  \bibinfo{author}{\bibfnamefont{R.}~\bibnamefont{Scherm}}, \bibnamefont{and}
  \bibinfo{author}{\bibfnamefont{A.}~\bibnamefont{Stunault}},
  \bibinfo{journal}{J. Low Temp. Phys.} \textbf{\bibinfo{volume}{97}},
  \bibinfo{pages}{445} (\bibinfo{year}{1994}).

\bibitem[{\citenamefont{Glyde et~al.}(2000)\citenamefont{Glyde, F\aa{}k, van
  Dijk, Godfrin, Guckelsberger, and Scherm}}]{GFvDG00}
\bibinfo{author}{\bibfnamefont{H.~R.} \bibnamefont{Glyde}},
  \bibinfo{author}{\bibfnamefont{B.}~\bibnamefont{F\aa{}k}},
  \bibinfo{author}{\bibfnamefont{N.~H.} \bibnamefont{van Dijk}},
  \bibinfo{author}{\bibfnamefont{H.}~\bibnamefont{Godfrin}},
  \bibinfo{author}{\bibfnamefont{K.}~\bibnamefont{Guckelsberger}},
  \bibnamefont{and} \bibinfo{author}{\bibfnamefont{R.}~\bibnamefont{Scherm}},
  \bibinfo{journal}{Phys. Rev. B} \textbf{\bibinfo{volume}{61}},
  \bibinfo{pages}{1421} (\bibinfo{year}{2000}).

\bibitem[{\citenamefont{Sternemann et~al.}(2005)\citenamefont{Sternemann,
  Huotari, Vank{\'o}, Volmer, Monaco, Gusarov, Sturm, and Sch{\"u}lke}}]{SHV05}
\bibinfo{author}{\bibfnamefont{C.}~\bibnamefont{Sternemann}},
  \bibinfo{author}{\bibfnamefont{S.}~\bibnamefont{Huotari}},
  \bibinfo{author}{\bibfnamefont{G.}~\bibnamefont{Vank{\'o}}},
  \bibinfo{author}{\bibfnamefont{M.}~\bibnamefont{Volmer}},
  \bibinfo{author}{\bibfnamefont{G.}~\bibnamefont{Monaco}},
  \bibinfo{author}{\bibfnamefont{A.}~\bibnamefont{Gusarov}},
  \bibinfo{author}{\bibfnamefont{H.~L.~K.} \bibnamefont{Sturm}},
  \bibnamefont{and}
  \bibinfo{author}{\bibfnamefont{W.}~\bibnamefont{Sch{\"u}lke}},
  \bibinfo{journal}{Phys. Rev. Lett.} \textbf{\bibinfo{volume}{95}},
  \bibinfo{pages}{157401} (\bibinfo{year}{2005}).

\bibitem[{\citenamefont{Huotari et~al.}(2008)\citenamefont{Huotari, Sternemann,
  Sch{\"u}lke, Sturm, Lustfeld, Sternemann, Volmer, Gusarov, M{\"u}ller, and
  Monaco}}]{HSS08}
\bibinfo{author}{\bibfnamefont{S.}~\bibnamefont{Huotari}},
  \bibinfo{author}{\bibfnamefont{C.}~\bibnamefont{Sternemann}},
  \bibinfo{author}{\bibfnamefont{W.}~\bibnamefont{Sch{\"u}lke}},
  \bibinfo{author}{\bibfnamefont{K.}~\bibnamefont{Sturm}},
  \bibinfo{author}{\bibfnamefont{H.}~\bibnamefont{Lustfeld}},
  \bibinfo{author}{\bibfnamefont{H.}~\bibnamefont{Sternemann}},
  \bibinfo{author}{\bibfnamefont{M.}~\bibnamefont{Volmer}},
  \bibinfo{author}{\bibfnamefont{A.}~\bibnamefont{Gusarov}},
  \bibinfo{author}{\bibfnamefont{H.}~\bibnamefont{M{\"u}ller}},
  \bibnamefont{and} \bibinfo{author}{\bibfnamefont{G.}~\bibnamefont{Monaco}},
  \bibinfo{journal}{Phys. Rev. B} \textbf{\bibinfo{volume}{77}},
  \bibinfo{pages}{in press} (\bibinfo{year}{2008}).

\bibitem[{\citenamefont{Krotscheck}(2000)}]{polish}
\bibinfo{author}{\bibfnamefont{E.}~\bibnamefont{Krotscheck}},
  \bibinfo{journal}{J. Low Temp. Phys.} \textbf{\bibinfo{volume}{119}},
  \bibinfo{pages}{103} (\bibinfo{year}{2000}).

\bibitem[{\citenamefont{Jackson et~al.}(1982)\citenamefont{Jackson, Lande, and
  Smith}}]{parquet1}
\bibinfo{author}{\bibfnamefont{A.~D.} \bibnamefont{Jackson}},
  \bibinfo{author}{\bibfnamefont{A.}~\bibnamefont{Lande}}, \bibnamefont{and}
  \bibinfo{author}{\bibfnamefont{R.~A.} \bibnamefont{Smith}},
  \bibinfo{journal}{Physics Reports} \textbf{\bibinfo{volume}{86}},
  \bibinfo{pages}{55} (\bibinfo{year}{1982}).

\bibitem[{\citenamefont{Morse and Feshbach}(1953)}]{MF1}
\bibinfo{author}{\bibfnamefont{P.~M.} \bibnamefont{Morse}} \bibnamefont{and}
  \bibinfo{author}{\bibfnamefont{H.}~\bibnamefont{Feshbach}},
  \emph{\bibinfo{title}{Methods of Theoretical Physics}},
  vol.~\bibinfo{volume}{I} (\bibinfo{publisher}{McGraw-Hill},
  \bibinfo{address}{New York - Toronto - London}, \bibinfo{year}{1953}).

\bibitem[{\citenamefont{Krotscheck and Clark}(1979)}]{CBF2}
\bibinfo{author}{\bibfnamefont{E.}~\bibnamefont{Krotscheck}} \bibnamefont{and}
  \bibinfo{author}{\bibfnamefont{J.~W.} \bibnamefont{Clark}},
  \bibinfo{journal}{Nucl. Phys. A} \textbf{\bibinfo{volume}{328}},
  \bibinfo{pages}{73} (\bibinfo{year}{1979}).

\bibitem[{\citenamefont{Krotscheck}(2002)}]{KroTrieste}
\bibinfo{author}{\bibfnamefont{E.}~\bibnamefont{Krotscheck}}, in
  \emph{\bibinfo{booktitle}{Introduction to Modern Methods of Quantum
  Many--Body Theory and their Applications}}, edited by
  \bibinfo{editor}{\bibfnamefont{A.}~\bibnamefont{Fabrocini}},
  \bibinfo{editor}{\bibfnamefont{S.}~\bibnamefont{Fantoni}}, \bibnamefont{and}
  \bibinfo{editor}{\bibfnamefont{E.}~\bibnamefont{Krotscheck}}
  (\bibinfo{publisher}{World Scientific}, \bibinfo{address}{Singapore},
  \bibinfo{year}{2002}), vol.~\bibinfo{volume}{7} of
  \emph{\bibinfo{series}{Advances in Quantum Many--Body Theory}}, pp.
  \bibinfo{pages}{267--330}.

\bibitem[{\citenamefont{Ceperley and Alder}(1980)}]{CeperleyGFMC}
\bibinfo{author}{\bibfnamefont{D.~M.} \bibnamefont{Ceperley}} \bibnamefont{and}
  \bibinfo{author}{\bibfnamefont{B.~J.} \bibnamefont{Alder}},
  \bibinfo{journal}{Phys. Rev. Lett.} \textbf{\bibinfo{volume}{45}},
  \bibinfo{pages}{566} (\bibinfo{year}{1980}).

\bibitem[{\citenamefont{Casulleras and Boronat}(2000)}]{BoronatHe3}
\bibinfo{author}{\bibfnamefont{J.}~\bibnamefont{Casulleras}} \bibnamefont{and}
  \bibinfo{author}{\bibfnamefont{J.}~\bibnamefont{Boronat}},
  \bibinfo{journal}{Phys. Rev. Lett.} \textbf{\bibinfo{volume}{84}},
  \bibinfo{pages}{3121} (\bibinfo{year}{2000}).

\bibitem[{\citenamefont{Krotscheck}(1984)}]{annals}
\bibinfo{author}{\bibfnamefont{E.}~\bibnamefont{Krotscheck}},
  \bibinfo{journal}{Ann. Phys. (NY)} \textbf{\bibinfo{volume}{155}},
  \bibinfo{pages}{1} (\bibinfo{year}{1984}).

\bibitem[{\citenamefont{Krotscheck}(1977)}]{EKVar}
\bibinfo{author}{\bibfnamefont{E.}~\bibnamefont{Krotscheck}},
  \bibinfo{journal}{Phys. Rev. A} \textbf{\bibinfo{volume}{15}},
  \bibinfo{pages}{397} (\bibinfo{year}{1977}).

\bibitem[{\citenamefont{Jackson and Smith}(1987)}]{parquet5}
\bibinfo{author}{\bibfnamefont{A.~D.} \bibnamefont{Jackson}} \bibnamefont{and}
  \bibinfo{author}{\bibfnamefont{R.~A.} \bibnamefont{Smith}},
  \bibinfo{journal}{Phys. Rev. A} \textbf{\bibinfo{volume}{36}},
  \bibinfo{pages}{2517} (\bibinfo{year}{1987}).

\bibitem[{\citenamefont{Apaja et~al.}(1997)\citenamefont{Apaja, Halinen,
  Halonen, Krotscheck, and Saarela}}]{bosegas}
\bibinfo{author}{\bibfnamefont{V.}~\bibnamefont{Apaja}},
  \bibinfo{author}{\bibfnamefont{J.}~\bibnamefont{Halinen}},
  \bibinfo{author}{\bibfnamefont{V.}~\bibnamefont{Halonen}},
  \bibinfo{author}{\bibfnamefont{E.}~\bibnamefont{Krotscheck}},
  \bibnamefont{and} \bibinfo{author}{\bibfnamefont{M.}~\bibnamefont{Saarela}},
  \bibinfo{journal}{Phys. Rev. B} \textbf{\bibinfo{volume}{55}},
  \bibinfo{pages}{12925} (\bibinfo{year}{1997}).

\bibitem[{\citenamefont{Campbell et~al.}(1996)\citenamefont{Campbell, Folk, and
  Krotscheck}}]{lowdens}
\bibinfo{author}{\bibfnamefont{C.~E.} \bibnamefont{Campbell}},
  \bibinfo{author}{\bibfnamefont{R.}~\bibnamefont{Folk}}, \bibnamefont{and}
  \bibinfo{author}{\bibfnamefont{E.}~\bibnamefont{Krotscheck}},
  \bibinfo{journal}{J. Low Temp. Phys.} \textbf{\bibinfo{volume}{105}},
  \bibinfo{pages}{13} (\bibinfo{year}{1996}).

\bibitem[{\citenamefont{Holas}(1986)}]{HolasReview}
\bibinfo{author}{\bibfnamefont{A.}~\bibnamefont{Holas}},
  \emph{\bibinfo{title}{Strongly Coupled Plasma Physics}}
  (\bibinfo{publisher}{Plenum Press}, \bibinfo{address}{New York},
  \bibinfo{year}{1986}), vol. \bibinfo{volume}{154}, pp.
  \bibinfo{pages}{463--482}.

\bibitem[{\citenamefont{Giuliani and Vignale}(2005)}]{VigGiulBook}
\bibinfo{author}{\bibfnamefont{G.}~\bibnamefont{Giuliani}} \bibnamefont{and}
  \bibinfo{author}{\bibfnamefont{G.}~\bibnamefont{Vignale}},
  \emph{\bibinfo{title}{Quantum Theory of the Electron Liquid}}
  (\bibinfo{publisher}{Cambridge University Press},
  \bibinfo{address}{Cambridge}, \bibinfo{year}{2005}).

\bibitem[{\citenamefont{Campbell and
  Krotscheck}(2010{\natexlab{b}})}]{QFS09_He4}
\bibinfo{author}{\bibfnamefont{C.~E.} \bibnamefont{Campbell}} \bibnamefont{and}
  \bibinfo{author}{\bibfnamefont{E.}~\bibnamefont{Krotscheck}},
  \bibinfo{journal}{J. Low Temp. Phys.} \textbf{\bibinfo{volume}{158}},
  \bibinfo{pages}{226} (\bibinfo{year}{2010}{\natexlab{b}}).

\bibitem[{\citenamefont{Glyde}(1994)}]{GlydeBook}
\bibinfo{author}{\bibfnamefont{H.}~\bibnamefont{Glyde}},
  \emph{\bibinfo{title}{Excitations in liquid and solid helium}}
  (\bibinfo{publisher}{Oxford University Press}, \bibinfo{address}{Oxford},
  \bibinfo{year}{1994}).

\bibitem[{\citenamefont{Albergamo et~al.}(2007)\citenamefont{Albergamo,
  Verbeni, Huotari, Vank\'{o}, and Monaco}}]{Albergamo}
\bibinfo{author}{\bibfnamefont{F.}~\bibnamefont{Albergamo}},
  \bibinfo{author}{\bibfnamefont{R.}~\bibnamefont{Verbeni}},
  \bibinfo{author}{\bibfnamefont{S.}~\bibnamefont{Huotari}},
  \bibinfo{author}{\bibfnamefont{G.}~\bibnamefont{Vank\'{o}}},
  \bibnamefont{and} \bibinfo{author}{\bibfnamefont{G.}~\bibnamefont{Monaco}},
  \bibinfo{journal}{Phys. Rev. Lett.} \textbf{\bibinfo{volume}{99}},
  \bibinfo{pages}{205301/1} (\bibinfo{year}{2007}).

\bibitem[{\citenamefont{Schmets and Montfrooij}(2008)}]{Albergamo_comment}
\bibinfo{author}{\bibfnamefont{A.~J.~M.} \bibnamefont{Schmets}}
  \bibnamefont{and}
  \bibinfo{author}{\bibfnamefont{W.}~\bibnamefont{Montfrooij}},
  \bibinfo{journal}{Phys. Rev. Lett.} \textbf{\bibinfo{volume}{100}},
  \bibinfo{pages}{239601} (\bibinfo{year}{2008}).

\bibitem[{\citenamefont{Albergamo et~al.}(2008)\citenamefont{Albergamo,
  Verbeni, Huotari, Vank\'{o}, and Monaco}}]{Albergamo_reply}
\bibinfo{author}{\bibfnamefont{F.}~\bibnamefont{Albergamo}},
  \bibinfo{author}{\bibfnamefont{R.}~\bibnamefont{Verbeni}},
  \bibinfo{author}{\bibfnamefont{S.}~\bibnamefont{Huotari}},
  \bibinfo{author}{\bibfnamefont{G.}~\bibnamefont{Vank\'{o}}},
  \bibnamefont{and} \bibinfo{author}{\bibfnamefont{G.}~\bibnamefont{Monaco}},
  \bibinfo{journal}{Phys. Rev. Lett.} \textbf{\bibinfo{volume}{100}},
  \bibinfo{pages}{239602} (\bibinfo{year}{2008}).

\bibitem[{\citenamefont{Pines}(Nov. 1981)}]{PinesPhysToday}
\bibinfo{author}{\bibfnamefont{D.}~\bibnamefont{Pines}},
  \bibinfo{journal}{Physics Today} \textbf{\bibinfo{volume}{34}},
  \bibinfo{pages}{106} (\bibinfo{year}{Nov. 1981}).

\bibitem[{\citenamefont{Mishra et~al.}(1983)\citenamefont{Mishra, Brown, and
  Pethick}}]{PethickMass}
\bibinfo{author}{\bibfnamefont{V.~K.} \bibnamefont{Mishra}},
  \bibinfo{author}{\bibfnamefont{G.~E.} \bibnamefont{Brown}}, \bibnamefont{and}
  \bibinfo{author}{\bibfnamefont{C.~J.} \bibnamefont{Pethick}},
  \bibinfo{journal}{J. Low Temp. Phys.} \textbf{\bibinfo{volume}{52}},
  \bibinfo{pages}{379} (\bibinfo{year}{1983}).

\bibitem[{\citenamefont{Brown et~al.}(1982)\citenamefont{Brown, Pethick, and
  Zaringhalam}}]{ZaringhalamMass}
\bibinfo{author}{\bibfnamefont{G.~E.} \bibnamefont{Brown}},
  \bibinfo{author}{\bibfnamefont{C.~J.} \bibnamefont{Pethick}},
  \bibnamefont{and}
  \bibinfo{author}{\bibfnamefont{A.}~\bibnamefont{Zaringhalam}},
  \bibinfo{journal}{J. Low Temp. Phys.} \textbf{\bibinfo{volume}{48}},
  \bibinfo{pages}{349} (\bibinfo{year}{1982}).

\bibitem[{\citenamefont{Friman and Krotscheck}(1982)}]{Bengt}
\bibinfo{author}{\bibfnamefont{B.~L.} \bibnamefont{Friman}} \bibnamefont{and}
  \bibinfo{author}{\bibfnamefont{E.}~\bibnamefont{Krotscheck}},
  \bibinfo{journal}{Phys. Rev. Lett.} \textbf{\bibinfo{volume}{49}},
  \bibinfo{pages}{1705} (\bibinfo{year}{1982}).

\bibitem[{\citenamefont{Krotscheck and Springer}(2003)}]{he3mass}
\bibinfo{author}{\bibfnamefont{E.}~\bibnamefont{Krotscheck}} \bibnamefont{and}
  \bibinfo{author}{\bibfnamefont{J.}~\bibnamefont{Springer}},
  \bibinfo{journal}{J. Low Temp. Phys.} \textbf{\bibinfo{volume}{132}},
  \bibinfo{pages}{281} (\bibinfo{year}{2003}).

\bibitem[{\citenamefont{Kwong}(1982)}]{NaiThesis}
\bibinfo{author}{\bibfnamefont{N.-H.} \bibnamefont{Kwong}}, Ph.D. thesis,
  \bibinfo{school}{California Institute of Technology} (\bibinfo{year}{1982}).

\bibitem[{\citenamefont{Panholzer et~al.}(2010)\citenamefont{Panholzer,
  B{\"o}hm, Holler, and Krotscheck}}]{QFS09_exch}
\bibinfo{author}{\bibfnamefont{M.}~\bibnamefont{Panholzer}},
  \bibinfo{author}{\bibfnamefont{H.~M.} \bibnamefont{B{\"o}hm}},
  \bibinfo{author}{\bibfnamefont{R.}~\bibnamefont{Holler}}, \bibnamefont{and}
  \bibinfo{author}{\bibfnamefont{E.}~\bibnamefont{Krotscheck}},
  \bibinfo{journal}{J. Low Temp. Phys.} \textbf{\bibinfo{volume}{158}},
  \bibinfo{pages}{135} (\bibinfo{year}{2010}).

\bibitem[{\citenamefont{Boronat et~al.}(2003)\citenamefont{Boronat, Casulleras,
  Grau, Krotscheck, and Springer}}]{2dmass}
\bibinfo{author}{\bibfnamefont{J.}~\bibnamefont{Boronat}},
  \bibinfo{author}{\bibfnamefont{J.}~\bibnamefont{Casulleras}},
  \bibinfo{author}{\bibfnamefont{V.}~\bibnamefont{Grau}},
  \bibinfo{author}{\bibfnamefont{E.}~\bibnamefont{Krotscheck}},
  \bibnamefont{and} \bibinfo{author}{\bibfnamefont{J.}~\bibnamefont{Springer}},
  \bibinfo{journal}{Phys. Rev. Lett.} \textbf{\bibinfo{volume}{91}},
  \bibinfo{pages}{085302} (\bibinfo{year}{2003}).

\bibitem[{\citenamefont{Aziz et~al.}(1987)\citenamefont{Aziz, McCourt, and
  Wong}}]{AzizII}
\bibinfo{author}{\bibfnamefont{R.~A.} \bibnamefont{Aziz}},
  \bibinfo{author}{\bibfnamefont{F.~R.~W.} \bibnamefont{McCourt}},
  \bibnamefont{and} \bibinfo{author}{\bibfnamefont{C.~C.~K.}
  \bibnamefont{Wong}}, \bibinfo{journal}{Molec. Phys.}
  \textbf{\bibinfo{volume}{61}}, \bibinfo{pages}{1487} (\bibinfo{year}{1987}).

\bibitem[{\citenamefont{Greywall}(1986)}]{GRE86}
\bibinfo{author}{\bibfnamefont{D.~S.} \bibnamefont{Greywall}},
  \bibinfo{journal}{Phys. Rev. B} \textbf{\bibinfo{volume}{33}},
  \bibinfo{pages}{7520} (\bibinfo{year}{1986}).

\bibitem[{\citenamefont{de~Bruyn~Ouboter and Yang}(1986)}]{OuY}
\bibinfo{author}{\bibfnamefont{R.}~\bibnamefont{de~Bruyn~Ouboter}}
  \bibnamefont{and} \bibinfo{author}{\bibfnamefont{C.~N.} \bibnamefont{Yang}},
  \bibinfo{journal}{Physica} \textbf{\bibinfo{volume}{144B}},
  \bibinfo{pages}{127} (\bibinfo{year}{1986}).

\bibitem[{\citenamefont{Moroni et~al.}(1992)\citenamefont{Moroni, Ceperley, and
  Senatore}}]{MCS92a}
\bibinfo{author}{\bibfnamefont{S.}~\bibnamefont{Moroni}},
  \bibinfo{author}{\bibfnamefont{D.~M.} \bibnamefont{Ceperley}},
  \bibnamefont{and} \bibinfo{author}{\bibfnamefont{G.}~\bibnamefont{Senatore}},
  \bibinfo{journal}{Phys. Rev. Lett.} \textbf{\bibinfo{volume}{69}},
  \bibinfo{pages}{1837} (\bibinfo{year}{1992}).

\bibitem[{\citenamefont{Caupin et~al.}(2008)\citenamefont{Caupin, Boronat, and
  Andersen}}]{CaupinJordi}
\bibinfo{author}{\bibfnamefont{F.}~\bibnamefont{Caupin}},
  \bibinfo{author}{\bibfnamefont{J.}~\bibnamefont{Boronat}}, \bibnamefont{and}
  \bibinfo{author}{\bibfnamefont{K.~H.} \bibnamefont{Andersen}},
  \bibinfo{journal}{J. Low Temp. Phys.} \textbf{\bibinfo{volume}{152}},
  \bibinfo{pages}{108} (\bibinfo{year}{2008}).

\bibitem[{\citenamefont{Moroni et~al.}(1995)\citenamefont{Moroni, Ceperley, and
  Senatore}}]{MCS95}
\bibinfo{author}{\bibfnamefont{S.}~\bibnamefont{Moroni}},
  \bibinfo{author}{\bibfnamefont{D.~M.} \bibnamefont{Ceperley}},
  \bibnamefont{and} \bibinfo{author}{\bibfnamefont{G.}~\bibnamefont{Senatore}},
  \bibinfo{journal}{Phys. Rev. Lett.} \textbf{\bibinfo{volume}{75}},
  \bibinfo{pages}{689} (\bibinfo{year}{1995}).

\bibitem[{\citenamefont{Singwi and Tosi}(1981)}]{SiT81}
\bibinfo{author}{\bibfnamefont{K.~S.} \bibnamefont{Singwi}} \bibnamefont{and}
  \bibinfo{author}{\bibfnamefont{M.~P.} \bibnamefont{Tosi}},
  \bibinfo{journal}{Solid State Phys.} \textbf{\bibinfo{volume}{36}},
  \bibinfo{pages}{177} (\bibinfo{year}{1981}).

\bibitem[{\citenamefont{Iwamoto et~al.}(1984)\citenamefont{Iwamoto, Krotscheck,
  and Pines}}]{IKP84}
\bibinfo{author}{\bibfnamefont{N.}~\bibnamefont{Iwamoto}},
  \bibinfo{author}{\bibfnamefont{E.}~\bibnamefont{Krotscheck}},
  \bibnamefont{and} \bibinfo{author}{\bibfnamefont{D.}~\bibnamefont{Pines}},
  \bibinfo{journal}{Phys. Rev. B} \textbf{\bibinfo{volume}{29}},
  \bibinfo{pages}{3936} (\bibinfo{year}{1984}).

\bibitem[{\citenamefont{Boehm et~al.}({2009})\citenamefont{Boehm, Holler,
  Krotscheck, and Panholzer}}]{Camerino2d}
\bibinfo{author}{\bibfnamefont{H.~M.} \bibnamefont{Boehm}},
  \bibinfo{author}{\bibfnamefont{R.}~\bibnamefont{Holler}},
  \bibinfo{author}{\bibfnamefont{E.}~\bibnamefont{Krotscheck}},
  \bibnamefont{and}
  \bibinfo{author}{\bibfnamefont{M.}~\bibnamefont{Panholzer}},
  \bibinfo{journal}{{JOURNAL OF PHYSICS A-MATHEMATICAL AND THEORETICAL}}
  \textbf{\bibinfo{volume}{{42}}} (\bibinfo{year}{{2009}}), ISSN
  \bibinfo{issn}{{1751-8113}}, \bibinfo{note}{{International Conference on
  Strongly Coupled Coulomb Systems, Camerino, ITALY, JUL 29-AUG 02, 2008}}.

\bibitem[{\citenamefont{Sturm and Gusarov}(2000)}]{StG00}
\bibinfo{author}{\bibfnamefont{K.}~\bibnamefont{Sturm}} \bibnamefont{and}
  \bibinfo{author}{\bibfnamefont{A.}~\bibnamefont{Gusarov}},
  \bibinfo{journal}{Phys. Rev. B} \textbf{\bibinfo{volume}{62}},
  \bibinfo{pages}{16474 } (\bibinfo{year}{2000}).

\bibitem[{\citenamefont{Corradini et~al.}(1998)\citenamefont{Corradini, Sole,
  Onida, , and Palummo}}]{CdSOP98}
\bibinfo{author}{\bibfnamefont{M.}~\bibnamefont{Corradini}},
  \bibinfo{author}{\bibfnamefont{R.~D.} \bibnamefont{Sole}},
  \bibinfo{author}{\bibfnamefont{G.}~\bibnamefont{Onida}}, , \bibnamefont{and}
  \bibinfo{author}{\bibfnamefont{M.}~\bibnamefont{Palummo}},
  \bibinfo{journal}{Phys. Rev. B} \textbf{\bibinfo{volume}{57}},
  \bibinfo{pages}{14569} (\bibinfo{year}{1998}).

\bibitem[{\citenamefont{Greywall}(1983)}]{GRE83}
\bibinfo{author}{\bibfnamefont{D.~S.} \bibnamefont{Greywall}},
  \bibinfo{journal}{Phys. Rev. B} \textbf{\bibinfo{volume}{27}},
  \bibinfo{pages}{2747} (\bibinfo{year}{1983}).

\bibitem[{\citenamefont{Krotscheck and Ristig}(1974)}]{Mistig}
\bibinfo{author}{\bibfnamefont{E.}~\bibnamefont{Krotscheck}} \bibnamefont{and}
  \bibinfo{author}{\bibfnamefont{M.~L.} \bibnamefont{Ristig}},
  \bibinfo{journal}{Phys. Lett. A} \textbf{\bibinfo{volume}{48}},
  \bibinfo{pages}{17} (\bibinfo{year}{1974}).

\end{thebibliography}

\end{document}